\documentclass[titlepage,a4paper,12pt]{report}
\usepackage{amsmath}
\usepackage{amsfonts}
\usepackage{amssymb}
\usepackage{amsthm}

\usepackage{dsfont}
\usepackage[utf8]{inputenc}
\usepackage{latexsym}
\usepackage{color}
\usepackage{graphicx}
\usepackage{graphics,color}
\usepackage{epsfig}
\usepackage{times}
\usepackage{pifont}
\usepackage{fancybox}
\usepackage{setspace}
\usepackage{fancyhdr}
\usepackage{comment}
\usepackage{epic}
\usepackage{appendix}

\usepackage[usenames,dvipsnames]{pstricks}
\usepackage{graphicx}
\usepackage{epsfig}
\usepackage{amssymb, amsmath,amsfonts}

\usepackage{cite}

\def\half{\frac{1}{2}}

\def\NN{\mathbb{N}}

\def\PP{\mathbb{P}}
\def\EE{\mathbb{E}}

\def\one{\mathbb{I}}

\def\tr{{\rm Tr}}
\def\dim{{\rm dim}}

\newcommand{\ben}{\begin{enumerate}}
\newcommand{\een}{\end{enumerate}}
\newcommand{\be}{\begin{equation}}
\newcommand{\ee}{\end{equation}}
\newcommand{\bes}{\begin{equation*}}
\newcommand{\ees}{\end{equation*}}
\newcommand{\bea}{\begin{eqnarray}}
\newcommand{\eea}{\end{eqnarray}}
\newcommand{\beas}{\begin{eqnarray*}}
\newcommand{\eeas}{\end{eqnarray*}}
\newcommand{\begth}{\begin{theorem}}
\newcommand{\enth}{\end{theorem}}
\newcommand{\blem}{\begin{lemma}}
\newcommand{\elem}{\end{lemma}}
\newcommand{\bpr}{\begin{proof}}
\newcommand{\epr}{\end{proof}}
\newtheorem{corollary}{Corollary}
\newcommand{\non}{\nonumber}

\newcommand{\uk}{\underline{k}}
\newcommand{\uj}{\underline{j}}

\newcommand{\ul}{\underline}
\def\disp{\displaystyle}

\def\EE{\mbox{\bf{E}}}

\newtheorem{theorem}{Theorem}[chapter]
\newtheorem{lemma}{Lemma}[section]

\newtheorem{remark}{Remark}

\numberwithin{equation}{section}

\begin{document}
\newcommand{\bra}[1]{\langle #1|}
\newcommand{\ket}[1]{|#1\rangle}

\markboth{Ciara Morgan} {The information-carrying capacity of certain quantum channels}

\begin{titlepage}
\begin{center}
{\large THE INFORMATION-CARRYING CAPACITY OF CERTAIN QUANTUM CHANNELS}\\

\vspace{10mm}

{\bf Ciara Morgan}\\

\vspace{20mm}

{\scshape The thesis is submitted to\\
University College Dublin\\
for the degree of PhD\\
in the College of\\
Engineering, Mathematical and Physical Sciences}\\

\vspace{10mm}

{\it January 2010}

\vspace{15mm}

Based on research conducted in the\\
Dublin Institute for Advanced Studies\\
and the School of Mathematical Sciences, \\ University College Dublin\\
{\footnotesize {\it (Head of School: Dr. Miche\'al \'O Searc\'oid )}}

\vspace{10mm}

under the supervision of\\
{\bf Prof. Tony Dorlas \, \& \, Prof. Joe  Pul\'e}

\end{center}

\end{titlepage}

\pagenumbering{roman}

\tableofcontents

\chapter*{} \emph{It is a very sad thing that nowadays there is so little useless information.}\\
-Oscar Wilde.
\chapter*{} \emph{Ba mhaith liom an tr\'achtas seo a tiomn\'u do mo mh\'athair, Helen} \\
\emph{(I dedicate this thesis to my mother, Helen)}
\chapter*{Acknowledgements} I thank my supervisor Tony Dorlas for giving me the opportunity to carrying out research in quantum information theory. I am grateful for his support and encouragement.
I thank Matthias Christandl for reading and examining the thesis and for his comments and suggestions.
I would also like to acknowledge and thank Andreas Winter for numerous interesting discussions over the last number of years, from which I learned many interesting and useful things, and also for the invitation to visit the University of Bristol and the National University of Singapore. I thank Nilanjana Datta for her helpful comments. I would also like to thank Christopher King for his encouragement.

I am very fortunate to have such a close and supportive family. I have dedicated this thesis to my Mom, who
instilled in me a love of learning and who taught me the value of perseverance.
My Dad, Brian, and Uncle Kieron, though no longer with us, are a constant source of inspiration to me.
I thank my brother Brian and his wife Niamh for their steadfast support and for their help with the execution of the thesis. I really could not have wished for a more caring and protective brother. I also thank the O' Reilly and Whale families for their encouragement.

My time as a PhD student has been very enjoyable. I am fortunate to have made wonderful friends both in DIAS and in UCD and I would like to thank them for their camaraderie. I would especially like to thank Sin\'ead Keegan for her advice and support. I also thank my oldest and closest friend Jonathan Tynan.

Above all, I could not have completed this thesis without the constant friendship, encouragement, love and support of my partner Mark Whale.
Is tusa mo chro\'i.

\chapter*{Abstract} In this thesis we consider the classical capacity of certain quantum channels, that is, the
maximum rate at which classical information, encoded as quantum states, can be transmitted reliably over a
quantum channel.

We first concentrate on the \emph{product-state} capacity of a particular
quantum channel, that is, the capacity which is achieved by encoding the output
states from a source into codewords comprising of states taken from ensembles of
non-entangled (i.e. separable) states and sending them  over copies of the
quantum channel. Using the ``single-letter" formula proved by Holevo
\cite{Hol98} and Schumacher and Westmoreland \cite{SW97} we obtain the
product-state capacity of the qubit quantum amplitude-damping channel, which is
determined by a transcendental equation in a single real variable and can be
solved numerically. We demonstrate that the product-state capacity of this
channel can be achieved using a minimal ensemble of \emph{non-orthogonal} pure
states. We also consider the \emph{generalised} amplitude-damping channel and
show that the technique used to calculate the product-state capacity for the ``traditional'' amplitude damping channel also
holds for this channel.

In the following chapter we consider the \emph{classical capacity} of two quantum channels with
memory namely, a periodic channel with quantum depolarising channel
branches and a convex combination of quantum channels. The classical capacity is
defined as the limit of the capacity of a channel, using a block of states which
are permitted to be entangled over $n$ channel uses and divided by $n$, as $n$
tends to infinity.

We prove that the classical capacity  for each of the
classical memory channels mentioned above is, in fact, equal to the respective
product-state capacities. For those channels this means that the classical
capacity is achieved without the use of entangled input-states. We also
demonstrate that the method used in the proof of the classical capacity of a
periodic channel with depolarising channels does not hold for a periodic channel
with \emph{amplitude-damping} channel branches. This is due to the fact that, unlike
the depolarising channel, the maximising ensemble for a qubit amplitude-damping
channel is not the same for all amplitude-damping channels.

We also investigate the product-state capacity of a convex combination of two
memoryless channels, which was shown in \cite{DD07Per} to be given by the
supremum of the minimum of the corresponding Holevo quantities, and we show in
particular that the product-state capacity of a convex combination of a
depolarising and an amplitude-damping channel, is not equal to the minimum of
their product-state capacities.

Next we introduce the channel coding theorem for memoryless quantum channels,
providing a known proof \cite{Winter99} for the strong converse of the theorem. We then consider the
strong converse to the channel coding theorem for a periodic quantum channel.


\chapter*{Notation} \begin{center}
\begin{tabular}{|l|l|}
  \hline
  Symbol & Interpretation \\
  \hline\hline
  $A^*$ & Conjugate transpose of $A$ \\ \hline
  $\tr(A)$ & Trace of operator $A$ \\ \hline
  $\big| \big| A \big| \big|_1$ & Trace-norm of operator $A$, given by $\tr \sqrt{A^*A}$ \\ \hline
  $\log$ & Binary logarithm \\ \hline
  $\ln$ & Natural logarithm \\ \hline
  $H(p)$ & Shannon entropy of probability distribution $p$\\ \hline
  $\ket{\psi}$ & Complex vector \\ \hline
  $\rho$ & Quantum state (density operator)\\ \hline
  $S(\rho)$ & von Neumann entropy of state $\rho$\\ \hline
  $\mathcal{N}$ & Classical channel \\ \hline
  $\Phi$ & Quantum channel \\ \hline
  $\mathcal{B}(\mathcal{H})$ & Set of quantum states on Hilbert space $\mathcal{H}$\\ \hline
  $\chi(\{ p_j, \rho_j \})$ & Holevo quantity with respect to the ensemble $\{ p_j, \rho_j \}$\\ \hline
  $\chi^*(\Phi)$ & Product-state capacity of $\Phi$, given by $\max_{\{ p_j, \rho_j \}} \chi(\{ p_j, \Phi(\rho_j) \})$\\ \hline
\end{tabular}
\end{center} 

\newpage

\setcounter{page}{1}
\pagenumbering{arabic}

\pagestyle{fancy}
\fancyhf{}

\fancyhead[LO]{\rightmark}
\fancyfoot[CO]{\thepage}

\chapter{Introduction} Informally, the capacity of a channel can be considered to be a measure of the
channels usefulness for sending information faithfully from source (or sender)
to receiver.
The capacity of a quantum channel, can be thought of as a measure of the
closeness of that channel to the quantum identity channel, which itself sends
quantum information with perfect fidelity. Throughout the thesis we concentrate
on the
case where \emph{classical} messages (or output from a classical information
source) are encoded into quantum states and
sent over quantum channels.

Classical information can be encoded into different
types of quantum states, i.e. orthogonal or non-orthogonal, separable or
entangled. Note that the latter is a purely quantum mechanical phenomenon.  We
are interested in the type of states and ensembles which
achieve the capacity of certain quantum channels and pay particular attention to the
capacity of noisy quantum channels with \emph{memory}, i.e. channels which have correlations
between successive uses.

We first introduce the concept of classical information entropy and
classical channel capacity (see \cite{CoverThomas} and \cite{Khinchin}, for
example). We do so because a great deal of what has been
achieved in the field of quantum information theory to has date been inspired by
results in classical information theory, most notably Claude Shannon's seminal
article
\cite{Shannon48} on classical channel capacity, published in 1948.
The brief review of Shannon's work on channel capacities is also justified in order to demonstrate that not all of his results on classical channels have been
successfully generalised to the quantum setting.
Unlike classical channels, there are a number of different types of \emph{quantum} channel capacities, namely, the classical capacity, quantum capacity and the private capacity.
These capacities have not yet been fully resolved.
Moreover, entangled input states, mentioned above, have recently been shown to improve the classical capacity of quantum
channels. Hastings \cite{Hastings09}, building on a result by Hayden and Winter
\cite{HW08}, recently presented a violation of one of the longest standing
conjectures in quantum information theory, namely the additivity conjecture
involving the Holevo quantity \cite{Hol98, SW97}.
This counterexample implies that the conjectured formula for the classical capacity of a quantum
channel is disproved and that a simple ``single-letter" formula for the capacity
remains to be discovered.
The classical capacity of a quantum channel can therefore only be determined asymptotically.
The question of whether this is an {\em intrinsic} property of the classical capacity of a quantum channel or
whether there is some missing element which has not yet been understood remains open.

Smith and Yard \cite{SY08} also recently proved the non-additivity of the quantum channel
capacity, disproving the operational interpretation of the additivity the of
quantum capacity of quantum channels, by showing that two channels, each with
zero quantum capacity, when used together can give rise to a non-zero capacity. This is known as ``superactivation" of channel capacity.
Cubitt, Chen and Harrow \cite{CCH09} have also demonstrated a similar result for the \emph{zero-error classical capacity} of a quantum channel.
Smith and Smolin \cite{SS09} and Li, Winter, Zou and Guo \cite{LWZG09} have
proved the non-additivity of the private capacity for a family of quantum
channels.

\section{Classical information theory}

In information theory, entropy measures the amount of uncertainty in the state
of a system before measurement. Shannon entropy measures the entropy of a
variable associated with a classical probability distribution. More formally,
the Shannon entropy of a random variable $X$ with probability distribution
$p(x)$ is given by $H(X) = -\sum_x p(x) \log\, p(x)$. As entropy is measured in
bits, $\log$ is taken to the base $2$, and $0\log(0) = 0$.

Mutual information, $H(X:Y)$, measures the amount of information two
random variables $X$ and $Y$ have in common,
\be\label{mutInfo}
H(X:Y) = H(X) + H(Y) - H(X,Y),
\ee
where, $H(X,Y)$ is the joint entropy~\cite{NC}.

Let $\mathcal{I}$ and $\mathcal{O}$ be the respective input and output alphabets
for a classical channel and let $X_n$ and $Y_n$ both be sequences of random variables such that $x \in \mathcal{I}$ and
$y \in \mathcal{O}$. A channel can be described in terms of the conditional
probabilities $p\left(y|x \right)$ i.e. the probabilities of obtaining different outcomes,
$y$, given the input variable $x$.


\subsection{Shannon's noisy channel coding theorem}

The capacity of a classical channel $\mathcal{N}$ provides a limit on the number of classical bits which can be transmitted reliably per channel use. The direct part of the classical channel coding theorem \cite{Shannon48} states that using $n$ copies of the channel, $M$ bits of information can be sent reliably over the channel at a rate $R=\frac{M}{n}$ if and only if $R \leq C$ in the asymptotic limit.

The strong converse of the channel coding theorem states that if the rate at which classical information is transmitted over a classical channel exceeds the capacity of the channel, i.e. if $R>C$, then the probability of decoding the information correctly goes to zero in the number of channel uses.

The capacity of a noisy classical channel, $\mathcal{N}$, is given by the maximum of
the mutual information obtained over all possible input
distributions, $p(x)$, for $X_n$,
\be
\mathcal{C}(\mathcal{N}) = \max_{p(x)}\, H(X:Y),
\ee
where $H(X:Y)$ is given by Equation \ref{mutInfo}.

The first proof of Shannon's coding theorem is due to Feinstein \cite{Feinstein}.

\section{Quantum information theory}

Quantum communication promises to allow unconditionally secure communication
\cite{BB84}. Techniques to protect quantum information from noise are therefore
of great importance.
A simple ``single-letter" formula which could be used to calculate the classical
capacity of a quantum channel, would lead to a better understanding of optimal encodings used to protect
quantum information from errors. Whether such a formula can be found remains an
open question.

The capacities of quantum channel with {\em memory}, widely considered to be
more realistic than memoryless channels, are being explored \cite{LGM09, BDF09, DM09}.

The quantum analogue of Shannon entropy is von Neumann entropy. It is defined as
follows. The entropy of a quantum state, $\rho$, is given by the von Neumann
entropy, $S(\rho) = -\tr \left( \rho \, \log
\, \rho\right)$. If $\rho$ has eigenvalues $\lambda_i$, then $S(\rho) = -
\sum_i \lambda_i \, \log(\lambda_i)$.

\subsection{Noisy quantum channel coding}

\begin{figure}[h]
\begin{center}
\scalebox{1} 
{
\begin{pspicture}(0,-1.2210938)(12.02,1.2610937)
\psframe[linewidth=0.04,dimen=outer](8.56,0.5989063)(6.78,-1.1810937)
\psframe[linewidth=0.04,dimen=outer](5.14,0.5989063)(3.38,-1.1610937)
\usefont{T1}{ptm}{m}{n}
\rput(0.8953125,-0.16109376){\Large $\rho_s$}
\usefont{T1}{ptm}{m}{n}
\rput(2.7253125,0.31890625){\Large $\mathcal{C}$}
\usefont{T1}{ptm}{m}{n}
\rput(4.2553124,-0.14109375){\Large $\rho_c$}
\usefont{T1}{ptm}{m}{n}
\rput(5.9453125,0.27890626){\Large $\Phi$}
\usefont{T1}{ptm}{m}{n}
\rput(7.6453123,-0.18109375){\Large $\rho_o$}
\usefont{T1}{ptm}{m}{n}
\rput(11.085313,-0.18109375){\Large $\rho_r$}
\usefont{T1}{ptm}{m}{n}
\rput(9.555312,0.33890626){\Large $\mathcal{D}$}
\usefont{T1}{ptm}{m}{n}
\rput(0.8565625,1.0939063){\footnotesize Source}
\usefont{T1}{ptm}{m}{n}
\rput(2.4639063,1.0739063){\footnotesize Encoding}
\usefont{T1}{ptm}{m}{n}
\rput(4.230625,1.0539062){\footnotesize Input}
\psframe[linewidth=0.04,dimen=outer](1.76,0.61890626)(0.0,-1.1410937)
\psframe[linewidth=0.04,dimen=outer](11.98,0.61890626)(10.22,-1.1410937)
\usefont{T1}{ptm}{m}{n}
\rput(5.9354687,1.0739063){\footnotesize Channel}
\usefont{T1}{ptm}{m}{n}
\rput(9.324375,1.0739063){\footnotesize Decoding}
\usefont{T1}{ptm}{m}{n}
\rput(11.131562,1.0939063){\footnotesize Receiver}
\usefont{T1}{ptm}{m}{n}
\rput(7.7190623,1.0939063){\footnotesize Output}
\psline[linewidth=0.04cm,arrowsize=0.05291667cm 2.0,arrowlength=1.4,arrowinset=0.4]{->}(5.16,-0.16109376)(6.86,-0.16109376)
\psline[linewidth=0.04cm,arrowsize=0.05291667cm 2.0,arrowlength=1.4,arrowinset=0.4]{->}(8.58,-0.16109376)(10.26,-0.16109376)
\psline[linewidth=0.04cm,arrowsize=0.05291667cm 2.0,arrowlength=1.4,arrowinset=0.4]{->}(1.76,-0.16109376)(3.46,-0.16109376)
\end{pspicture}
}
\end{center}
\caption{Transmitting classical information over a single quantum channel.}
\label{diagram}
\end{figure}
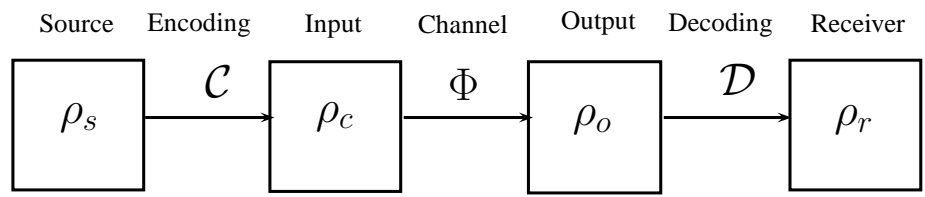

Figure \ref{diagram} depicts a quantum information transmission process from source to receiver \cite{BNS98}.

\begin{figure}[h]
\begin{center}
\scalebox{1} 
{
\begin{pspicture}(0,-2.31)(11.1,2.31)
\psframe[linewidth=0.04,dimen=outer](3.4,2.31)(1.4,-2.31)
\psframe[linewidth=0.04,dimen=outer](9.62,2.31)(7.62,-2.31)
\psframe[linewidth=0.04,dimen=outer](6.26,1.87)(4.8,1.25)
\psframe[linewidth=0.04,dimen=outer](6.24,0.89)(4.78,0.27)
\psline[linewidth=0.04cm,arrowsize=0.05291667cm 2.0,arrowlength=1.4,arrowinset=0.4]{->}(3.42,1.69)(4.86,1.69)
\usefont{T1}{ptm}{m}{n}
\rput(5.5114064,1.58){$\Phi$}
\usefont{T1}{ptm}{m}{n}
\rput(5.4914064,0.6){$\Phi$}
\usefont{T1}{ptm}{m}{n}
\rput(8.645312,0.01){\Large $E^n$}
\usefont{T1}{ptm}{m}{n}
\rput(2.3953125,0.03){\Large $C^n$}
\psline[linewidth=0.04cm,arrowsize=0.05291667cm 2.0,arrowlength=1.4,arrowinset=0.4]{->}(3.38,0.69)(4.82,0.69)
\psline[linewidth=0.04cm,arrowsize=0.05291667cm 2.0,arrowlength=1.4,arrowinset=0.4]{->}(6.24,1.67)(7.68,1.67)
\psline[linewidth=0.04cm,arrowsize=0.05291667cm 2.0,arrowlength=1.4,arrowinset=0.4]{->}(6.2,0.67)(7.64,0.67)
\psframe[linewidth=0.04,dimen=outer](6.22,-0.11)(4.76,-0.73)
\psframe[linewidth=0.04,dimen=outer](6.2,-1.09)(4.74,-1.71)
\psline[linewidth=0.04cm,arrowsize=0.05291667cm 2.0,arrowlength=1.4,arrowinset=0.4]{->}(3.38,-0.29)(4.82,-0.29)
\usefont{T1}{ptm}{m}{n}
\rput(5.4714065,-0.4){$\Phi$}
\usefont{T1}{ptm}{m}{n}
\rput(5.4514065,-1.38){$\Phi$}
\psline[linewidth=0.04cm,arrowsize=0.05291667cm 2.0,arrowlength=1.4,arrowinset=0.4]{->}(3.34,-1.29)(4.78,-1.29)
\psline[linewidth=0.04cm,arrowsize=0.05291667cm 2.0,arrowlength=1.4,arrowinset=0.4]{->}(6.2,-0.31)(7.64,-0.31)
\psline[linewidth=0.04cm,arrowsize=0.05291667cm 2.0,arrowlength=1.4,arrowinset=0.4]{->}(6.16,-1.31)(7.6,-1.31)
\psline[linewidth=0.04cm,arrowsize=0.05291667cm 2.0,arrowlength=1.4,arrowinset=0.4]{->}(0.0,1.27)(1.44,1.27)
\psline[linewidth=0.04cm,arrowsize=0.05291667cm 2.0,arrowlength=1.4,arrowinset=0.4]{->}(0.0,0.31)(1.44,0.31)
\psline[linewidth=0.04cm,arrowsize=0.05291667cm 2.0,arrowlength=1.4,arrowinset=0.4]{->}(0.02,-0.73)(1.46,-0.73)
\psline[linewidth=0.04cm,arrowsize=0.05291667cm 2.0,arrowlength=1.4,arrowinset=0.4]{->}(9.62,1.27)(11.06,1.27)
\psline[linewidth=0.04cm,arrowsize=0.05291667cm 2.0,arrowlength=1.4,arrowinset=0.4]{->}(9.62,0.31)(11.06,0.31)
\psline[linewidth=0.04cm,arrowsize=0.05291667cm 2.0,arrowlength=1.4,arrowinset=0.4]{->}(9.64,-0.73)(11.08,-0.73)
\end{pspicture}
}
\end{center}
\caption{Transmitting classical information over copies of a quantum channel.}
\label{multiChannel}
\end{figure}
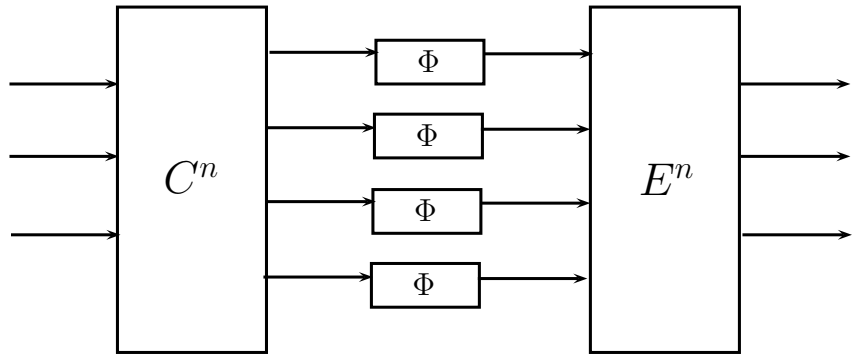

The sender encodes their message into a block of quantum states. This \emph{codeword} can then be transmitted over copies of a quantum channel, $\Phi$, see Figure \ref{multiChannel}.

The above encodings $C^n$ and $E^n$ will be important in a later chapter when we investigate the channel coding theorem for quantum channels.

\section{Thesis layout}

In Chapter 2 we introduce some mathematical preliminaries and discuss concepts
fundamental to the understanding
of quantum information theory.

We obtain a maximiser for the quantum mutual information
for classical information sent over the qubit amplitude-damping
and depolarising channels in Chapter 3. This is achieved by limiting the
ensemble of input states to antipodal states, in the calculation
of the product state capacity for the channels.
In Section \ref{ampDampSection} we evaluate the capacity of the amplitude-damping
channel and plot a graph of this capacity versus the damping parameter. We discuss the ``generalised'' amplitude damping channel in Section \ref{genAmpSect}, and show that the
approach taken to calculate the product state capacity of the conventional amplitude damping channel can also be taken for this channel.
We introduce the depolarising channel in Section \ref{Depol} and discuss the
maximising ensemble of the corresponding Holevo quantity.
The contents of this chapter have been published by T.C.
Dorlas and the Author in \cite{DM08}.

Chapter 4 is based on an article published by Dorlas together with the Author
\cite{DM09}.
Here we investigate the classical capacity of two quantum channels with memory, that is,
a periodic channel with depolarising channel branches, and
a convex combination of depolarising channels. We prove that the
capacity is additive in both cases. As a result, the channel
capacity is achieved without the use of entangled input states. In
the case of a convex combination of depolarising channels the
proof provided can be extended to other quantum channels whose
classical capacity has been proved to be additive in the
memoryless case.

In Section \ref{PerSect} we
introduce the periodic channel and investigate the product-state
capacity of the channel with depolarising channel branches. We
derive a result based on the invariance of the maximising ensemble
of the depolarising channel, which enables us to prove that the
capacity of such a periodic channel is additive. In Section \ref{CapCC} the
additivity of the classical capacity
of a convex combination of depolarising channels is proved. This is
done independently of the result derived in Section \ref{PerSect} and can
therefore be generalised to a class of other quantum channels.

In Section \ref{CC} we state the theorem proved by Datta and Dorlas in
\cite{DD07Per} concerning the product-state capacity
of a convex combination of memoryless channels and
we show that in the case of two (or more)
depolarising channels or two (or more) amplitude-damping channels,
this is in fact equal to the minimum of the individual capacities. We show
however in the case of a depolarising channel and an amplitude-damping
channel, that this is not the case.

The channel coding theorem and strong converse is discussed in Chapter 5 and we provide the
proof by Winter \cite{Winter99}. We then consider the strong converse for a periodic quantum channel, in light of
a result shown in Section \ref{ampDampBranch} for the periodic channel with amplitude damping channel branches.

Appendix \ref{Cara} states Carath\'eodory's Theorem which is used
in Chapter 3. A proof provided by N. Datta, which states that it is sufficient
to consider ensembles consisting
of at most $d^2$ pure states in the maximisation of the Holevo quantity for a
CPT map, is given in Appendix \ref{Datta}.

The proof of the product-state capacity of a periodic quantum channel, provided
by Datta and Dorlas, is given in Appendix \ref{ProofPSC}. The periodic channel
is a special case of a channel with \emph{arbitrary} Markovian noise
correlations. The proof of the formula for the product-state capacity of such a
channel, i.e. one with noise given by \emph{arbitrary} Markovian correlations,
is given by Datta and Dorlas in \cite{DD07Per}.

\chapter{Preliminaries} We begin by establishing some concepts fundamental to the study of quantum
information theory. We build on the framework of quantum information
transmission introduced in
the previous chapter by making these ideas mathematically concrete. We introduce
quantum states and channels and describe the operator sum representation, a tool
widely used
in quantum information theory to describe the behavior of an input state with a given
quantum channel.
The definition of quantum entanglement is provided and we discuss the mutual information between
different quantum states and the Holevo-Schumacher-Westmoreland theorem, which is
used to determine the
product-state capacity for classical information sent over quantum channels.
The theory of Markov processes is introduced, providing definitions necessary
for Chapter 4, where we introduce two channels which have memory that can be described using Markov chains.

\section{Quantum states and quantum channels}\label{stateChannel}

We now provide definitions for quantum states and quantum channels.

\subsection{Quantum state}

A quantum state is given by a positive semi-definite Hermitian operator of unit
trace on a Hilbert space. We now define the terms used in this
definition.

A Hilbert space $\mathcal{H}$ is a complex vector space equipped with an
inner product. Note that we will only consider finite dimensional Hilbert spaces.
An element of a Hilbert space, known as a vector, is denoted $\ket{v}$. An element of the dual space $\mathcal{H}^*$, the conjugate transpose of $\ket{v} \in \mathcal{H}$, is denoted $\bra{v}$, where
\be
\ket{v} = \left( \begin{array}{c}
 v_1 \\
\vdots \\
 v_d \\
\end{array} \right) \in \mathcal{H}, \qquad
\bra{v} = (\bar{v_1} \cdots \bar{v_d}) \in \mathcal{H}^*,
\ee
and $\bar{v}$ is the complex conjugate of $v$.
In fact, due to the correspondence between $\mathcal{H}$ and its dual space $\mathcal{H}^*$, given by the inner product, we can consider each element $\ket{v} \in \mathcal{H}$ as an element of $\mathcal{H}^*$.

The norm of the vector $\ket{u} \in \mathcal{H}$ is defined as follows,
\be
\big|\big| u \big|\big| = \sqrt{\langle{u} | u\rangle}.
\ee

Note that positive operators are a subclass of Hermitian operators, both are defined below.
An operator $A \in \mathcal{B}(\mathcal{H})$ is \emph{Hermitian} if $A^*=A$
where $A^*$
is the adjoint of the operator $A$, defined by $\langle{Au}| v\rangle = \langle{u}|{A^*v}\rangle$, for all vectors
$\ket{u}$ and $\ket{v}$ in the state space of $A$.

An operator $A \in \mathcal{B}(\mathcal{H})$ is called \emph{positive} (semi-definite), if for all
vectors $\ket{v} \neq 0 \in \mathcal{H}$, the following holds
$\langle{v}|{Av}\rangle \geq 0$.

Since a density operator is defined to be a positive (Hermitian) operator with trace one
on a Hilbert space $\mathcal{H}$ a \emph{quantum state} can be represented by a density operator.
We now define a quantum state to be a positive operator of unit trace
$\rho \in \mathcal{B}(\mathcal{H})$, where
$\mathcal{B}(\mathcal{H})$ denotes the algebra of
linear operators acting on a finite dimensional Hilbert space
$\mathcal{H}$.

\subsubsection{Pure and mixed states}

According to the first postulate of quantum mechanics, a quantum system is
completely described by its state vector, $\ket{\psi}$. A state vector is a unit
vector in the state space of the system. A system whose state is completely
known is said to be in a \emph{pure state}. The density operator for that system
is given by the projection $\rho = \ket{\psi} \bra{\psi}$. If, however a system
is in one of a number of states, then the system is said to be in a \emph{mixed
state}. If a system is in one of the states $\ket{\psi_i}$ with respective
probabilities $p_i$, then $\{p_i, \ket{\psi_i}\}$ is called an ensemble of pure
states. The corresponding density operator is given by $\rho = \sum_{i=1}^{\dim(\mathcal{H})} p_i \ket{\psi_i}
\bra{\psi_i}$.

\subsubsection{Composite quantum systems}

A composite of two quantum systems ${\cal H}$ and ${\cal K}$ can be described by the tensor product of the two Hilbert spaces $\mathcal{H} \otimes \mathcal{K}$. Note that $\dim \, \mathcal{H} \otimes \mathcal{K} = \dim{\cal H} \times \dim{\cal K}$. The state of one of the Hilbert spaces can be extracted from the product state of the two Hilbert spaces by performing the partial trace on the composite system.

The tensor product and partial trace are both defined in the following section.

\subsubsection{The tensor product and partial trace}

The tensor product of two vectors, $\ket{h} \in \mathcal{H}$ and $\ket{k} \in \mathcal{K}$,
is defined as
\be
\ket{h} \otimes \ket{k}  = \sum_i \sum_j  x_i y_j \ket{h_i} \otimes \ket{k_j},
\ee
where $\mathcal{H}$ and $\mathcal{K}$ represent Hilbert spaces with respective bases $\{ h_i\}$ and $\{ k_j\}$.

The following demonstrates how the tensor product of two vectors and two matrices is computed, respectively
\be
\left(
     \begin{array}{c}
       a \\
       b \\
     \end{array}
   \right)
   \otimes
\left(
  \begin{array}{c}
    c \\
    d \\
  \end{array}
\right)
=
\left(
  \begin{array}{c}
    a c \\
    a d \\
    b c \\
    b d \\
  \end{array}
\right),
\ee
\be
\left(
  \begin{array}{cc}
    a & b \\
    c & d \\
  \end{array}
\right)
\otimes
\left(
  \begin{array}{cc}
    e & f \\
    g & h \\
  \end{array}
\right)
=
\left(
  \begin{array}{cccc}
    a e & a f & b e & b f \\
    a g & a h & b g & b h \\
    c e & c f & d e & d f \\
    c g & c h & d g & d h \\
  \end{array}
\right).
\ee
Properties of the tensor product include,
\be
(\Phi \otimes \Psi)(\rho \otimes \rho') = \Phi(\rho) \otimes \Psi(\rho').
\ee

The trace of an operator $A \in \mathcal{B}(\mathcal{H})$, with the orthonormal basis
$\ket{\phi_i}$, is given by
\be\label{trace} \tr(A) = \sum_{i=1}^{\dim(\mathcal{H})}
\bra{\phi_i}A \ket{\phi_i}.
\ee
We now introduce partial trace. Let $\mathcal{H}$ and $\mathcal{K}$
represent two Hilbert spaces with orthonormal bases $\{ \ket{h_i}
\}_{i=1}^{\dim(\mathcal{H})}$ and  $\{ \ket{k_j}
\}_{j=1}^{\dim(\mathcal{K})}$, respectively. Let $\rho$ be a state
defined on the composite system such that $\rho \in
\mathcal{B}( \mathcal{H} \otimes \mathcal{K})$. The
state of the subsystem $\mathcal{H}$ is given by the reduced density
operator $\rho_{\mathcal{H}}$ and is defined by
\be
\bra{\Phi} \rho_{\mathcal{H}} \ket{\Psi} = \sum_j \bra{\Phi \otimes k_j} \rho \ket{\Psi \otimes k_j},
\ee
where $\ket{\Phi}$ and $\ket{\Psi}$ are states on $\mathcal{H}$.
Inserting the basis elements for $\Phi$ and $\Psi$, the partial trace can be calculated as follows
\be\label{Parttrace}
\rho_{\mathcal{H}} = \tr_{\cal{K}} (\rho) = \sum_{l,m} \sum_{j} \bra{ h_l \otimes k_j } \rho \ket{h_m \otimes k_j} \, \ket{h_l} \bra{h_m},
\ee
where $\tr_{\mathcal{K}}$ is the partial trace
operation from $\mathcal{B}(\mathcal{H} \otimes \mathcal{K})$ onto $\mathcal{B}(\mathcal{H})$.
The state of the subsystem $\mathcal{K}$ is similarly defined.

\section{Quantum channel}

A map $\Phi: \mathcal{B}(\mathcal{H})
\rightarrow  \mathcal{B}({\mathcal{K}})$ is said to be
\emph{completely positive} if
\be
\left(\Phi \otimes \mathcal{I}\right)
\left( A \right) \geq 0,
\ee
where $A \geq 0$, is an operator defined on the
Hilbert space $\mathcal{H} \otimes \hat{\mathcal{H}}$, where $\mathcal{\hat{H}}$
is an arbitrary space, ${\mathcal{K}}$ is the Hilbert space of
the output state and $\mathcal{I}$ is the identity operator.

A \emph{quantum channel} is defined as a completely positive, trace preserving map,
which maps density operators from one Hilbert space to another.

In general, when a pure input state is transmitted through a noisy quantum channel,
the output state is not known with absolute certainty i.e. it is no longer a pure state. The corresponding state is said to be mixed.

The initial state to a channel is given by the tensor product of the information
state, $\rho$, defined on the Hilbert space $\mathcal{H}$ and the initial
state of the environment $\rho_{env} = \ket{\psi_0}\bra{\psi_0}$,
assumed to be in a pure state and defined on the Hilbert space
$\mathcal{H}_{env}$.

\begin{remark}
It may be assumed that the initial state of a system is in a pure state as the
state of the system can always be defined in terms of a larger composite system
which can be chosen to be in a pure state. This is known as state purification.
\end{remark}

During transmission over a channel, the composite state, $\rho
\otimes \rho_{env}$ will evolve unitarily such that $U\left( \rho
\otimes \rho_{env} \right)U^*$, for the unitary operator $U$
on $\mathcal{H} \otimes \mathcal{H}_{env}$.

After the interaction of the channel with the state $\rho \otimes
\rho_{env}$, the output state $\rho_{out} \in
\mathcal{B}(\mathcal{H})$ is given by
\be\label{output_state} \rho_{out} = \tr_{\mathcal{H}_{env}}
\left[ U \left( \rho \otimes
\rho_{env} \right)U^* \right]. \ee
This corresponds to a measurement operator on the information state
alone after it has evolved in interaction with the environment.

Note that, a unitary operator is defined as $U^{*}U = I$.
A \emph{unital} channel is a channel where the following identity holds,
\be
\Phi(I) = I.
\ee
Next we introduce operator sum representation, a way of describing the action of a quantum channel on an input state.

\subsection{Operator sum representation}

Quantum channels can be represented using operator-sum, or Kraus
representation. By tracing over the state space of the environment,
the dynamics of the principal system alone are extracted and
represented explicitly. We will now show that this representation is a
re-statement of equation (\ref{output_state}).

Let $\{ \psi_k\}_{k=1}^{\dim(\mathcal{H}_{env})}$ denote an orthonormal basis for
the Hilbert space of the environment $\mathcal{H}_{env}$ and recall
the definition for the partial trace of an operator given by
Equation (\ref{Parttrace}).
Equation (\ref{output_state}) now becomes,
\bea
\Phi(\rho) &=& \sum_{k=1}^{\dim(\mathcal{H}_{env})} \sum_{l,l'} \,
\bra{ h_l \otimes \psi_k} U\left(\rho \otimes \ket{\psi_0}\bra{\psi_0}
\right)U^*
\ket{h_{l'} \otimes \psi_k} \ket{h_l} \bra{h_{l'}} \non \\
&=& \sum_{k=1}^{\dim(\mathcal{H}_{env})} E_k \rho E_k^*, \eea
where $E_k =  \sum_{l,l'} \bra{h_l \otimes \psi_k} U \ket{h_{l'} \otimes \psi_0} \ket{h_l} \bra{h_{l'}}$, $\{h_l\}_{l=1}^{\dim( \mathcal{H})}$ an orthonormal basis for $\cal H$ and $\rho \in \mathcal{B}(\mathcal{H})$. The operators $\{E_k
\}_{k=1}^{\dim(\mathcal{H})} \in \mathcal{B}(\mathcal{H})$ are known as operation
elements.

The operator $\Phi(\rho)$ represents the output
state, and therefore must satisfy a completeness relation such that
$\tr \ \Phi( \rho)= 1$.
Using the operation elements defined above,
\be
1 = \tr \ \Phi
( \rho ) = \tr\left( \sum_{k=1}^{\dim(\mathcal{H})} E_k \rho E_k^*
\right).
\ee
Using the cyclic property of trace,
\be  \tr\left( \sum_{k=1}^{\dim(\mathcal{H})} E_k \rho E_k^* \right) = \tr\left(
 \sum_{k=1}^{\dim(\mathcal{H})} E_k^*E_k  \rho \right) = 1 .\ee
since this must hold \emph{for all} $\rho$, it follows that $\sum_k E_k^*E_k = I$.

A map $\Phi: \mathcal{B}(\mathcal{H})
\rightarrow \mathcal{B}(\mathcal{K})$ is completely
positive and trace preserving if it admits a Kraus Representation
\cite{Kraus83}
\be\label{krausRep}
\Phi(\rho) = \sum_i E_i
\rho E_i^*, \qquad
\sum_i E_i^* E_i = I. \ee
A memoryless channel is given by a completely positive map
$\Phi:{\cal B}({\cal H}) \to {\cal B}({\cal K})$, where ${\cal B}({\cal H})$
and ${\cal B}({\cal K})$ denote the states on the input and output Hilbert spaces
${\cal H}$ and ${\cal K}$.

\section{Positive operator-valued measure}

Measurement of a quantum system can be described by a set of Hermitian matrices,
$\{ E_k\}$,
satisfying $E_k \geq 0$ and $\sum_k E_k = I$ (\cite{NC, Hayashi}). The set $\{ E_k\}$ is called a positive operator-valued measure (POVM).

If measurement, described by the set $\{ E_k\}$, is performed on a system in a state $\rho$, then the probability of obtaining outcome label $k$ is given by $\tr (\rho E_k)$.

\section{Quantum entanglement}

A state $\rho \in \mathcal{B}( \mathcal{H} \otimes \mathcal{K})$ is said to be separable if it can be written as a probabilistic mixture of product states
\be
\rho = \sum_i p_i \ket{h_i}\bra{h_i} \otimes  \ket{k_i}\bra{k_i},
\ee
where $\ket{h_i} \in \mathcal{H}$, $\ket{k_i} \in \mathcal{K}$, $\sum_i p_i = 1$, and $p_i \geq 0$. Otherwise the state is said to be an \emph{entangled} state.

Entanglement is an important resource in quantum information processing and plays an essential role in quantum teleportation, quantum cryptography, quantum computation, quantum error correction \cite{HHHH09}.

\section{Classical information over a quantum channel}

The transmission of classical information over a quantum channel is
achieved by encoding the information into quantum states. To
accomplish this, a set of possible input states $\rho_j  \in
\mathcal{B}(\mathcal{H})$ with probabilities $p_j$ are
prepared, describing the ensemble $\{p_j,\rho_j\}$. The average
input state to the channel is expressed as $\rho = \sum_j p_j
\rho_j.$ The average \emph{output} state is $\tilde{\rho} = \sum_j
p_j \Phi(\rho_j)$ \cite{SW98}.

\subsection{Holevo bound}\label{HBound}

When a state is sent through a noisy quantum channel, the amount of
information about the input state that can be inferred from the
output state is called the \emph{accessible information}. 
For any ensemble $\{ p_j, \rho_j \}$, the Holevo quantity is defined as 
\be\label{HolQ}
\chi\left( \right \{ p_j, \rho_j \left\}\right) \mathrel{\mathop:}= S\left(\sum_j p_j\, \rho_{j}\right) - \sum_j p_j \, S (\rho_{j}).
\ee
The Holevo bound \cite{Hol73} provides an
upper bound on the accessible information and is given by,
\bea\label{holevo_bound} H(X:Y) \leq \, \chi\left( \right \{ p_j, \Phi(\rho_j) \left\}\right), \eea 
where $\chi\left( \right \{ p_j, \Phi(\rho_j) \left\}\right)$ is the Holevo quantity of channel $\Phi$.
Here $X$ is the random variable
representing the classical input to the channel. The possible
values $x_j$ are mapped to states $\rho_j$ which are transformed
to $\Phi(\rho_j)$ by the channel. Then, a generalised
measurement with corresponding POVM $\{E_j\}$ allows the
determination of the output random variable $Y$ with conditional
probability distribution given by
\be \PP(Y=x_k\,|\,X=x_j) = \tr
(\Phi(\rho_j) E_k).
\ee 
The second term in the Holevo bound is often referred to as the
\emph{output entropy}. This term represents the joint entropy of
the system $\mathcal{H} \otimes \mathcal{H}_{env}$ after evolution
and can be interpreted as the final entropy of the environment,
assuming that the environment was initially in a pure state. This
is the amount of information that the information state, or
principal system, has exchanged with the environment. We therefore
want to minimize the output entropy and maximise the entropy of
the expected state. This justifies the definition of the capacity
of the channel as the maximum of the mutual information. When
quantum information is sent down a noisy quantum channel, the
output entropy is known as \emph{entropy exchange} \cite{SN96}.

Holevo \cite{Hol73} has introduced a measure of
the amount of classical information remaining in a state that has
been sent over a noisy quantum channel. The product-state capacity of a channel
is given by the maximisation of this Holevo quantity over an
ensemble of input states, and can be interpreted as the amount of
information that can be sent, in the form of product-states, reliably over the channel.

In this case the fact that the capacity is given by the maximum of the
Holevo quantity is known as the Holevo-Schumacher-Westmoreland
(HSW) Theorem.

\section{The Holevo-Schumacher-Westmoreland theorem}\label{HSWSect}

If the possible input states to a channel are prepared as product
states of the form $ \rho_1 \otimes \rho_2 \otimes \cdots$, then the
associated capacity is known as the product state capacity. This
implies that the input states have not been entangled over multiple
uses of the channel. The capacity for channels with entangled input
states has been studied \cite{Schumacher96}, and it has been shown that for
certain channels the use of entangled states can enhance the
inference of the output state and increase the capacity
(e.g. \cite{FBS96}).

The HSW theorem, proved independently by Holevo \cite{Hol98} and
by Schumacher and Westmoreland \cite{SW97}, provides an
expression to calculate the \emph{product state capacity} for
classical information sent through a quantum channel,
$\Phi$, and can be calculated using the following
expression,
\be\label{HSWEq}
\chi^*(\Phi) = \max_{\{p_j,\, \rho_{j}\}} \, \chi\left( \right \{ p_j, \Phi(\rho_j) \left\}\right)
\ee
where $S$ is the von Neumann entropy, $S(\rho) = -\tr \left( \rho
\, \log \, \rho\right)$. If $\rho$ has eigenvalues $\lambda_i$,
then $S(\rho) = - \sum_i \lambda_i \, \log(\lambda_i)$. The
capacity is given by the maximum mutual information calculated
over all ensembles $\{p_j,\rho_j\}$~\cite{NC}. Properties characterising optimal input ensembles for have been studied \cite{SW01}.

\begin{remark}
Prior to the HSW theorem, Holevo \cite{Holevo73PD} developed a formula for calculating the product-state capacity of a quantum channel where a maximisation of the accessible information is taken explicitly over both the input ensemble and over \emph{product} measurements performed on the output of the channel. It has been shown that, in certain cases, more information can be transmitted per use of a quantum channel using collective measurements rather than separable ones (see \cite{Hol98, PW91, HJSWW96}).
\end{remark}

\subsection{Optimal input enembles}\label{MaxEnsemble}

By concavity of the entropy, the maximum in Equation \ref{HSWEq} is always
attained for an
ensemble of \emph{pure states} $\rho_j$. Indeed, we can decompose each
$\rho_j$ as convex combinations of pure states: $\rho_j = \sum_k
q_k |\psi_{j,k} \rangle \langle \psi_{j,k}|$. This does not change
the first term of (\ref{HSWEq}), but by concavity of the entropy,
\be
S(\Phi(\rho_j)) \geq \sum_k q_k
S(\Phi(|\psi_{j,k}\rangle \langle \psi_{j,k}|)).
\ee
Moreover, it follows from Carath\'eodory's theorem \cite{Davies78, Eggleston, Grunbaum}, that the ensemble can always be
assumed to contain no more than $d^2$ pure states, where $d ={\rm
dim}\,({\cal H})$.

A statement of Carath\'eodory's Theorem is provided in Appendix \ref{Cara} along
with a proof by N. Datta in Appendix \ref{Datta} which states that it is
sufficient to consider ensembles consisting
of $d^2$ pure states in the maximisation of the Holevo quantity $\chi(\Phi)$,
for some CPT map $\Phi$.

Next we introduce two models for quantum memory.
\section{Models for quantum memory channels}\label{models}

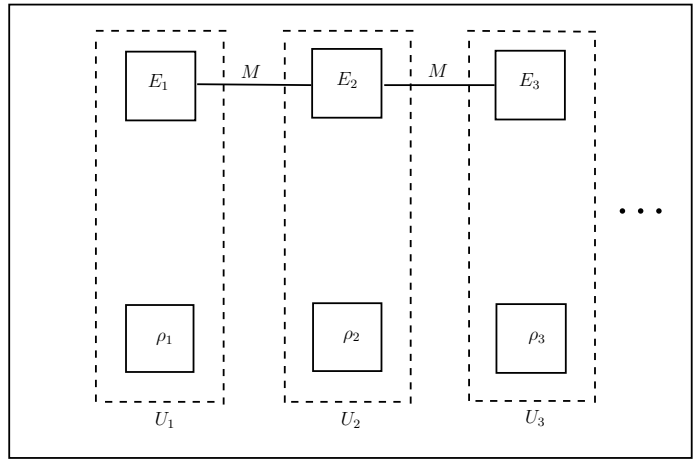
\begin{figure}[ht]
\begin{center}
\scalebox{0.6} 
{
\begin{pspicture}(0,-5.03)(15.0,5.03)
\psframe[linewidth=0.04,dimen=outer](4.12,3.99)(2.56,2.43)
\psframe[linewidth=0.04,dimen=outer](8.2,4.05)(6.64,2.49)
\psframe[linewidth=0.04,dimen=outer](12.22,4.01)(10.66,2.45)
\usefont{T1}{ptm}{m}{n}
\rput(3.2914062,3.3){$E_1$}
\usefont{T1}{ptm}{m}{n}
\rput(7.431406,3.34){$E_2$}
\usefont{T1}{ptm}{m}{n}
\rput(11.431406,3.32){$E_3$}
\psframe[linewidth=0.04,dimen=outer](4.08,-1.61)(2.56,-3.13)
\psframe[linewidth=0.04,dimen=outer](8.2,-1.57)(6.66,-3.11)
\psframe[linewidth=0.04,dimen=outer](12.24,-1.59)(10.68,-3.15)
\usefont{T1}{ptm}{m}{n}
\rput(3.4314063,-2.38){$\rho_1$}
\usefont{T1}{ptm}{m}{n}
\rput(7.5314064,-2.32){$\rho_2$}
\usefont{T1}{ptm}{m}{n}
\rput(11.591406,-2.36){$\rho_3$}
\psline[linewidth=0.04cm,linestyle=dashed,dash=0.16cm 0.16cm](1.92,4.43)(4.78,4.43)
\psline[linewidth=0.04cm,linestyle=dashed,dash=0.16cm 0.16cm](10.06,4.43)(12.88,4.41)
\psline[linewidth=0.04cm,linestyle=dashed,dash=0.16cm 0.16cm](4.72,4.19)(4.72,-3.83)
\psline[linewidth=0.04cm,linestyle=dashed,dash=0.16cm 0.16cm](1.92,4.19)(1.92,-3.77)
\psline[linewidth=0.04cm,linestyle=dashed,dash=0.16cm 0.16cm](2.12,-3.77)(4.52,-3.77)
\psline[linewidth=0.04cm,linestyle=dashed,dash=0.16cm 0.16cm](6.02,4.43)(8.88,4.43)
\psline[linewidth=0.04cm,linestyle=dashed,dash=0.16cm 0.16cm](8.82,4.21)(8.82,-3.81)
\psline[linewidth=0.04cm,linestyle=dashed,dash=0.16cm 0.16cm](6.2,-3.77)(8.6,-3.77)
\psline[linewidth=0.04cm,linestyle=dashed,dash=0.16cm 0.16cm](6.06,4.17)(6.06,-3.79)
\psline[linewidth=0.04cm,linestyle=dashed,dash=0.16cm 0.16cm](12.86,4.23)(12.86,-3.79)
\psline[linewidth=0.04cm,linestyle=dashed,dash=0.16cm 0.16cm](10.1,4.21)(10.1,-3.75)
\psline[linewidth=0.04cm,linestyle=dashed,dash=0.16cm 0.16cm](10.24,-3.75)(12.64,-3.75)
\psdots[dotsize=0.12](13.46,0.45)
\psdots[dotsize=0.12](13.86,0.45)
\psdots[dotsize=0.12](14.26,0.45)
\usefont{T1}{ptm}{m}{n}
\rput(3.4314063,-4.16){$U_1$}
\usefont{T1}{ptm}{m}{n}
\rput(7.5114064,-4.18){$U_2$}
\usefont{T1}{ptm}{m}{n}
\rput(11.551406,-4.12){$U_3$}
\psline[linewidth=0.04cm](4.14,3.25)(6.64,3.23)
\psline[linewidth=0.04cm](8.24,3.23)(10.66,3.23)
\usefont{T1}{ptm}{m}{n}
\rput(5.3214064,3.54){$M$}
\usefont{T1}{ptm}{m}{n}
\rput(9.421406,3.56){$M$}
\psframe[linewidth=0.04,dimen=outer](15.0,5.03)(0.0,-5.03)
\end{pspicture}
}
\end{center}
\caption{A model for quantum channel memory: each input state $\rho_j$ interacts with its own environment, which is itself correlated with the other environments \cite{Mancini06}.}
\label{memModel1}
\end{figure}

\begin{figure}[ht]
\begin{center}
\scalebox{.6} 
{
\begin{pspicture}(0,-4.91)(14.58,4.91)
\psframe[linewidth=0.04,dimen=outer](5.44,3.89)(3.88,2.33)
\psframe[linewidth=0.04,dimen=outer](8.62,3.93)(7.06,2.37)
\psframe[linewidth=0.04,dimen=outer](11.82,3.89)(10.26,2.33)
\usefont{T1}{ptm}{m}{n}
\rput(4.6114063,3.2){$E_1$}
\usefont{T1}{ptm}{m}{n}
\rput(7.851406,3.22){$E_2$}
\usefont{T1}{ptm}{m}{n}
\rput(11.031406,3.2){$E_3$}
\psframe[linewidth=0.04,dimen=outer](5.44,1.09)(3.88,-0.47)
\psframe[linewidth=0.04,dimen=outer](8.62,1.09)(7.06,-0.47)
\psframe[linewidth=0.04,dimen=outer](11.82,1.09)(10.26,-0.47)
\usefont{T1}{ptm}{m}{n}
\rput(4.581406,0.4){$M$}
\usefont{T1}{ptm}{m}{n}
\rput(7.8014064,0.44){$M$}
\usefont{T1}{ptm}{m}{n}
\rput(10.981406,0.4){$M$}
\psframe[linewidth=0.04,dimen=outer](5.4,-1.71)(3.88,-3.23)
\psframe[linewidth=0.04,dimen=outer](8.62,-1.69)(7.08,-3.23)
\psframe[linewidth=0.04,dimen=outer](11.84,-1.71)(10.28,-3.27)
\usefont{T1}{ptm}{m}{n}
\rput(4.751406,-2.48){$\rho_1$}
\usefont{T1}{ptm}{m}{n}
\rput(7.9514065,-2.44){$\rho_2$}
\usefont{T1}{ptm}{m}{n}
\rput(11.191406,-2.48){$\rho_3$}
\psline[linewidth=0.04cm,arrowsize=0.05291667cm 2.0,arrowlength=1.4,arrowinset=0.4]{->}(5.46,0.33)(7.1,0.33)
\psline[linewidth=0.04cm,arrowsize=0.05291667cm 2.0,arrowlength=1.4,arrowinset=0.4]{->}(8.68,0.33)(10.36,0.31)
\psline[linewidth=0.04cm,linestyle=dashed,dash=0.16cm 0.16cm](3.24,4.33)(6.1,4.33)
\psline[linewidth=0.04cm,linestyle=dashed,dash=0.16cm 0.16cm](9.66,4.31)(12.48,4.29)
\psline[linewidth=0.04cm,linestyle=dashed,dash=0.16cm 0.16cm](6.04,4.09)(6.04,-3.93)
\psline[linewidth=0.04cm,linestyle=dashed,dash=0.16cm 0.16cm](3.24,4.09)(3.24,-3.87)
\psline[linewidth=0.04cm,linestyle=dashed,dash=0.16cm 0.16cm](3.44,-3.87)(5.84,-3.87)
\psline[linewidth=0.04cm,linestyle=dashed,dash=0.16cm 0.16cm](6.44,4.31)(9.3,4.31)
\psline[linewidth=0.04cm,linestyle=dashed,dash=0.16cm 0.16cm](9.24,4.09)(9.24,-3.93)
\psline[linewidth=0.04cm,linestyle=dashed,dash=0.16cm 0.16cm](6.62,-3.89)(9.02,-3.89)
\psline[linewidth=0.04cm,linestyle=dashed,dash=0.16cm 0.16cm](6.48,4.05)(6.48,-3.91)
\psline[linewidth=0.04cm,linestyle=dashed,dash=0.16cm 0.16cm](12.46,4.11)(12.46,-3.91)
\psline[linewidth=0.04cm,linestyle=dashed,dash=0.16cm 0.16cm](9.7,4.09)(9.7,-3.87)
\psline[linewidth=0.04cm,linestyle=dashed,dash=0.16cm 0.16cm](9.84,-3.87)(12.24,-3.87)
\psdots[dotsize=0.12](13.06,0.33)
\psdots[dotsize=0.12](13.46,0.33)
\psdots[dotsize=0.12](13.86,0.33)
\psframe[linewidth=0.04,dimen=outer](2.16,1.09)(0.6,-0.47)
\usefont{T1}{ptm}{m}{n}
\rput(1.3014063,0.4){$M$}
\psline[linewidth=0.04cm,arrowsize=0.05291667cm 2.0,arrowlength=1.4,arrowinset=0.4]{->}(2.18,0.33)(3.82,0.33)
\usefont{T1}{ptm}{m}{n}
\rput(4.751406,-4.26){$U_1$}
\usefont{T1}{ptm}{m}{n}
\rput(7.931406,-4.3){$U_2$}
\usefont{T1}{ptm}{m}{n}
\rput(11.151406,-4.24){$U_3$}
\psframe[linewidth=0.04,dimen=outer](14.58,4.91)(0.0,-4.91)
\end{pspicture}
}
\end{center}
\caption{Model for quantum channel memory: correlations between each error operator and input state are determined by the relevant unitary operator and the memory state \cite{Mancini06}.}
\label{mem2}
\end{figure}
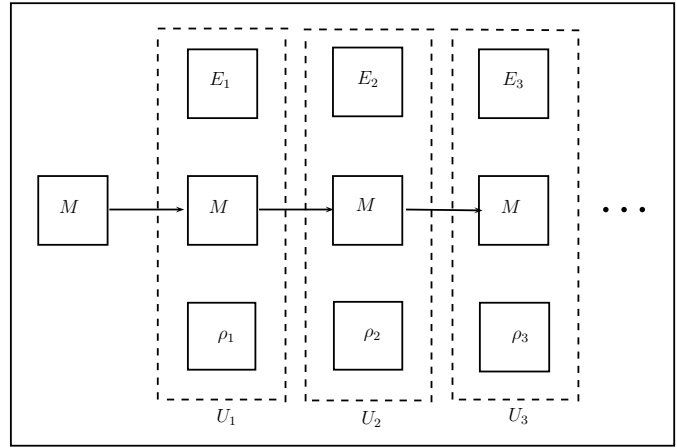

Bowen and Mancini \cite{BM04, GM04} introduced two models for quantum channels with memory. 
The model shown in Figure \ref{memModel1} depicts an interaction between each memory state $\rho_j$ and its environment $E_j$. The environments $\{E_j\}_{j=1}^n$ are correlated, which leads to a memory effect at each stage of the evolution.
In contrast to the previous model, Figure \ref{mem2} depicts the input state $\rho_j$ interacting with its own environment \emph{and} with the memory state. The error operators at each stage of the evolution are correlated, and may be determined using the relevant unitary operator and the input state. Both process will be described in the following subsections.

\subsection{Several uses of a memoryless quantum channel}

Recall that it is known that any quantum channel, described by a completely positive trace preserving (CPT) map, can be represented by a unitary operation on the input state to the channel and the initial (known) state of the environment \cite{Kraus83}.
The output state following a {\em sequence of $n$ uses} of the \emph{memoryless} channel, $\Phi$ is given by
\bea
\Phi^{(n)} \left( \rho^{(n)} \right) &=&
Tr_{E} \biggl[  U_{n,E_n} \cdots U_{1,E_1} \left( \rho^{(n)} \otimes \ket{0_{E_1} \cdots 0_{E_n}} \bra{0_{E_1} \cdots 0_{E_n}} \right) \times \non \\
&\times& U^*_{1,E_1} \cdots U^*_{n,E_n}\biggr]
\eea
where, $\rho^{(n)} \in \mathcal{H}^{\otimes n}$ is a (possibly entangled) input state codeword and \\$\rho_E = \ket{0_{E_1} \cdots 0_{E_n}} \bra{0_{E_1} \cdots 0_{E_n}}$ represents the (product) state of the environment. Note that the trace is taken over each state comprising the state of the environment.

\subsection{Several uses of a quantum channel with memory}

The action of a quantum \emph{memory} channel on a sequence of input states can be viewed in the following two ways.

\subsubsection{Model 1}
The action of the channel described by Figure \ref{memModel1} can be described as follows
\be
\Phi^{(n)} \left( \rho^{(n)} \right) =
Tr_{E} \biggl[U_{n,E_n} \cdots U_{1,E_1} \bigl( \rho^{(n)} \otimes \omega_{E} \bigr) U^*_{1,E_1} \cdots U^*_{n, E_n}\biggr]
\ee
where $\omega_{E}  = \Omega_E \rho_E \Omega_E^*$ and $\Omega_E$ is a unitary operator on $E$ which introduces correlations between the environments $E_j$. Here we are replacing the separable state $\rho_E$ introdcued in Section \ref{models}, with a correlated state $\omega_E$.

\subsubsection{Model 2}

Each input state $\rho_i$, to the channel will act with a unitary interaction on the {\em channel memory state}, denoted $\ket{M}\bra{M}$, and also an independent environment $E_i$. This process is depicted in Figure \ref{mem2}.

The output state from such a quantum memory channel can be expressed as follows
\bea
\Phi^{(n)} \left( \rho^{(n)} \right) &=&
Tr_{ME} \biggl[U_{n,ME_n} \cdots U_{1,ME_1} \bigl( \rho^{(n)} \otimes \ket{M}\bra{M}  \non \\
&\otimes& \ket{0_{E} \cdots 0_{E}} \bra{0_{E} \cdots 0_{E}} \bigr) U^*_{1,ME_1} \cdots U^*_{n,ME_n}\biggr].
\eea
Note that if the unitaries acting on the state memory and environment can be written as $U_{k,ME_k} = U_{kE_k} U_{M}$ then the memory can be traced out and we recover the memoryless channel.

Quantum channels which have \emph{Markovian} noise correlations are a class of channel which can be represented by the above model. This class of channel is of particular interest to us and is discussed below.

\section{Markov processes and channel memory}\label{Markov}

Next, we provide definitions \cite{Norris} needed to describe quantum channels with classical memory.

\subsection{Definitions and basic properties}

Let $I$ denote a countable set and let $\lambda_i = \mathbb{P}(X=i)$,
where $X$ is a random variable taking values in the state space $I$. Let $P$ denote a transition matrix, with entries labeled $p_{j|i}$.

A Markov chain is given  by a sequence of random variables $X_0, \dots, X_{n-1}$ with the following property,
\bea
\mathbb{P} (X_{n-1} = i_{n-1} | X_{n-2} = i_{n-2}, \dots, X_0 = x_0) &=& \mathbb{P} (X_{n-1} = i_{n-1} | X_{n-2} =i_{n-2}) \non \\
&=& p_{i_{n-1}|i_{n-2}}.
\eea
Equivalently, a discrete time random process denoted $X_n$ can be considered to be a Markov chain with transition matrix $P$ and initial distribution $\lambda$, if and only if the following holds for $i_0, \dots, i_{n-1} \in I$, (see Norris \cite{Norris} Theorem 1.1.1)
\be
\mathbb{P}(X_0 =i_0, X_1 = i_1, \dots , X_{n-1} = i_{n-1}) = \lambda_{i_{0}} p_{i_{1}|i_{0}} p_{i_{2}|i_{1}} \cdots p_{i_{n-1}|i_{n-2}}.
\ee
A state $j$ is said to be accessible from state $i$, and can be written $i \rightarrow j$, if there exists $n \geq 0$ such that
\be
\mathbb{P}(X_n=j | X_0=i) >0.
\ee
The state $i \in I$ is said to \emph{communicate} with state $j \in I$ if $i \rightarrow j$ and $j \rightarrow i$. This relation, denoted $i \leftrightarrow j$, partitions the state space $I$ into \emph{communicating classes}. A Markov chain is said to be \emph{irreducible} if the state space $I$ is a single class.

A state $i$ has period $L$ if any return to the state $i$ occurs in multiples of $L$ time steps, i.e.
\be
L=\gcd \{n : \mathbb{P}(X_n =i | X_0 =i) >0\}.
\ee
Next we use the concepts to describe classical memory.

\subsection{Classical memory}

A channel, of length $n$, with Markovian noise correlations can be described as follows \cite{DD07Per}
\be
\Phi^{(n)}(\rho^{(n)}) = \sum_{i_0 \dots i_{n-1}} (q_{i_{n-1}|i_{n-2}} \dots q_{i_1|i_0}) \; \lambda_{i_{0}} (\Phi_{i_0} \otimes \cdots \otimes \Phi_{i_{n-1}}) (\rho^{(n)}),
\ee
where $q_{j|i}$ denotes the elements of the transition matrix of a discrete-time Markov chain, and $\{ \lambda_i\}$ represents an invariant distribution on the Markov chain.

In later chapters we analyse two particular channels with classical memory, the periodic channel and a convex combination of memoryless channels. Both are described below.

A periodic channel can be described as follows
\be
\Phi^{\left(n \right)} \left( \rho^{\left(n \right)} \right) = \frac{1}{L} \sum_{i=0}^{L-1} \left(
\Phi_i \otimes \Phi_{i+1} \otimes \cdots\otimes \Phi_{i+n-1}
\right) \left( \rho ^{\left(n \right)} \right),
\ee
where $\Phi_i$ are CPT maps and the index is cyclic modulo the period $L$. In this case $q_{j|i} = \theta_{i,j}$, where
\be
\theta_{i,j}=
\begin{cases} 1, & \text{if $j=(i+1) \mod L$}
\\
0, &\text{otherwise.}
\end{cases}
\ee
A convex combination of product channels is defined by the following channel
\be
\Phi^{(n)}\left(\rho^{(n)} \right)= \sum_{i=1}^M \gamma_i
\,\Phi_i^{\otimes n}(\rho^{(n)}), \ee where $\gamma_i,\
(i=1,\dots,M)$ is a probability distribution over channels
\\ $\Phi_1,\dots,\Phi_M$. The action of the channel can be interpreted as follows. With probability $\gamma_i$ a given input state $\rho^{(n)} \in
{\mathcal{B}}({\cal H}^{\otimes n})$ is transmitted through one of the
memoryless channels. The corresponding Markov chain is aperiodic but not irreducible.

In this case the elements of the transition matrix are $q_{j|i} = \delta_{ij}$, i.e. the transition matrix is equal to the identity matrix. Note that Ahlswede \cite{Ahlswede68} introduced the classical version of this channel and its capacity was proved by Jacobs \cite{Jacobs62}.

\chapter{Deriving a minimal ensemble for the quantum amplitude damping \\channel} In this chapter we focus on obtaining the maximiser for classical information transmitted
in the form of product-states over a noisy quantum channel. We consider in particular the problem of determining this maximiser in the case of the amplitude damping channel.
The amplitude damping channel models the loss of energy in a system and is an example of a non-unital channel (see Section \ref{stateChannel}). The effect of the qubit amplitude damping channel on the
Bloch (Poincar\'e) sphere is to ``squash'' the sphere to the $\ket0$ pole, resulting in an ellipsoid. The Bloch sphere will be discussed in more detail in Section \ref{PSCamp} and the resulting space of the output states from an example amplitude-damping channel, using the optimal input ensemble for that channel, can be seen in Figure \ref{ellPlot} of this chapter.

It is known in general that the maximising ensemble can
always be assumed to consist of at most $d^2$ pure states if $d$
is the dimension of the state space, but we show that in the case
of the qubit amplitude-damping channel, the maximum is in fact obtained
for an ensemble of \emph{two} pure states. Moreover, these states are in
general \emph{not orthogonal}. This result is rather surprising, since
nonorthogonal quantum states cannot be distinguished with
perfect reliability.

Note that Fuchs \cite{Fuchs97} has also described a particular channel, the so-called ``splaying" channel, whose product state capacity is
maximised using an ensemble of non-orthogonal states.

\section{The amplitude-damping channel}\label{ampDampSection}

The qubit amplitude-damping channel models the loss of energy in a qubit
quantum system and is described, with error parameter $0 \leq \gamma \leq 1$, by the following
operation elements \cite{NC}
\be E_0 = \left(
\begin{array}{cc}
1 & 0  \\
0 & \sqrt{1 -\gamma}   \\
\end{array} \right), \; \;
E_1 = \left( \begin{array}{cc}
0 & \sqrt{\gamma}  \\
0 & 0   \\
\end{array} \right).\ee
Using the operation elements above, the qubit amplitude-damping channel can be expressed as follows
\be \Phi_{amp}(\rho) = E_0 \, \rho
\, E_0^* + E_1 \, \rho \, E_1^*. \ee
Note that since $E^*_0 E_0 + E^*_1 E_1 = I$, the operator $\Phi_{amp}$ is a CPT map and therefore a legitimate quantum channel.

Acting on the general qubit state $\rho$, given by \be
\rho = \left(
\begin{array}{cc} a & b \\ {\bar b} & 1-a \end{array} \right) \label{gen_state},
\ee
the output of the channel $\Phi_{amp}$ is given by \be\label{ampChannel}
\Phi_{amp}(\rho) = \left(\begin{array}{cc}
a + (1-a) \gamma & b\sqrt{1 - \gamma}  \\
\bar{b}\sqrt{1- \gamma} & (1-a)(1- \gamma)   \\
\end{array} \right).
 \ee
The amplitude-damping channel can be interpreted as follows. Evaluating
\be
E_0 \rho E_0^* =  \left(
\begin{array}{cc}
a & b \sqrt{1 - \gamma}  \\
\bar{b} \sqrt{1 - \gamma}  & (1 -\gamma)(1-a)   \\
\end{array} \right),
\ee
we can easily see that if the input state is given by $\rho = \ket{0}\bra{0}$, then the state is left unchanged by $E_0 \rho E_0^*$, with probability $1$.
However, if the input state is $\rho = \ket{1}\bra{1}$, the amplitude of the state is multiplied by a factor $1- \gamma$.

On the other hand,
\be
E_1 \rho E_1^* =  \gamma \left(
\begin{array}{cc}
1-a & 0   \\
0  & 0  \
\end{array} \right).
\ee
In this case, the input state $\rho = \ket{1}\bra{1}$ is replaced with the state $\ket{0}\bra{0}$ with probability $\gamma$.

Therefore,
\be\label{rho0}
\Phi_{amp} (\ket{0}\bra{0}) = \ket{0}\bra{0},
\ee
and
\be
\Phi_{amp} (\ket{1}\bra{1}) = \gamma \ket{0}\bra{0} + (1 - \gamma) \ket{1}\bra{1}.
\ee
The eigenvalues of $\Phi_{amp}(\rho)$ are easily found to
be \be\label{amp_eig} \lambda_{amp\pm} = \frac{1}{2} \left(1
\pm \sqrt{\left(1+2a(\gamma -1) -2\gamma \right)^2 -4|b|^2 (\gamma
-1)} \right).\ee

Next we derive the product-state capacity of the qubit amplitude damping channel.

\subsection{Product-state capacity of the qubit amplitude damping \\channel}\label{PSCamp}

Recall (Section \ref{HBound}) that the Holevo quantity for a channel $\Phi$ is defined as
\be\label{holevoEqn}
\chi(\{p_j, \Phi(\rho_j)\}) = S\left( \Phi\left(
\sum_j p_j \rho_j\right)  \right) - \sum_j p_j S(
\Phi(\rho_j) ).
\ee

In the case of the amplitude-damping channel, given by Equation (\ref{ampChannel}), the Holevo quantity is given as follows,
\bea\label{amp}
\chi(\{p_j, \Phi_{amp}(\rho_j) \}) &=& S \left[ \sum_j \left(
\begin{array}{cc}
p_j\left(a_j+(1-a_j)\gamma \right) & p_j b_j\sqrt{(1 - \gamma)}  \non \\
p_j {\bar b}_j \sqrt{(1 - \gamma)} & p_j(1-a_j)(1 - \gamma)   \\
\end{array} \right) \right]\\
&-& \sum_j p_j\, S \left(
\begin{array}{cc}
a_j+(1-a_j)\gamma & b_j\sqrt{1 - \gamma}  \\
{\bar b}_j \sqrt{1 - \gamma} & (1-a_j)(1 - \gamma)   \\
\end{array} \right).
\eea
To maximise Equation (\ref{amp}) we will show that the first term is
increased, while \\keeping the second term fixed, if each pure state
$\rho_j$ is replaced by itself and its mirror image in the real
$b$-axis. In other words, replacing $ \rho_j = \left(
\begin{array}{cc}
a_j & b_j  \\
\bar{b}_j & (1-a_j)   \\
\end{array} \right)$ associated with probability $p_j$, with the states
$\rho_j = \left( \begin{array}{cc}
a_j & b_j  \\
\bar{b_j} & (1-a_j)   \\
\end{array} \right)$ and $\rho_j' = \left(\begin{array}{cc}
a_j & -b_j  \\
-\bar{b}_j & (1-a_j)  \\
\end{array} \right) $, both with probabilities $p_j/2$, will increase Equation (\ref{amp}).

The Bloch sphere (also known as the Poincar\'e sphere) is a representation of the state space of a two-level quantum system i.e. a qubit. Pure states (corresponding to the extreme points in the (convex) set of density operators) are given by points on the surface of the sphere.

An example of antipodal states is shown in Figure \ref{antipodalStates} below, which depicts a two-dimensional cross-section of the Block sphere. Here, the state $\rho_1$ has been replaced by itself and $\rho_1'$, similarly for $\rho_2'$.

\begin{figure}[ht]
          \centerline{
             \epsfig{file=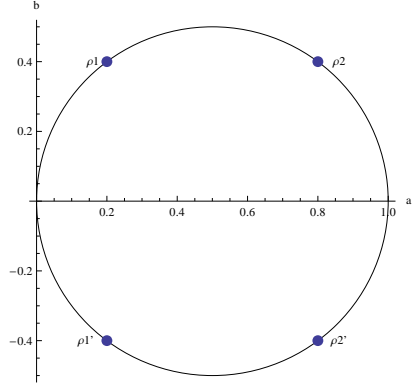, scale=.5}}
             \caption{An example of two pairs of antipodal pure states.}
             \label{antipodalStates}
       \end{figure}

As remarked above, the maximum in Equation (\ref{HSWEq}) can be
achieved by a pure state ensemble of (at most) $d^2$ states, where
$d$ is the dimension of the input to the channel.
In general, the states $\rho_j$ must lie inside the Poincar\'e sphere
\be
\left(a-\half\right)^2 + |b|^2 \leq \frac{1}{4}
\ee
and so, the
pure states will lie on the boundary
\begin{equation}\label{eqncircle} \left(a -
\frac{1}{2}\right)^2 + |b|^2 = \frac{1}{4} \\
\Rightarrow |b|^2 = a(1-a).
\end{equation}
We first show that the second term in Equation (\ref{holevoEqn})
remains unchanged when the states are replaced in the way
described above. Indeed, since the eigenvalues (\ref{amp_eig})
depend only on $|b|$, we have $S\left(\Phi(\rho_j)\right) =
S\left(\Phi(\rho_j')\right)$ and therefore,
\be
\sum_j \frac{p_j}{2} \left[ S(\Phi(\rho_j)) +
S(\Phi(\rho_j'))\right] = \sum_j p_j
S\left( \Phi(\rho_j)\right). \ee
Secondly, we prove that $S\left( \Phi\left( \sum_j p_j
\rho_j\right) \right)$ is in fact increased by replacing each
state with itself and its mirror image, each with half their
original weight. Indeed, as $S$ is a concave function,
\be
S\left(\sum_j \frac{p_j}{2} \Phi(\rho_j + \rho_j') \right)
\geq \frac{1}{2} \left[ S \left(\Phi \left(\sum_j p_j
\rho_j\right) \right) +   S \left( \Phi\left(\sum_j p_j
\rho_j'\right) \right) \right]
\ee
and again, since
$S\left(\Phi(\sum_j p_j \rho_j)\right) =
S\left(\Phi(\sum_j p_j \rho_j')\right)$,
\be S\left(\sum_j
\frac{p_j}{2} \Phi\left(\rho_j + \rho_j'\right) \right)
\geq S\left( \Phi \left( \sum_j p_j \rho_j \right)
\right).
\ee
We can conclude that the first term in Equation
(\ref{holevoEqn}) is increased with the second term fixed if each
state $\rho_j$ is replaced by itself together with its mirror
image.

\begin{remark} It follows,in particular, that we can assume from
now on that all $b_j$ are real as the average state $\sum_j
\frac{p_j}{2} (\rho_j + \rho'_j)$ has zero off-diagonal elements,
whereas the eigenvalues of $\Phi(\rho_j)$ only depend on
$|b_j|$. \end{remark}

\subsubsection{Convexity of the output entropy}

We concentrate here on proving that, in the case of the amplitude-
damping channel, the second term in the equation for the
Holevo quantity is convex as a function of the parameters
$a_j$, when $\rho_j$ is taken to be a pure state, i.e. $b_j =
\sqrt{a_j(1-a_j)}$. Thus $S\left( \Phi(\rho_j)\right)$ is a
function of one variable only, i.e. $S(a_j)$. It is given by, \be
S(a_j) = S (\Phi_{amp}(\rho_{a_j})), \ee where \be \rho_a =
\left(\begin{array}{cc}
a  & \sqrt{a(1-a)} \\
\sqrt{a(1-a)} & 1-a
\end{array} \right), \ee that is, \be \sigma(a) =
\Phi_{amp}(\rho_a) = \left(\begin{array}{cc}
a + (1-a) \gamma & \sqrt{a(1-a)}\sqrt{1 - \gamma}  \\
\sqrt{a(1-a)}\sqrt{1- \gamma} & (1-a)(1- \gamma)
\end{array} \right). \ee
Inserting $b^2 = a(1-a)$ into Equation (\ref{amp_eig}) the eigenvalues for the amplitude-damping channel can be written as \be
\lambda_{amp\pm} = \frac{1}{2} \left(1 \pm
\sqrt{1-4\gamma(1-\gamma)(1-a)^2} \right). \ee
Denote \be
x=\sqrt{1-4\gamma(1-\gamma)(1-a)^2}. \ee Then \be S(a) = -
\left(\frac{1 + x}{2} \right) \log \left(\frac{1 + x}{2}\right) -
\left(\frac{1 - x}{2} \right) \log \left(\frac{1 - x}{2}\right).
\ee
We prove that $S''(a)$ is positive. A straightforward
calculation yields
\bea S''(a) \ln 2 &=&
\frac{2\gamma(1-\gamma)}{x^3} \ln \left( \frac{1+x}{1-x} \right) -
\frac{4\gamma(1-\gamma)}{x^2} \\
&=& \left( \frac{2\gamma(1-\gamma)}{x^2}\right) \left( \frac{1}{x}
\ln \left( \frac{1+x}{1-x}\right) -2 \right). \eea
Since the first
term in the above equation is positive, the problem of proving the
convexity of $S(a)$ reduces to proving that, \be \ln \left(
\frac{1+x}{1-x} \right)  \geq 2x. \ee
This is easily shown. Note that $0 \leq x \leq 1$. Both functions are plotted in Figure \ref{GreaterEqual} below. We conclude that $S''(a)$ is positive and therefore $S(a)$ is
convex.
\begin{figure}[h]
          \centerline{
             \epsfig{file=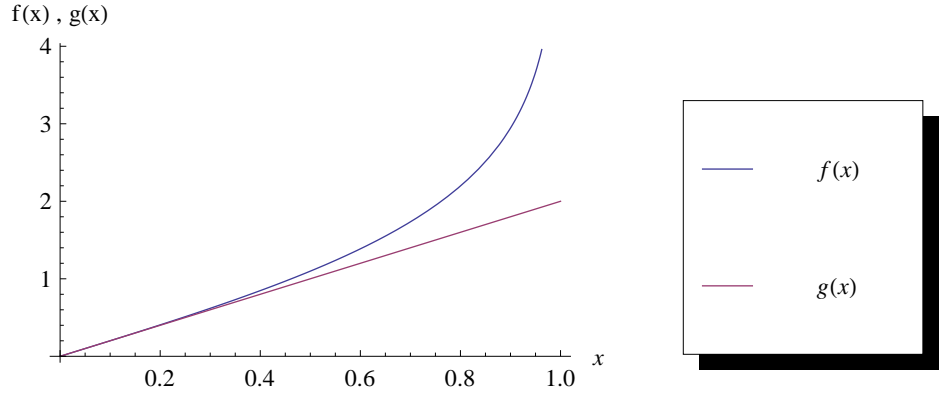, scale=1}}
             \caption{The functions $f(x) = \ln \left(
\frac{1+x}{1-x} \right)$ and $g(x) = 2x $ plotted for $0 \leq x \leq 1$.}
             \label{GreaterEqual}
       \end{figure}
Writing ${\bar \rho_a} = \sum_j p_j\, \rho_{a_j}$, with ${\bar
a} = \sum_j p_j\, a_j$ and since the entropy function $S$ is convex in $a$ we have
\bea \chi(\{p_j, \Phi_{amp}(\rho_j)\}) &=&
S(\Phi_{amp}({\bar \rho_a})) - \sum_j p_j\,S(a_j) \non \\ &\leq&
S(\Phi_{amp}({\bar \rho_a})) - S({\bar a}).
\eea
The capacity is therefore given by \be \chi^*(\Phi_{amp}) =
\max_{a \in [0,1]} \left[ S \left( \frac{1}{2} (\sigma(a) +
\sigma'(a)) \right) - S(\sigma(a)) \right], \ee where $\sigma(a)$
is given by
\be\sigma(a) =
\Phi_{amp}(\rho_a) = \left(\begin{array}{cc}
a + (1-a) \gamma & \sqrt{a(1-a)}\sqrt{1 - \gamma}  \\
\sqrt{a(1-a)}\sqrt{1- \gamma} & (1-a)(1- \gamma)
\end{array} \right),
\ee
and hence
\be
\half(\sigma(a) + \sigma(a)')
= \left(\begin{array}{cc}
a + (1-a) \gamma & 0  \\
0 & (1-a)(1- \gamma)
\end{array} \right).
\ee
We have proved that $S(a)$ is \emph{convex}. Therefore
$-S(a)$ is concave. On the other hand, it follows from the
concavity of $S$ that the first term is also a concave function of
$a$.

\begin{figure}[ht]
          \centerline{
             \epsfig{file=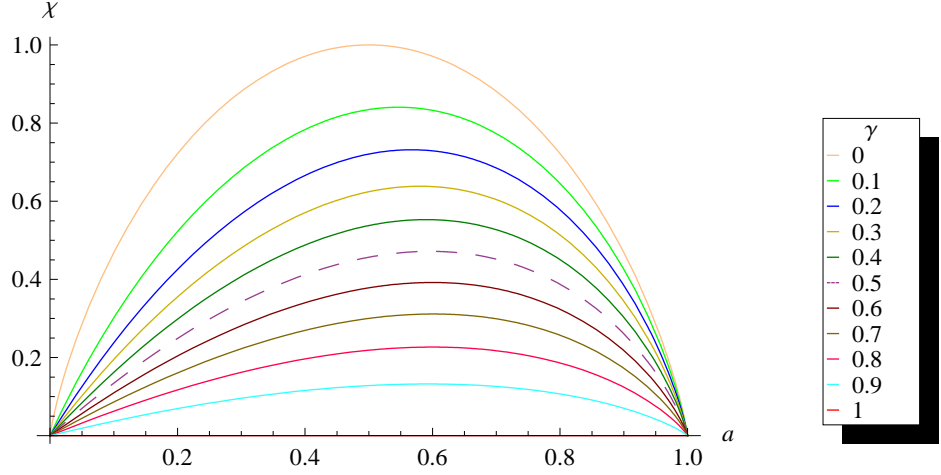, scale=1}}
             \caption{$\chi_{AD}(a)$ for $0 \leq \gamma \leq 1$ plotted over
$a$.}
             \label{chi_graphs}
       \end{figure}

It follows that $\chi_{AD}(a) = S \left( \frac{1}{2}
(\sigma(a) + \sigma'(a)) \right) - S(\sigma(a))$ is a concave
function, and its maximum is achieved at a single point. The
maximising value of $a$ is given by the transcendental equation
$\chi'_{AD}(a)=0$ and can only be computed numerically.
In  Figure \ref{chi_graphs} we plot $\chi_{AD}(a)$ as a function
of $a$.

\begin{figure}[ht]
\centerline{ \epsfig{file=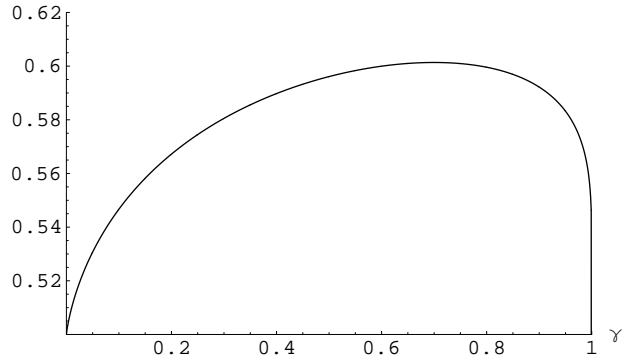, scale=.8}}
\caption{Maximising $a$'s for $0 \leq \gamma \leq 1$ for the
amplitude-damping channel.}\label{dchiPlot}
\end{figure}

The maximising $a$ for fixed $\gamma \in [0,1]$ is
plotted as a function of $\gamma$ in Figure \ref{dchiPlot}.

\begin{figure}[h]
\centerline{ \epsfig{file=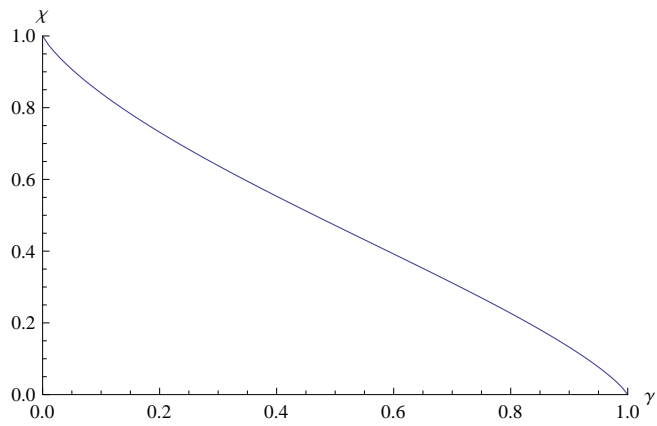, scale=.8}}
\caption{$\chi_{AD}(a_{max})$ vs. $\gamma$.}\label{maxchi}
\end{figure}

Note that $a_{\rm max} \geq 0.5$ for all $\gamma$. This is easily
proved: The determining equation is \be \chi'(a) = (1-\gamma) \ln
\frac{(1-\gamma)(1-a)}{a + \gamma (1-a)} + \frac{2 \gamma
(1-\gamma) (1-a)}{x} \ln \frac{1+x}{1-x} = 0. \label{amax} \ee
Since $\chi(a)$ is concave, the statement follows if we show that
$\chi'(0) > 0$ and \\ $\chi'(\half) > 0$. But, if $a=0$ then
$x=\sqrt{1-4\gamma (1-\gamma)} = |1-2 \gamma|$ so
\be
\chi'(0) =
(1-\gamma) \ln \frac{1-\gamma}{\gamma} + \frac{2 \gamma
(1-\gamma)}{|1-2\gamma|} \ln \frac{1+|1-2\gamma|}{1-|1-2\gamma|} =
\frac{1-\gamma}{1-2\gamma} \ln \frac{1-\gamma}{\gamma} > 0.
\ee
For $a=\half$ we have $x = \sqrt{1-\gamma+\gamma^2}$ and
\be
\chi'(0.5)
= - (1-\gamma) \ln \frac{1+\gamma}{1-\gamma} + \frac{\gamma
(1-\gamma)}{x} \ln \frac{1+x}{1-x}.
\ee
This is also positive because $x > \gamma$ and the function
\be
\frac{1}{2x} \ln
\frac{1+x}{1-x} = \frac{\tanh^{-1}(x)}{x}
\ee
is increasing. The resulting capacity is plotted in Figure \ref{maxchi}.

\subsection{Non-orthogonality of the maximising ensemble}

We have proved that the Holevo quantity for the amplitude-damping channel can be maximised using an ensemble of just \emph{two} pure states. Concentrating on the first term, the average of two mutually orthogonal states will lie in the centre of the Bloch sphere, i.e. at $a=\frac{1}{2}$.
However, we have proved that $a_{max} \geq 0.5$ for all $\gamma$. This implies that the product-state capacity of each amplitude-damping channel is achieved for an ensemble of \emph{non-orthogonal} states.

In \cite{Fuchs97}, Fuchs compares the product-state capacities for the ``splaying'' channel with both orthogonal and non-orthogonal input states, for certain values of the error parameter. We will now compare the Holevo quantity of the qubit amplitude-damping channel using orthogonal and non-orthogonal input states.

\textbf{Example:} We choose $\gamma = 0.2$, indeed any choice of parameter $\gamma \in (0,1]$ will do, and we first take $a=0.5$, representing an orthogonal ensemble. We find that the corresponding Holevo quantity is
\be
\chi_{\perp} \approx 0.720726.
\ee
Again choosing $\gamma = 0.2$ but this time solving the transcendental equation $\chi'_{AD} (a) =0$ numerically to find $a_{max} \approx 0.567214$, we get
\be
\chi^* = \chi_{non\perp} \approx 0.731645.
\ee
Since $a_{max} \approx 0.567214$, this implies that the optimal input states $\rho_{\pm} = \ket{\sigma_{\pm}}\bra{\sigma_{\pm}}$, where \be
\ket{\sigma_{\pm}} = \left(
\begin{array}{cc}
\sqrt{a_{max}}\\
\pm \sqrt{1-a_{max}}  \\
\end{array} \right)
=\left(
\begin{array}{cc}
0.753133\\
\pm 0.657862  \\
\end{array} \right)
\ee

The angle between the two states $\rho$ and $\rho'$ is approximately $83$ degrees. This demonstrates that the optimal ensemble to the qubit amplitude-damping channel, with error-parameter $\gamma=0.2$, consists of non-orthogonal states.

\begin{figure}[h]
\centerline{ \epsfig{file=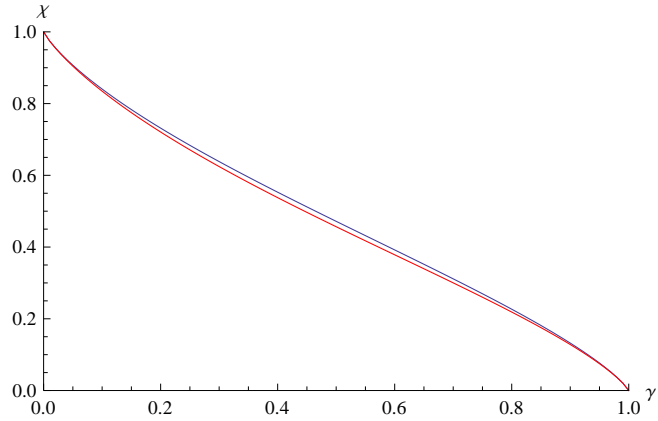, scale=.8}}
\caption{$\chi_{AD}(a_{max})$ vs. $\gamma$ in blue and $\chi_{AD}(a=0.5)$ vs. $\gamma$ in red.}\label{supchi}
\end{figure}

Figure \ref{supchi} shows that $\chi(a_{max})$, i.e. the \emph{actual} product-state capacity, differs from $\chi(0.5)$ except when $\gamma = 0$ and $\gamma = 1$. This is due to the fact that $\gamma = x$ at $\gamma = 1$ and since the error parameter $\gamma$ is at its maximum value we have $\chi(a_{max}) = 0$. At $\gamma = 0$, the first term of the Holevo quantity for the amplitude-damping channel $S\left(\Phi_{AD}(\sum_j p_j \rho_j)\right) = S \left( \begin{array}{cc}
a & 0 \\
0 & (1-a)   \\
\end{array} \right)$ which is maximised at $a=0.5$.

\begin{figure}[ht]
\centerline{ \epsfig{file=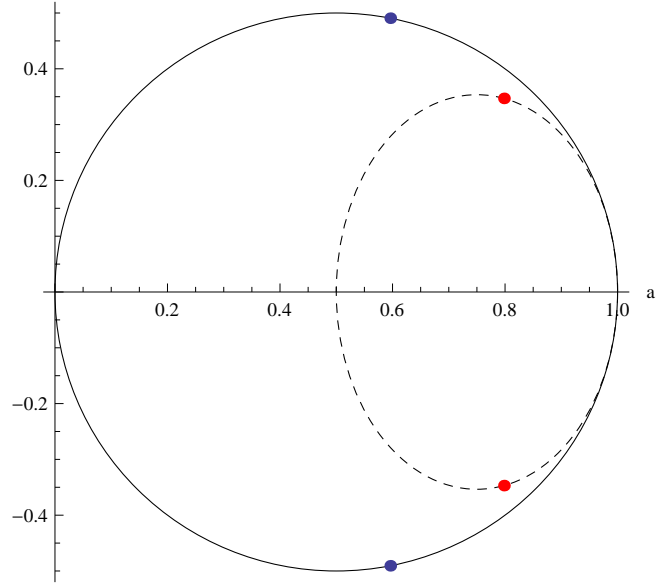, scale=.8}}
\caption{Optimal input states (blue) to the amplitude-damping channel with $\gamma = 0.5$ and the resulting output states from the channel (red).}\label{ellPlot}
\end{figure}

Figure \ref{ellPlot} below demonstrates the amplitude damping channel with $\gamma = \half$ with the optimal input states represented in blue and the corresponding output state in red.

Note that a qubit channel maps the Bloch sphere to an ellipsoid and the action of a \emph{unital} channel on the Bloch sphere results in an ellipsoid which is centered at the origin of the Bloch sphere \cite{KR01, RSW02}.

\section{The generalised amplitude-damping channel}\label{genAmpSect}

The operation elements for the \emph{generalised} amplitude-damping channel \cite{NC} are as follows
\be
E_0 = \sqrt{p}
\left(
\begin{array}{cc}
 1 & 0 \\
 0 & \sqrt{1-\gamma }
\end{array}
\right), \qquad
E_1 = \sqrt{p}
\left(
\begin{array}{cc}
 0 & \sqrt{\gamma } \\
 0 & 0
\end{array}
\right)
\ee

\be
E_2 = \sqrt{1-p}
\left(
\begin{array}{cc}
 \sqrt{1-\gamma } & 0 \\
 0 & 1
\end{array}
\right), \qquad
E_3 = \sqrt{1-p} \left(
\begin{array}{cc}
 0 & 0 \\
 \sqrt{\gamma } & 0
\end{array}
\right).
\ee
\begin{remark}
Note that we recover the traditional amplitude-damping channel for $p=1$.
\end{remark}
Therefore the generalised amplitude-damping channel, $\Phi_{GAD}$, acting on the qubit state
\be
\rho = \left(
\begin{array}{cc} a & b \\ {\bar b} & 1-a \end{array} \right),
\ee
can be written as
\bea\label{GAD}
\Phi_{GAD} (\rho) &=& E_0 \rho E_0^* + E_1 \rho E_1^* + E_2 \rho E_2^* + E_3 \rho E_3^* \non \\  \vspace{4mm}
&=& \left(
\begin{array}{cc}
 a+ p\gamma - a \gamma & \sqrt{a(1-a)}
   \sqrt{1-\gamma } \\
 \sqrt{a(1-a)} \sqrt{1-\gamma } & a (\gamma
   -1)-p \gamma +1
\end{array}
\right).
\eea
The eigenvalues of $\Phi_{GAD} (\rho)$ are,
\be
\lambda_{\pm} = \frac{1}{2} \left(1 \pm \sqrt{1 + 4 a^2 \gamma
   ^2-4 a^2 \gamma -8 a p \gamma ^2+8 a
   p \gamma +4 p^2 \gamma ^2-4 p \gamma}\right).
\ee
Again, letting
\be
x = \sqrt{1 + 4 a^2 \gamma^2-4 a^2 \gamma -8 a p \gamma ^2+8 a
   p \gamma +4 p^2 \gamma ^2-4 p \gamma},
   \ee
we have
\be
S(a) = -
\left(\frac{1 + x}{2} \right) \log \left(\frac{1 + x}{2}\right) -
\left(\frac{1 - x}{2} \right) \log \left(\frac{1 - x}{2}\right).
\ee
Now,
\be
S^{\, \prime}(a) \ln(2) = -\frac{2\gamma (1-\gamma)(p-a)}{x} \ln
\frac{1+x}{1-x} \ee
and the second derivative
\be S(a)^{\,\prime \prime} \ln(2) =  \frac{2 \gamma (1-\gamma)}{x^3}
\left\{ (1-c) \ln \frac{1+x}{1-x} - 2 x \frac{1-c-x^2}{1-x^2} \right\},
\ee
where $c=4 \gamma p (1-p)$.

Since $S''(a)$ can be shown to be positive, we can conclude that $S(a)$ is a convex function of $a$ and since the first term in the Holevo quantity is concave we deduce that $\chi(a)$ is concave. Applying the technique used for the amplitude damping channel in Section \ref{ampDampSection}, i.e. replacing each state in the ensemble by itself and it's antipode with half the original probability, we can maximise $\chi(a)$ over an ensemble containing two pure states which are specified completely, for each fixed $\gamma$ and $p$, by a \emph{single} value $a$.

\begin{figure}[ht]
          \centerline{
             \epsfig{file=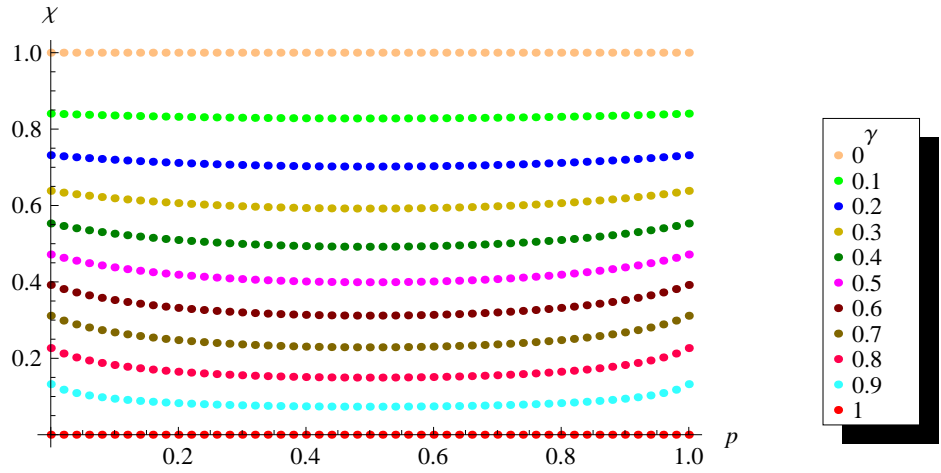, scale=1}}
             \caption{The Holevo capacity $\chi^*(p)$ for the generalised amplitude-damping channel as a function of $p$ and for various values of $0 \leq \gamma \leq 1$.}
             \label{GADFixGammaPoints}
       \end{figure}

Figure \ref{GADFixGammaPoints} depicts the Holevo capacity $\chi^*$ as a function of $p$ for values of $\gamma \in [0,1]$. We can see that $\chi(p)$ varies more at larger values of $\gamma$. This is due to the fact that $p$ only occurs with $\gamma$ in the expression for $\Phi_{GAD}$ in Equation (\ref{GAD}).
Figure \ref{GADGraphs} below shows the Holevo $\chi(a)$ quantity for fixed $\gamma=0.5$ and for $p$ fixed at various values between $0$ and $1$. The symmetry between $p$ and $1-p$ about $p=0.5$ can clearly be seen.
       \begin{figure}[ht]
          \centerline{
             \epsfig{file=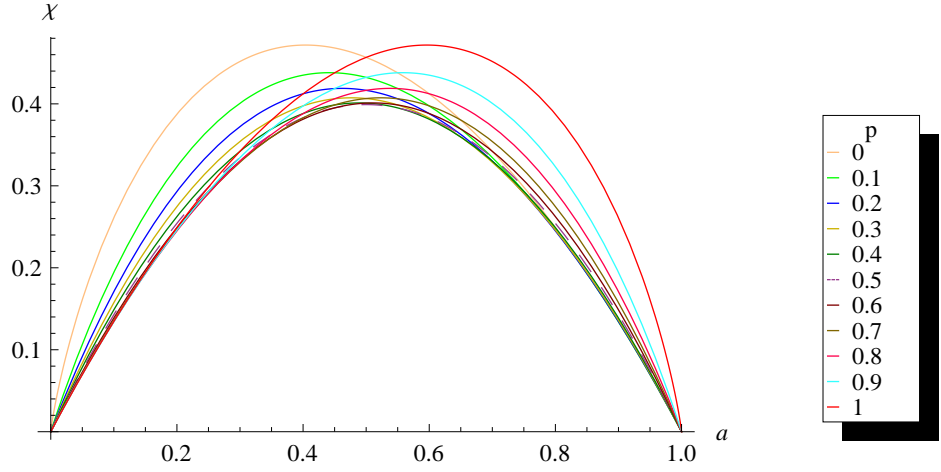, scale=1}}
             \caption{The Holevo $\chi$ quantity for the generalised amplitude damping channel, with $\gamma = 0.5$ plotted for various values of $0 \leq p \leq 1$.}
             \label{GADGraphs}
       \end{figure}

\section{The depolarising channel and the Holevo \\quantity}\label{Depol}

The state of a qubit sent through the depolarising channel is replaced by a
completely mixed state with probability $\lambda$, as follows
\be
\Delta_{\lambda}(\rho) = (1 - \lambda) \rho + \lambda
\left(\frac{I}{2} \right).\ee
We demonstrate a method of obtaining the product-state capacity of a qubit
depolarising channel using a minimal input ensemble, analogous to the above argument
for the amplitude-damping channel.
The depolarising channel acting on the qubit state $\rho = \left(
\begin{array}{cc} a & b \\ \bar{b} & 1-a \end{array} \right)$ can be written as,
\be\label{depChannel}
\Delta_{\lambda}(\rho) = \left(
\begin{array}{cc}
(1 - \lambda)a + \frac{\lambda}{2} & b(1 - \lambda)  \\
\bar{b}(1 - \lambda) & (1 - \lambda)(1-a) + \frac{\lambda}{2}   \\
 \end{array} \right). \ee
Note that for pure input states $b=\sqrt{a(1-a)}$ and the corresponding
eigenvalues are
\be\label{eig_Dep}
\Lambda_+ = \frac{\lambda}{2}, \qquad  \Lambda_- = 1 - \frac{\lambda}{2}.
\ee
For a pure input state $\rho_j$, the second term in the Holevo quantity $S \left(
\Phi(\rho_j) \right)$, given by Equation \ref{holevoEqn}, is a function of one
variable only, in other words $S(\Phi(\rho_{a_j}))= S\left( a_j \right)$, where
\be
\rho_a = \left(\begin{array}{cc}
a  & \sqrt{a(1-a)} \\
\sqrt{a(1-a)} & 1-a
\end{array} \right).
\ee
Denoting $\chi_{Dep}(\{p_j, \rho_j\})$ by $\chi_{Dep}$, the Holevo quantity for
the depolarising channel with a pure state ensemble is given by
\bea\label{chiDep}
\chi_{Dep} &=& S \left[ \sum_j \left(\begin{array}{cc}
p_j  \left( (1 - \lambda)a_j + \frac{\lambda}{2}\right)  & p_j\sqrt{a(1-a)}(1 -
\lambda)  \\
p_j\sqrt{a(1-a)}(1 - \lambda) & p_j \left( (1 - \lambda)(1-a_j) + \frac{\lambda}{2}
\right) \non  \\ 
 \end{array} \right) \right] \\
&-& \sum_j p_j \, S\left( a_j \right).
\eea
The product-state capacity of the channel is achieved by maximising Equation \\
(\ref{chiDep}). We replace each state in the manner described in Section \ref{PSCamp}.
Since the eigenvalues given in Equation (\ref{eig_Dep}) do not depend on the
state, the second term in Equation (\ref{chiDep}) is also independent of the
input state. Then, by concavity of the entropy,
the first term in Equation (\ref{chiDep}) is increased and hence Equation
(\ref{chiDep}) increases.
From the concavity of the entropy $S$, it follows that the first
term in the above equation is a concave function of $a_j$. The Holevo quantity
$\chi_{Dep}(\{p_j, \rho_j\})$ is concave in the input state $\rho_a$ and its
maximum can therefore be achieved at a single point. The Holevo quantity now
becomes
\bea\label{Dep_Quantity}
\chi_{Dep}(\{p_j, \rho_j\}) &=& S \left( \half \left( \Delta_{\lambda}\left(
\rho \right) + \Delta_{\lambda}\left( \rho' \right) \right) \right) - H\left(
\frac{\lambda}{2}\right) \non \qquad \qquad \\  &=&
H \left( (1 - \lambda)\,a + \frac{\lambda}{2}\right)
- H\left( \frac{\lambda}{2}\right)
\eea where $H(p)=-p \log (p) - (1-p) \log (1-p)$ is the binary entropy.

As the second term above is independent of the input ensemble, we concentrate on
maximising the first term to obtain the product-state capacity.
The average of any two mutually orthogonal states will lie in the center of the
Bloch sphere, in other words at $a=\half$. We therefore obtain equal
eigenvalues,  $\Lambda_0=\Lambda_1=\half$, for the first term and the entropy is
a maximum.
The product-state capacity can therefore be achieved with an ensemble
containing any pair of orthogonal pure states $\rho_a$ and $\rho_a'$ with equal
probability $\half$.

\begin{figure}[ht]
          \centerline{
             \epsfig{file=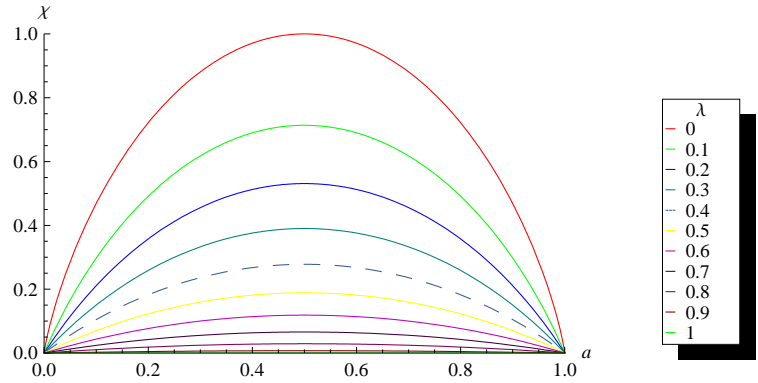, scale=.8}}
             \caption{The Holevo quantity for the depolarising
             channel as a function of $a$,
             for $\lambda \in [0,1]$.}
             \label{dep}
       \end{figure}

\begin{figure}[ht]
          \centerline{
             \epsfig{file=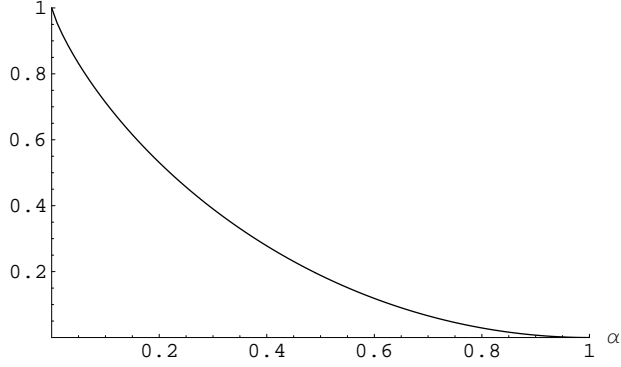, scale=.8}}
             \caption{$\chi^*(a_{max})$ vs. $\lambda$.}
             \label{maxDep}
       \end{figure}

We can therefore again maximise over a minimal ensemble of two
mirror image states with equal probability $\half$. The maximum is
clearly attained at $a = \half$ as can be seen in Figure~\ref{dep}.
The value for $\chi^*(\Delta_{\lambda})$ is shown in Figure
\ref{maxDep} as a function of $\lambda$.
The depolarising channel can therefore be considered to be rotationally
invariant.
The product-state capacity for the qubit depolarising channel is $\chi^*_{Dep} =
1 - H\left( \frac{\lambda}{2}\right)$. In fact, it was proved by King
\cite{King03}, that this is also the classical capacity of the channel.
In $d$ dimensions the capacity of the depolarising channel is
\be
\chi^*\left( \Delta_{\lambda} \right) = \log\left( d\right) - S_{min} \left(
\Delta_{\lambda} \right)
\ee
where $S_{min}$ is defined as,
\be S_{min} \left( \Delta_{\lambda} \right) = \inf_{\rho} S\left(
\Delta_{\lambda} \right). \ee

\section{Summary}

To summarise, we have introduced a method for calculating the product state capacity for the qubit amplitude damping channel using a minimal ensemble containing two antipodal states.
We analysed the behaviour of the product state capacity of the channel as a function of its error parameter and also showed that the product state capacity of the channel is achieved using non-orthogonal states.

Next we discussed the generalised amplitude damping channel and the depolarising channel. We have shown that the technique used to calculate the product state capacity of the qubit amplitude damping channel can also be used to calculate the product state capacity of these channels. 
\chapter{The classical capacity of two quantum channels with memory} \section{Introduction}

The problem of determining the classical information-carrying
capacity of a quantum channel is one which has not been fully
resolved to date. In the case where the input to the channel is
prepared in the form of non-entangled states, the classical
capacity can be determined using a simple formula, that is, the supremum of the Holevo quantity introduced in Chapter 2.

However, if entanglement between multiple uses of the channel is permitted,
then the channel capacity can only be determined asymptotically.
Moreover, the proposed additivity conjecture which promised to
provide such a ``single-letter'' formula for the classical capacity of a
channel with general input states, has recently been disproved even in the
case of \emph{memoryless} channels \cite{Hastings09}.

Note, that a channel is said to be memoryless if the noise acts
independently on each state sent over the channel. If $\Phi^{(n)} = \Phi^{\otimes n}$ is a memoryless
channel, then the product-state capacity is given by the supremum of
the Holevo quantity evaluated over all possible input state
ensembles. This is also known as the Holevo capacity
$\chi^*(\Phi)$ of the channel.

We remark that, Shor \cite{Shor04} (see also Pomeransky \cite{Pomeransky03} and Fukuda \cite{Fukuda07})
proved that the additivity conjectures involving the entanglement
of formation \cite{BDiVSWK96}, the minimum output entropy \cite{KR01}, the
strong superadditivity of the entanglement of formation and the Holevo capacity \cite{Hol98, SW97} are in fact equivalent.

The additivity conjecture of the Holevo capacity, discussed in detail in Section \ref{classCapSect}, states that the rate at which classical information can be transmitted over a quantum channel cannot be improved by sending \emph{entangled} codewords over copies of the channel.

Additivity of the Holevo capacity has been proved for unital qubit channels
\cite{KingUnital}, entanglement-breaking channels \cite{Shor02},
and the depolarising channel \cite{King03}. However Hastings \cite{Hastings09}
recently provided a counter example to the above conjecture using random unitary
channels, thereby disproving the conjecture for memoryless channels.


We are concerned here with the classical capacity of two quantum channels with \emph{memory}. Note that it has been shown that the capacity of certain channels with memory can be enhanced using entangled state inputs. See \cite{Keyl02, MP02, Ham02, HN03, MPV03, DWMcI03, CJCM04, RSGM05}.

In particular, Macchiavello and Palma \cite{MP02} proved that, although the
product-state capacity of the \emph{memoryless} depolarising channel is additive
\cite{King03}, the product-state capacity of the depolarising channel with partial noise
correlations is in fact \emph{non-additive}.
This result contributes to our motivation for considering the classical capacity
of quantum channels with memory.

Memoryless channels, i.e. channels which have no correlation between noise acting on
successive channel inputs, can also be seen to be unrealistic, since real-world
quantum channels may not exhibit this independence and correlations between
errors are common.
Noise correlations are also necessary for certain models of quantum
communication (see \cite{Bose03}, for example). These channels are known as bosonic channels and such channels have received much attention in recent years.
Moreover, the classical and quantum capacities of lossy bosonic channels were recently evaluated \cite{LGM09}. Note that the classical capacity of a bosonic memory channel with Gauss-Markov noise has also been recently investigated  \cite{SDK09}.

In this Chapter we consider the classical capacity of two
particular types of channels with \emph{memory} consisting of depolarising
channel branches, namely a periodic channel and a convex
combination of memoryless channels.

In \cite{DD07Per} Datta and Dorlas derived a general expression
for the classical capacity of a quantum channel with
\emph{arbitrary} Markovian correlated noise.

We consider two special cases of this channel, that is, a periodic channel with
depolarising channel branches and a convex combination of
memoryless channels, and we prove that the corresponding
capacities are additive in the sense that they are equal to the
product-state capacities.

To prove the additivity of the product-state capacity of the periodic channel with depolarising channel branches, we use two properties of the memoryless channel, namely the additivity of the memoryless depolarising channel \cite{King03}, and the fact that the product-state capacity of the memoryless depolarising channel can be achieved using an ensemble containing any pair of orthogonal pure states. The latter is demonstrated in Section \ref{Depol}.


We demonstrate in Section \ref{ampDampBranch} that we cannot extend the technique used in the proof for the periodic channel with depolarising channel branches to the amplitude-\\damping channel.

On the other hand, to prove the additivity of the product-state capacity of a convex combination of depolarising channels we only need the additivity of the Holevo capacity for the memoryless depolarising channel. This result can therefore be extended to include convex combinations of channels which have been proved to be additive in the memoryless case. These include the unital qubit channels \cite{KingUnital} and the entanglement-breaking channel \cite{Shor02}.

A convex combination of memoryless channels was discussed in \cite{DD07CC} and can be described by a
Markov chain which is aperiodic but not irreducible. Both channels
are examples of a channel with long-term memory.
See note on Markov chains and channel memory in Section \ref{Markov}.

We also consider the product-state capacity of a convex combination of a two depolarising channels,
two amplitude-damping channels and a
depolarising channel and an amplitude-damping channel. We show
in the case of one depolarising channel and one amplitude-damping channel that the corresponding product-state capacity, which was shown in
\cite{DD07Per} to be given by the supremum of the minimum of the
corresponding Holevo quantities, is not equal to the minimum of
their product-state capacities.

\section{Classical capacity}\label{classCapSect}

Using product-state encoding, i.e. encoding a message into
a tensor product of $n$ quantum states on a finite-dimensional
Hilbert space $\cal H$, each state can be transmitted over a
quantum channel given by a completely positive trace-preserving (CPT)
map $\Phi^{(n)}$ on ${\cal B}({\cal H}^{\otimes n})$. The
associated capacity is known as the product-state capacity of the
channel. See Section \ref{HSWSect}.

On the other hand, a block of input states could be permitted to
be entangled over $n$ channel uses. The classical capacity is
defined as the limit of the capacity for such $n$-fold entangled
states divided by $n$, as $n$ tends to infinity. If the Holevo
capacity of a memoryless channel is additive, then it is equal to
the classical capacity of that channel and there is no advantage
to using entangled input state codewords. The additivity
conjecture for the Holevo capacity of most classes of memoryless
channel remains open. However, the classical capacity of certain
memoryless quantum channels have been shown to be additive (see
\cite{KingUnital, Shor02, King03}, for example). On
the other hand, there now exists an example of a memoryless
channel for which the conjecture does not hold, see \cite{Hastings09}.

It was first shown in \cite{FBS96} that for some channels, it is
possible to gain a higher rate of transmission by sending
entangled states across multiple copies of a quantum channel. In
general, allowing both entangled \emph{input} states and output
measurements and with an unlimited number of copies of the
channel, the classical capacity of $\Phi$ is given by \cite{Hol72}
\be\label{ult} C \left( \Phi \right) = \lim_{n \rightarrow
\infty} \frac{1}{n} \, \chi^* \left( \Phi^{(n)}\right), \ee where
\be \chi^*(\Phi^{(n)}) =
\sup_{\{p_j^{(n)},\rho_{j}^{(n)}\}}\left[ S \left(\Phi^{(n)}\left(
\sum_j p_j^{(n)}\rho_j^{(n)} \right)\right) - \sum_j
p_j^{(n)}S\left(\Phi^{(n)}\left(\rho_j^{(n)}\right)\right)\right]
\ee
denotes the Holevo capacity of the channel $\Phi^{(n)}$ with
an $n$-fold input state ensemble.

The Holevo capacity of a channel $\Phi$ is said to be
\emph{additive} if the following holds for an arbitrary channel
$\Psi$
\be \chi^*\left( \Phi \otimes \Psi \right) = \chi^*\left(
\Phi \right) + \chi^*\left( \Psi \right). \ee
In particular, if we can prove that the Holevo capacity of a
particular channel is additive then \be \chi^* \left(
\Phi^{\otimes n} \right) = n \; \chi^*\left( \Phi \right), \ee
which implies that the classical capacity of the memoryless channel $\Phi^{\otimes n}$ is equal
to the product-state capacity, that is,
\be\label{classCap} C
\left( \Phi \right) = \chi^*\left( \Phi \right).
\ee
This will imply that the classical capacity of that channel cannot be
increased by entangling inputs across two or more uses of the
channel.

Here we use the additivity of the memoryless depolarising channel to prove
Equation (\ref{classCap}) for a periodic channel
with depolarising channel branches and for a convex combination of depolarising
channels (replacing the $\chi^*(\Phi)$ term in Equation (\ref{classCap}) with the appropriate formula
for calculating the product-state capacity for the particular channel).

\section{The periodic channel}\label{PerSect}

A periodic channel acting on an $n$-fold density operator has the
form \be\label{periodic} \Omega_{per}^{\left(n \right)} \left( \rho
^{\left(n \right)} \right) = \frac{1}{L} \sum_{i=0}^{L-1} \left(
\Omega_i \otimes \Omega_{i+1} \otimes \cdots\otimes \Omega_{i+n-1}
\right) \left( \rho ^{\left(n \right)} \right), \ee
where $\Omega_i$ are CPT maps and the index is cyclic modulo the period $L$.

We denote the Holevo quantity for the $i$-th component $\Omega_i$ of the channel
by \\$\chi_i(\{p_j,\rho_j\})$, i.e. \be\label{Hquantity}
\chi_i(\{p_j, \rho_j\}) = S \left( \sum_j p_j \Omega_i\left(
\rho_j \right)\right) - \sum_j p_j S\left(\Omega_i(\rho_j)\right).
\ee
Since there is a correlation between the noise affecting
successive input states to the periodic channel (\ref{periodic}),
the channel is considered to have memory and the product-state
capacity of the channel is no longer given by the supremum of the
Holevo quantity. Instead, the product -state capacity of this
channel is given by the following expression
\be\label{cap_Per}
C_p \left( \Omega \right) = \frac{1}{L} \sup_{\{p_j, \rho_j\}}
\sum_{i=0}^{L-1} \chi_i(\{p_j, \rho_j\}).
\ee
The proof of the above formula (direct part) is provided in Appendix B. The strong converse is discussed in Chapter 5.

Next, we introduce the depolarising channel and investigate the product-state
capacity of a periodic channel with depolarising channel branches.

\subsection{A periodic channel with depolarising channel
branches}\label{sect_Dep}

Recall that the $d$-dimensional quantum depolarising channel can be written as follows \be
\Delta_{\lambda} \left( \rho \right)= \lambda \rho + \frac{1 -
\lambda}{d} I \ee where $\rho \in \mathcal{B}\left( \mathcal{H}
\right)$ and $I$ is the $d \times d$ identity matrix. Note that in
order for the channel to be completely positive the parameter
$\lambda$ must lie within the range
\be - \frac{1}{d^2 -1} \leq
\lambda \leq 1.
\ee
Output states from this channel have
eigenvalues $\left(\lambda + \frac{1-\lambda}{d}\right)$ with
multiplicity $1$ and $\left(\frac{1-\lambda}{d}\right)$ with
multiplicity $d-1$.

The minimum output entropy of a channel $\Phi$  is defined by
\be
S_{min} \left( \Phi \right) = \inf_{\rho} S\left(\Phi \left( \rho
\right) \right).
\ee
Using an ensemble containing orthogonal pure states, with uniform distribution, results in the average input state $\bar{\rho} = \frac{I}{d}$ and the product-state capacity of the depolarising channel is given by \be\label{depCap}
\chi^*\left( \Delta_{\lambda} \right) = \log\left( d\right) -
S_{min} \left( \Delta_{\lambda} \right), \ee where the minimum
entropy is also attained for any set of orthonormal vector states, and
is given by
\bea
 S_{min} \left( \Delta_{\lambda} \right) &=& - \left(
\lambda + \frac{1-\lambda}{d}\right) \, \log\left( \lambda +
\frac{1-\lambda}{d} \right) \non \\ &-& \hspace{-2mm} (d-1) \left(
\frac{1-\lambda}{d}\right) \, \log
\left(\frac{1-\lambda}{d}\right).
\eea
Next we show that the product-state capacity of a periodic channel
with $L$ depolarising channel branches is given by the sum of the
maximum of the Holevo quantities of the individual depolarising
channels, in other words we show that
\be\label{conj} \frac{1}{L}
\sup_{\{p_j, \rho_j\}} \sum_{i=0}^{L-1} \chi_i(\{p_j,
\rho_j\})  = \frac{1}{L} \sum_{i=0}^{L-1}
\sup_{\{p_j, \rho_j\}} \chi_i(\{p_j, \rho_j\}). \ee
Let
$\Delta_{\lambda_1}, \Delta_{\lambda_2}, \cdots,
\Delta_{\lambda_L}$ denote $d$-dimensional depolarising channels
with respective error parameters $\lambda_1, \lambda_2, \cdots,
\lambda_L$. Using the product-state capacity given by Equation (\ref{depCap})
and since every depolarising channel can be maximised using a
single ensemble of orthogonal pure states independently of the
error parameter (as shown in Section \ref{Depol}), the right-hand side of Equation (\ref{conj}) can
be written as
\be\label{sumChi} \frac{1}{L} \sum_{i=0}^{L-1}
\sup_{\{p_j, \rho_j\}} \chi_i(\{p_j, \rho_j\})  = 1 - \frac{1}{L}  \biggl[
S_{min}\left( \Delta_{\lambda_1}\right)  + \cdots  +
S_{min}\left( \Delta_{\lambda_L} \right)\biggr]. \ee
Clearly, the left-hand side of Equation (\ref{conj}) is bounded
above by the right-hand side
\be \frac{1}{L}
\sup_{\{p_j, \rho_j\}} \sum_{i=0}^{L-1} \chi_i(\{p_j,
\rho_j\})  \leq \frac{1}{L} \sum_{i=0}^{L-1}
\sup_{\{p_j, \rho_j\}} \chi_i(\{p_j, \rho_j\}). \ee
On the other hand, choosing the
ensemble to be an orthogonal basis of states with uniform
probabilities, i.e. taking $\{p_j, \rho_j\}$ to be the optimal-ensemble, we have
\be \frac{1}{L} \sum_{i=0}^{L-1} \chi_i
(\{p_j, \rho_j\})  = 1 - \frac{1}{L} \sum_{i=0}^{L-1} S_{min}
\left( \Delta_{\lambda_i}\right). \ee We can now conclude that
Equation (\ref{conj}) holds for a periodic channel with $L$
depolarising branches of arbitrary dimension.

\subsection{The classical capacity of a periodic channel}

We now consider the classical capacity of the periodic channel,
$\Omega_{per}$, given by Equation (\ref{periodic}), where $\Omega_i =
\Delta_{\lambda_i}$ are depolarising channels with dimension $d$.
Denote by $\Psi_0^{(n)} , \dots, \Psi_{L-1}^{(n)}$ the following
product-channels \be\label{psi} \Psi_i^{(n)} = \Delta_{\lambda_i}
\otimes \dots \otimes \Delta_{\lambda_{i+n-1}}, \ee where the
index $i$ is taken modulo $L$.

We define a \emph{single use} of the periodic channel,
$\Omega_{per}$, to be the application of one of the depolarising
maps $\Delta_{\lambda_i}$. If $n$ copies of the channel are
available, then with probability $\frac{1}{L}$ one of the product
branches $\Psi_i^{(n)}$ will be applied to an $n$-fold input
state.

We aim to prove the following theorem. \begth\label{theorem1} The
classical capacity of the periodic channel $\Omega_{per}$ with
depolarising channel branches is equal to its product-state
capacity, \bes C \left( \Omega_{per} \right) = \; C_p \left(
\Omega_{per} \right) = 1-\frac{1}{L} \sum_{i=0}^{L-1}
S_{min}(\Delta_{\lambda_i}). \ees \enth

To prove Theorem \ref{theorem1} we first need a relationship
between the supremum of the Holevo quantity $\chi^*$ and the
channel branches $\Psi_i^{(n)}$. King \cite{King03} proved that
the supremum of the Holevo quantity of the product channel
$\Delta_{\lambda} \otimes \Psi$ is additive, where
$\Delta_{\lambda}$ is a depolarising channel and $\Psi$ is a
completely arbitrary channel, i.e., \be\label{addpol} \chi^*
\left( \Delta_{\lambda} \otimes \Psi \right) = \chi^* \left(
\Delta_{\lambda} \right) + \chi^*\left( \Psi \right). \ee It
follows immediately that
\bea\label{add} \chi^* \left(
\Psi_i^{(n)} \right) &=& \chi^* \left( \Delta_{\lambda_i} \right) +
\chi^* \left( \Psi_{i+1}^{(n-1)} \right) \non \\ &=& \sum_{i=0}^{L-1} \chi^*
\left( \Delta_{\lambda_i} \right) + \chi^* \left( \Psi_i^{(n-L)}
\right). \eea
Next, we use this result to prove Theorem 1.

\begin{proof}
The classical capacity of an arbitrary \emph{memoryless} quantum channel $\Omega$ is
given by
\be\label{asympCap} C \left( \Omega \right) = \lim_{ n
\rightarrow \infty}\; \frac{1}{n} \, \sup_{\{ p_j^{(n)}, \, \rho_j^{(n)} \}}
\chi \left( \left\{p_j, \Omega^{(n)} \left( \rho_j^{(n)}
\right)\right \} \right). \ee
In Section \ref{sect_Dep} we showed that
the product-state capacity of the periodic channel $\Omega_{per}$, with depolarising channel branches denoted $\Delta_{\lambda_i}$,
can be written as \be C_p \left( \Omega_{per} \right) =
\frac{1}{L} \sum_{i=0}^{L-1} \chi^* \left( \Delta_{\lambda_i}
\right). \ee Using the product channels $\Psi_i^{(n)}\left(
\rho_j^{(n)}\right)$ defined by Equations (\ref{psi}), the
periodic channel $\Omega_{per}$ can be written as
\be\label{perChannel} \Omega_{per}^{(n)} \left( \rho_j^{(n)}
\right) = \frac{1}{L} \sum_{i=0}^{L-1} \Psi_i^{(n)}\left(
\rho_j^{(n)} \right). \ee
Since it is clear that \be\label{greatEq} C \left( \Omega_{per}
\right) \geq C_p \left( \Omega_{per}\right), \ee we concentrate
on proving the inequality in the other direction.

First suppose that \be\label{assumption} C \left( \Omega_{per}
\right) \geq \frac{1}{L} \sum_{i=0}^{L-1}  \chi^* \left(
\Delta_{\lambda_i} \right) + \epsilon, \ee
for some $\epsilon
>0$. Then $\exists n_0$ such that if $n \geq n_0$, then \be\label{intermediate}
\frac{1}{n} \, \sup_{\{  p_j^{(n)}, \, \rho_j^{(n)} \}} \chi \left(
\left \{ p_j^{(n)}, \, \Omega_{per}^{(n)} \left( \rho_j^{(n)} \right) \right \} \right) \geq
\frac{1}{L} \sum_{i=0}^{L-1}  \chi^* \left( \Delta_{\lambda_i}
\right) + \frac{\epsilon}{2}. \ee The supremum in Equation
(\ref{intermediate}) is taken over all possible input ensembles \\ ${\{
\rho_j^{(n)}, \, p_j^{(n)} \}}$. Therefore, for $n \geq n_0$,
there exists an ensemble ${\{ p_j^{(n)},  \rho_j^{(n)} \}}$ such
that \be\label{ineqEn} \frac{1}{n} \,  \chi \left( \left\{ p_j,
\Omega_{per}^{(n)} \left( \rho_j^{(n)} \right) \right\} \right)
\geq \frac{1}{L} \sum_{i=0}^{L-1}  \chi^* \left(
\Delta_{\lambda_i} \right) + \frac{\epsilon}{2}. \ee  The Holevo quantity can be expressed as the average of the relative entropy of members of the ensemble with respect to the average ensemble state, \be \chi\left( \{ p_k,\, \rho_k \}\right) = \sum_k \,
p_k \, S \left( \rho_k \,\big|\big|\, \sum_k \, p_k \,
\rho_k\right), \ee where, $S\left(A \,||\,B \right) = \tr \left(
A \, \log A \right) - \tr \left(A \, \log B\right)$, represents
the relative entropy of $A$ with respect to $B$.

Vedral \cite{Vedral02} has argued that the distinguishability of quantum
states can be measured by the quantum relative entropy. Since the relative entropy is jointly convex in its arguments \cite{NC}, it
follows that the Holevo quantity of the periodic channel
$\Omega_{per}$ is also convex.

Therefore, by (\ref{perChannel}),
\be \chi\left( \left\{
p_j^{(n)}, \Omega_{per}^{(n)} \left( \rho_j^{(n)} \right)
\right\}\right) \leq \frac{1}{L} \sum_{i=0}^{L-1}  \chi \left(
\left\{ p_j^{(n)}, \Psi_i^{(n)}\left( \rho_j^{(n)} \right) \right\}
\right). \ee
Using Equation (\ref{ineqEn}) we thus have
\be
\frac{1}{L} \sum_{i=0}^{L-1}  \chi^* \left( \Delta_{\lambda_i}
\right) + \frac{\epsilon}{2} \leq \frac{1}{nL} \sum_{i=0}^{L-1}
\chi \left( \left\{ p_j^{(n)}, \Psi_i^{(n)}\left( \rho_j^{(n)}
\right) \right\} \right). \ee
It follows that there is an index
$i$ such that \be\label{ineq-i} \frac{1}{L} \sum_{i=0}^{L-1}
\chi^* \left( \Delta_{\lambda_i} \right) + \frac{\epsilon}{2} \leq
\frac{1}{n} \chi \left( \left\{ p_j^{(n)}, \Psi_i^{(n)}\left(
\rho_j^{(n)} \right) \right\} \right). \ee
But Equation
(\ref{add}) implies that \be \chi \left( \left\{ p_j^{(n)},
\Psi_i^{(n)}\left( \rho_j^{(n)} \right) \right\} \right) \leq
\frac{n}{L} \sum_{i=0}^{L-1}  \chi^* \left( \Delta_{\lambda_i}
\right). \ee
Therefore the inequality (\ref{ineq-i}) and hence
the assumption made in Equation \\(\ref{assumption}) cannot hold,
and \be\label{lesseq} C \left( \Omega_{per} \right) \leq C_p
\left( \Omega_{per} \right). \ee
The above equation together with Equation (\ref{greatEq}) yields the required
result.
\end{proof}

\section{The classical capacity of a convex combination \\ of memoryless
channels}\label{CapCC}

In \cite{DD07CC} the product-state capacity of a convex
combination of memoryless channels was determined. Given a finite
collection of memoryless channels $\Phi_1, \dots,\Phi_M$ with
common input Hilbert space $\cal H$ and output Hilbert space $\cal
K$, a convex combination of these channels is defined by the map
\be \Phi^{(n)}\left(\rho^{(n)} \right)= \sum_{i=1}^M \gamma_i
\,\Phi_i^{\otimes n}(\rho^{(n)}), \ee where $\gamma_i,\
(i=1,\dots,M)$ is a probability distribution over the channels
\\ $\Phi_1,\dots,\Phi_M$. Thus, a given input state $\rho^{(n)} \in
{\mathcal{B}}({\cal H}^{\otimes n})$ is sent down one of the
memoryless channels with probability $\gamma_i$. This introduces
long-term memory, and as a result the (product-state) capacity of
the channel $\Phi^{(n)}$ is no longer given by the supremum of the
Holevo quantity. Instead, it was proved in \cite{DD07CC} that the
product-state capacity is given by \be\label{convexcap} C_p(\Phi) =
\sup_{\{p_j,\rho_{j}\}} \left[ \bigwedge_{i=1}^M
\chi(\{p_j,\Phi_i(\rho_j)\}) \right].\ee
Again, let  $\Delta_{\lambda_i}$ be depolarising channels with parameters
$\lambda_i$, and let $\Phi_{rand}$ denote the channel whose
memoryless channel branches are given by $\Lambda_i^{(n)}$ where
\be \Lambda_i^{(n)} = \Delta_{\lambda_i}^{\otimes n}. \ee  Since
the capacity of the depolarising channel decreases with the error
parameter the product-state capacity of $\Phi_{rand}$ is given by
\be C_p (\Phi_{rand}) = \bigwedge_{i=1}^M \chi^*
(\Delta_{\lambda_i}) = \chi^* \left(\bigvee_{i=1}^M \lambda_i
\right), \ee
where
\be
\chi^* (\lambda) \mathrel{\mathop:}= \chi^*(\Delta_{\lambda}).
\ee
We aim to prove the following theorem.

\begth The classical capacity of a convex combination of
depolarising channels is equal to its product-state capacity \bes
C \left(\Phi_{rand} \right) = C_p \left( \Phi_{rand}\right). \ees
\enth

\bpr According to \cite{DD07CC} the classical capacity of this
channel can be written as follows \be \label{ClassCapCC}
C\left( \Phi_{rand} \right)  =  \lim_{ n \rightarrow
\infty }\; \frac{1}{n} \, \sup_{\{ p_j^{(n)}, \rho_j^{(n)} \}}
\bigwedge_{i=1}^M \chi \left( \left\{ p_j^{(n)}, \Lambda_i^{(n)}
\left(\rho_j^{(n)}\right) \right\} \right). \ee Suppose that \be
\label{assumptionConvex} C\left( \Phi_{rand} \right)
\geq  \bigwedge_{i=1}^M  \chi^*(\Delta_{\lambda_i}) + \epsilon,
\ee for some $\epsilon >0$.

Then $\exists\, n_0$, such that if $n \geq n_0$, then
\be
\frac{1}{n} \, \sup_{\{ p_j^{(n)},  \rho_j^{(n)} \}}
\bigwedge_{i=1}^M \chi \left( \left\{ p_j^{(n)}, \Lambda_i^{(n)}
\left(\rho_j^{(n)} \right)  \right\} \right) \geq
\bigwedge_{i=1}^M \chi^*(\Delta_{\lambda_i}) + \epsilon.
\ee
Hence, for $n \geq n_0$ there exists an ensemble $\{p_j^{(n)}, \,
\rho_j^{(n)}\}$ such that \be \frac{1}{n} \bigwedge_{i=1}^M \chi
\left( \left\{ p_j^{(n)},\Lambda_i^{(n)} \left(\rho_j^{(n)}\right)
\right\} \right) \geq \bigwedge_{i=1}^M \chi^*(\Delta_{\lambda_i})
+ \epsilon. \ee But King \cite{King03} proved that the product
state capacity of the depolarising channel is equal to its
classical capacity, therefore \be\label{Additivity} \chi^* \left(
\Lambda_i^{(n)} \right) = n \, \chi^* \left( \Delta_{\lambda_i}
\right). \ee
In other words, $\chi \left( \left\{ p_j^{(n)},
\Lambda_i^{(n)} \left(\rho_j^{(n)}\right) \right\} \right)$ is
bounded above by $n \, \chi^*\left(\Delta_{\lambda_i}\right)$. Now,
if $i_0$ is such that \be \bigwedge_{i=1}^M
\chi^*(\Delta_{\lambda_i}) = \chi^*(\Delta_{\lambda_{i_0}}), \ee
then
\bea \frac{1}{n} \bigwedge_{i=1}^M \chi \left( \left\{
p_j^{(n)}, \Lambda_i^{(n)}\left(\rho_j^{(n)}\right) \right\}
\right) &\leq& \chi \left( \left\{ p_j^{(n)}, \Lambda_{i_0}^{(n)}
\left(\rho_j^{(n)}\right) \right\} \right) \non \\ &\leq&
\chi^*(\Delta_{\lambda_{i_0}}). \eea
Therefore \be \frac{1}{n}
\bigwedge_{i=1}^M \chi \left( \left\{ p_j^{(n)},
\Lambda_i^{(n)}\left(\rho_j^{(n)}\right)\right\} \right) \leq
\bigwedge_{i=1}^M \chi^*(\Delta_{\lambda_i}). \ee
This contradicts the assumption made by Equation (\ref{assumptionConvex}) and
therefore \be \label{lessEQ} C\left(
\Phi_{rand}\right) \leq \bigwedge_{i=1}^M
\chi^*(\Delta_{\lambda_i}) = C_p \left( \Phi_{rand}\right). \ee On
the other hand, it is clear that $ C
\left(\Phi_{rand}\right) \geq C_p\left(\Phi_{rand}\right), $ and
therefore \\ $ C\left( \Phi_{rand}\right) = C_p \left(
\Phi_{rand}\right). $ \epr

\begin{remark}
Note that, in contrast to the proof of
Theorem 1, the proof above does not rely on the invariance of the
maximising ensemble of the depolarising channel. The proof uses
the additivity of the Holevo capacity of the depolarising channel
(see Equation (\ref{Additivity})) and the result can therefore be
generalised to all channels for which the additivity of the Holevo
capacity has been proved.
\end{remark}


\section{Convex combinations of two memoryless \\ channels}\label{CC}

Recall that it was shown in \cite{DD07Per} that the product-state
capacity of a convex combination of memoryless channels, denoted $\Phi^{(n)}$, is given by (\ref{convexcap}). Note that the following always holds
\be C_p(\Phi^{(n)}) \leq \wedge_{i=1}^{M} \chi_i^*.
\ee
We investigate whether equality holds for the expression above in the following three cases: a convex combination of two depolarising
channels, two \\amplitude-damping channels, and a convex combination of one depolarising and
one amplitude-damping channel.

\subsection{Two depolarising channels}

In the case of a convex combination of two depolarising qubit
channels $\Delta_{\lambda_i}(\rho) = (1 - \lambda_i) \rho + \lambda_i
\left(\frac{I}{2} \right)$ with parameters $\lambda_1$ and
$\lambda_2$, we have \be C(\Phi_{\lambda_1,\lambda_2}^{(n)}) =
\chi^*(\lambda_1) \wedge \chi^*(\lambda_2) = \chi^*(\lambda_1 \vee
\lambda_2). \ee Indeed, since the maximising ensemble for both
channels is the same (see \ref{Depol}), namely two projections onto orthogonal
states, this also maximises the minimum $\chi_1 \wedge \chi_2$.

\subsection{Two amplitude-damping channels}\label{twoAmp}

A convex combination of amplitude-damping channels is similar. In
that case, the maximising ensemble does depend on the parameter
$\gamma$, but as can be seen from Figure \ref{chi_graphs}, for any $a$,
$\chi_{AD}(a)$ decreases with $\gamma$, so $\chi(\gamma_1) \wedge
\chi(\gamma_2) = \chi(\gamma_1 \vee \gamma_2)$ and we have again,
\be
C(\Phi_{\gamma_1,\gamma_2}^{(n)}) = \chi^*(\gamma_1)
\wedge \chi^*(\gamma_2) = \chi^*(\gamma_1 \vee \gamma_2).
\ee
The fact that $\chi_{AD}(a)$ decreases with $\gamma$ can be seen as follows.
The derivative with respect to $\gamma$ is given by
\be
\frac{\partial
\chi}{\partial \gamma} = -(1-a) \ln \frac{a + \gamma
(1-a)}{(1-\gamma)(1-a)} + \frac{(2\gamma-1)(1-a)^2}{x} \ln
\frac{1+x}{1-x}.
\ee
For $\gamma \leq \half$, both terms are negative if $\frac{a}{1-a} > 1-2\gamma$. Otherwise, the first term is positive and we remark that
\be x \geq
(1-2\gamma)(1-a).
\ee
So that it suffices if
\be x > y = 1-2\gamma - 2a(1-\gamma).
\ee
This is easily checked.

In the case $\gamma > \frac{1}{2}$, we need to show that
\be
f(a,\gamma) = \ln
\frac{a + \gamma(1-a)}{(1-\gamma)(1-a)} -
\frac{(2\gamma-1)(1-a)}{x} \ln \frac{1+x}{1-x} \geq 0.
\ee
Now, if $a=0$, then $f(0,\gamma) = 0$. The derivative is easily
computed to be
\be \frac{\partial f(a,\gamma)}{\partial a} =
\frac{1-\gamma}{a+\gamma(1-a)} + \frac{1}{1-a} +
\frac{2\gamma-1}{x^3} \ln \frac{1+x}{1-x} - \frac{2
(2\gamma-1)}{x^2}.
\ee
This is positive since the first two terms
are positive and the other two are bounded by
\be \frac{2\gamma-1}{x^3} \ln \frac{1+x}{1-x} - \frac{2
(2\gamma-1)}{x^2} \geq \frac{2\gamma-1}{x^2} \left\{ \frac{1}{x}
\ln \frac{1+x}{1-x} - 2 \right\} \geq 0. \ee

\subsection{A depolarising channel \& an amplitude-damping channel}

We now investigate the product-state capacity of a convex
combination of an \\ amplitude-damping and a depolarising channel.
Let $\chi_1$ and $\chi_2$ denote the Holevo quantity of the
amplitude-damping and depolarising channels respectively.
\begin{figure}[ht]
          \centerline{
             \epsfig{file=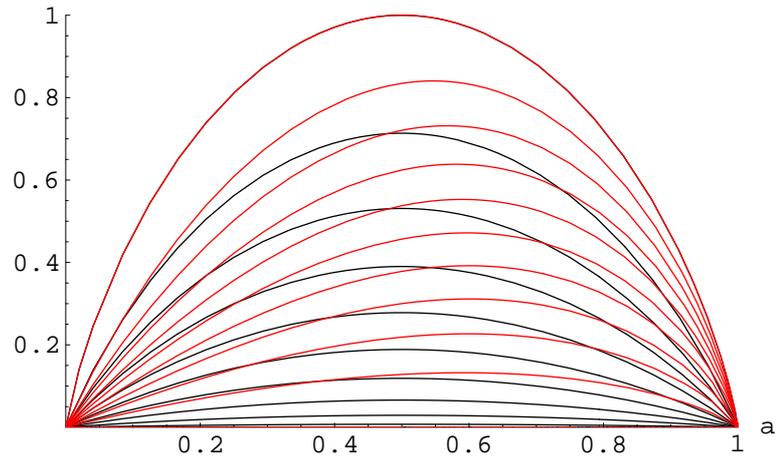, scale=1}}
             \caption{The Holevo $\chi$ quantity for the amplitude
             damping channel and the depolarising channel plotted
             as a function of $a$ in red and black respectively.}
             \label{solveBoth}
       \end{figure}
They are plotted in Figure \ref{solveBoth} for $0 \leq \gamma ,
\lambda \leq 1$. The plot in Figure \ref{solveBoth} indicates that,
for certain values of $\gamma$ and $\lambda$ the maximiser for the
amplitude-damping channel lies to the right of the intersection of
$\chi_1(a)$ and $\chi_2(a)$ , whereas
that for the depolarising channel lies to the left. Indeed, keeping $\lambda$
fixed, we can increase $\gamma$ until the maximum of $\chi_{AD}(\gamma)$ lies
above the graph of $\chi(\Delta_{\lambda})$. The two
graphs then intersect at a value of $a$ intermediate between
$\half$ and the maximiser for $\chi_{AD}$.
This proves that the maximum of the minimum of the channels does not equal the
minimum of the maximum of the channels.


\section{The periodic channel with amplitude-damping \\ channel branches}\label{ampDampBranch}

Recall that the amplitude-damping channel acting on the state $\rho = \left(
\begin{array}{cc} a & b \\ \bar{b} & 1-a \end{array} \right)$ is given by
\be
\Phi_{amp}(\rho) = \left(\begin{array}{cc}
a + (1-a) \gamma & b\sqrt{1 - \gamma}  \\
\bar{b}\sqrt{1- \gamma} & (1-a)(1- \gamma)   \\
\end{array} \right).
\ee
Recall also the expression for the product-state capacity of the amplitude-damping
channel,
\begin{eqnarray}\label{amp1}
\chi\left(\Phi_{amp}(\{p_j,\rho_j\})\right) &=& S \left[ \sum_j \left(
\begin{array}{cc}
p_j\left(a_j+(1-a_j)\gamma \right) & p_j b_j\sqrt{(1 - \gamma)}  \non \\
p_j {\bar b}_j \sqrt{(1 - \gamma)} & p_j(1-a_j)(1 - \gamma)   \\
\end{array} \right) \right]\\
&-& \sum_j p_j\, S \left(
\begin{array}{cc}
a_j+(1-a_j)\gamma & b_j\sqrt{1 - \gamma}  \\
{\bar b}_j \sqrt{1 - \gamma} & (1-a_j)(1 - \gamma)   \\
\end{array} \right).
\end{eqnarray}

In Section \ref{ampDampSection} we argued that the Holevo quantity for the
amplitude-damping channel can be increased by replacing each pure state $\rho_j$
in the ensemble by itself and its mirror image,
each with half the original probability.
Let
\be
\rho = \left(
\begin{array}{cc}
a & b \\
\bar{b} & (1-a)   \\
\end{array} \right)
\qquad
\rho' = \left(\begin{array}{cc}
a & -b  \\
-\bar{b} & (1-a)  \\
\end{array} \right).
\ee
We now investigate whether the following equation holds for a periodic channel
with two amplitude-damping channel branches
\be\label{conj_amp}
\half \sup_{\{p_j, \rho_j\}} \left( \sum_{i=0}^{1} \chi_i(\{p_j,
\rho_j\})\right)  = \half \sum_{i=0}^{1}  \left( \sup_{\{p_j, \rho_j\}}
\chi_i(\{p_j, \rho_j\})\right).
\ee
Note that, unlike the depolarising channel, the maximising ensemble for the
amplitude damping channel does not, in general, consist of \emph{orthogonal} pure
states.
Instead the maximising ensemble depends on the value of the error parameter.

Let $\gamma_0$ and $\gamma_1$ represent the error parameters for two
amplitude-damping channels $\Phi_0$ and $\Phi_1$ respectively. We have argued
that the Holevo quantity for the amplitude-damping channel can be increased
using an ensemble containing two mirror image pure states each with probability
$\half$. Using this minimal ensemble we investigate both sides of Equation
(\ref{conj_amp}), for a periodic channel with two amplitude-damping channel
branches.

Since the sum of two amplitude damping channels is convex, the corresponding Holevo quantity is maximised for a single parameter. The \emph{left hand side} of Equation (\ref{conj_amp}) will therefore be attained for a single $a_{max}$.

In other words, the maximising ensemble will contain two
equiprobable states, and can be written as
\bea\label{lhs_amp}
\half \sup_{\{p_j, \rho_j\}} \left( \sum_{i=0}^{1} \chi_i \right) &=& \half
\left[  \chi^* \left( \gamma_0, \gamma_1, a=a_{max}\right)\right]   \non \\
&=& \half\, H_{bin}((1-a_{max})(1 - \gamma_0))\non \\ &+&
\half\, H_{bin}((1-a_{max})(1 - \gamma_1)) \\ &-&
\half \left[ S\left( \Phi_0(\rho_{a_{max}})  \right) + S\left(
\Phi_1(\rho_{a_{max}})  \right) \right].
\eea
Note that given the eigenvalues for the amplitude-damping channel
\be
\lambda_{amp\pm} = \frac{1}{2} \left(1 \pm
\sqrt{1-4\gamma(1-\gamma)(1-a)^2} \right), \ee again, we denote \be
x=\sqrt{1-4\gamma(1-\gamma)(1-a)^2}. \ee
Let $\chi(\gamma_0,\gamma_1,a)$ denote the sum of the Holevo quantities of the
two channels.
The value for $a_{max}$ can be determined by solving the following equation
\bea\label{detLHS}
\frac{\partial \chi(\gamma_0,\gamma_1,a)}{ \partial a} &=& \half \left[ \frac{2
\gamma_0 (1-\gamma_0)(1-a)}{x_0} \ln\left( \frac{1+x_0}{1-x_0} \right) \right]
\non \\
&+& \half \left[ \frac{2 \gamma_1 (1-\gamma_1)(1-a)}{x_1} \ln\left(
\frac{1+x_1}{1-x_1} \right) \right] \non \\
&+& \half \left[ (1-\gamma_0) \ln \left( \frac{(1-a)(1-\gamma_0)}{a +
(1-a)\gamma_0}\right) \right] \non \\
&+& \half \left[ (1-\gamma_1) \ln \left( \frac{(1-a)(1-\gamma_1)}{a +
(1-a)\gamma_1}\right) \right] \non \\
&=& 0.
\eea
The \emph{right hand side} of Equation (\ref{conj_amp}) cannot be obtained by a
single $a_{max}$. Instead, the supremum for each channel will be attained at a
different value of the input state parameter $a$. If we denote by $a_{max_0}$
and $a_{max_1}$ the state parameter that achieves the product-state capacity for
the channels $\Phi_0$ and $\Phi_1$ respectively, then the right hand side of
Equation (\ref{conj_amp}) can be written as
\bea\label{rhs_amp}
\sum_{i=0}^{1}  \left( \sup_{\{p_j, \rho_j\}} \chi_i \right) &=&  \left( \chi^*
\left( \gamma_0, a_{max_0} \right) + \chi^* \left( \gamma_1, a_{max_1} \right)
\right) \non \\
&=& S  \left(\begin{array}{cc}
a_{max_0} + (1-a_{max_0}) \gamma_0 & 0  \\
0 & (1-a_{max_0})(1- \gamma_0)   \\
\end{array} \right) \non \\ &+&
 S  \left(\begin{array}{cc}
a_{max_1} + (1-a_{max_1}) \gamma_1 & 0  \\
0 & (1-a_{max_1})(1- \gamma_1)   \\
\end{array} \right) \non \\ &-&
S\left( \Phi_0(\rho_{a_{max_0}})  \right) + S\left( \Phi_1(\rho_{a_{max_1}})
\right).
\eea
Let $\chi_0(a)$ and $\chi_1(a)$ denote the Holevo quantities of the channels
$\Phi_0$ and $\Phi_1$, respectively.
Denoting $x_{0,1} = \sqrt{1-4\gamma_{0,1}\left(1 - \gamma_{0,1} \right)(1- a^2)
}$, the values for $a_{max_0}$ and $a_{max_1}$ can be determined by
separately solving the following two equations
\bea\label{detRHSa0}
\frac{d \chi_0 (a)}{d a } &=& ( 1- \gamma_0)  \ln \left(
\frac{(1-a)(1-\gamma_0)}{a + (1-a)\gamma_0} \right) \non \\
&+& \frac{2 \gamma_0 (1-\gamma_0)(1-a)}{x_0} \ln\left( \frac{1+x_0}{1-x_0}
\right) \non \\
&=& 0
\eea
\bea\label{detRHSa1}
\frac{d \chi_1 (a)}{d a } &=& (1- \gamma_1)  \ln \left(
\frac{(1-a)(1-\gamma_1)}{a + (1-a)\gamma_1} \right) \non \\
&+& \frac{2 \gamma_1 (1-\gamma_1)(1-a)}{x_1} \ln\left( \frac{1+x_1}{1-x_1}
\right) \non \\
&=& 0.
\eea
Let $\chi^*_{avg}(\gamma_0,\gamma_1,a_{max_0},a_{max_1})$ denote the average of
the supremum of the Holevo capacities of the channels
$\Phi_0$ and $\Phi_1$
\be
\chi^*_{avg}(\gamma_0,\gamma_1,a_{max_0},a_{max_1}) = \half
\left(\chi_0^*(a_{max_0}) + \chi_1^*(a_{max_1}) \right).
\ee

It is not difficult to show that
\be
\chi^*(\gamma_0=1,\gamma_1,a_{max}) =
\chi^*_{avg}(\gamma_0=1,\gamma_1,a_{max_0},a_{max_1}).
\ee
Similarly, we can show that
$$\chi^*(\gamma_0,\gamma_1=1,a_{max}) =
\chi^*_{avg}(\gamma_0,\gamma_1=1,a_{max_0},a_{max_1}).$$

Next, we let one of the error parameters equal zero. Taking $\gamma_0 = 0$, the
expression $\chi^*(\gamma_0,\gamma_1,a_{max})$ becomes
\bea
\chi^*(\gamma_0=0,\gamma_1,a_{max}) &=&  H_{bin}(a_{max1}) \non \\
&+&  H_{bin}((1-a_{max1})(1 - \gamma_1)) \non \\
&-&   S\left( \Phi_1\left( \rho_{amax} \right) \right).
\eea
Denoting $\chi^*_{avg}(\gamma_0, \gamma_1,a_{max_0},a_{max_1})$ by
$\chi^*_{avg}\left( \gamma_1 \right)$ the right hand side becomes
\bea
 \chi^*_{avg}\left( \gamma_1 \right) &=&  H_{bin}(a_{max1}) \non \\
&+&  H_{bin}((1-a_{max1})(1 - \gamma_1))  \non \\
&-&   S\left( \Phi_1\left( \rho_{amax_1} \right) \right).
\eea
We will now show that
\be
\chi^*(\gamma_0=0,\gamma_1,a_{max}) <
\chi^*_{avg}(\gamma_0=0,\gamma_1,a_{max_0},a_{max_1}).
\ee
Clearly, $a_{max_1} = \half$. To show that $a_{max} < a_{max_1}$, we must show
that $ \frac{d }{da}\sum_i \chi_i (a) <0$
at $a=a_{max_1}$.

In other words, we want to show that
\be
\frac{d \chi_0 (a)}{da} + \frac{d \chi_1
(a)}{da} < 0
\ee
at $a = \half$.

For $\gamma_0 = 0$ the Holevo quantity of the channel $\Phi_0$ becomes
\be
\chi_0 (a)= S  \left(\begin{array}{cc}
a  & 0  \\
0 & (1-a)   \\
\end{array} \right) - S (\rho).
\ee
But $\rho$ is a pure state and therefore $S(\rho) = 0$. Therefore, from Equation
(\ref{detRHSa0}),
\be
\frac{d \chi_0 (a)}{d a} = \ln \left( \frac{(1- a)}{a}
\right).
\ee
We have previously shown that the maximising state parameter for the
amplitude-damping channel is achieved at $a \geq \half$.
We are considering the case where $\gamma_0 \neq \gamma_1$, i.e. $\gamma_1 \neq
0$, therefore $a_{max_1} > \half$.
The expression $\chi_0(a)$ now represents the binary entropy, $H(a)$, and is
therefore maximised at $a= \half$.
It was shown above that the entropy $S(a)$ is a strictly concave function for
$\gamma_0 = 0$ and $\chi_0 (a)$ is therefore decreasing at $a = a_{max_1}$.

The capacity $\chi_1^*(a)$ is achieved at $a = a_{max_1}$. Therefore $\frac{d
\chi_1 (a)}{da}$ is equal to zero at this point.

We can  now conclude that $\frac{d }{da}\sum_i \chi_i (a) <0$ when $a=a_{max_1}$
and therefore
\be
\chi^*(\gamma_0=0,\gamma_1,a_{max}) <
\chi^*_{avg}(\gamma_0=0,\gamma_1,a_{max_0},a_{max_1}).
\ee
We now show that an inequality exists between the expressions
$\chi^*(\gamma_0,\gamma_1,a_{max})$  \\and
$\chi^*_{avg}(\gamma_0,\gamma_1,a_{max_0},a_{max_1})$ for fixed $\gamma_0$, such
that $0 < \gamma_0 <1$.

In Section \ref{twoAmp} we proved that if $\gamma_0 < \gamma_1$, then $\chi(\gamma_0) > \chi(\gamma_1)$ and therefore $a_{max_0} <
a_{max_1}$. Therefore, $\frac{d \chi_0 (a)}{da} <0$ at $a=a_{max_1}$ and $a_{max} <
a_{max_1}$.
Similarly, if $\gamma_0 > \gamma_1$, then $a_{max_0} > a_{max_1}$ and $\frac{d
 \chi_0 (a)}{da} >0$ at $a=a_{max_1}$ and $a_{max} > a_{max_1}$.

As a result, $a_{max}$ will always lie in between $a_{max_0}$ and $a_{max_1}$.
We have previously shown that the Holevo quantity for the amplitude-damping
channel is concave in its state parameter.
Therefore $a_{max} > \tilde{a}$, where $\tilde{a}$ is the parameter value
associated with $\chi^*_{avg}(\gamma,\gamma_1,a_{max_0},a_{max_1})$, i.e.
$\sum_i \sup_a \chi_i(a) = \chi^*_{\gamma_0,\gamma_1} (\tilde{a})$.
This proves that $\chi^*(\gamma_0,\gamma_1,a_{max}) <
\chi^*_{avg}(\gamma,\gamma_1,a_{max_0},a_{max_1})$.

In conclusion, if $\gamma_0 = 1$ or $\gamma_1=1$, then $a_{max} = a_{max_0}$ or
$a_{max} = a_{max_1}$ respectively and  $\chi^*(\gamma_0,\gamma_1,a_{max}) =
\chi^*_{avg}(\gamma,\gamma_1,a_{max_1},a_{max_1})$.
However, if $\gamma_0, \gamma_1 \neq 1$, then\\
$\chi^*(\gamma_0,\gamma_1,a_{max}) <
\chi^*_{avg}(\gamma,\gamma_1,a_{max_1},a_{max_1})$.
Therefore, in the case of a periodic channel with amplitude-damping channel
branches
\be
\half \sup_{\{p_j, \rho_j\}} \left( \sum_{i=0}^{1} \chi_i(\{p_j,
\rho_j\})\right) \neq \half \sum_{i=0}^{1}  \left( \sup_{\{p_j, \rho_j\}}
\chi_i(\{p_j, \rho_j\})\right).
\ee

\section{Summary}

In summary, we have investigated the classical capacity of two particular quantum channels with memory, namely a periodic quantum channel and a random quantum channel. We have shown that in both cases the product state capacity of each channel is equal to its classical capacity. We can therefore conclude that entangled input state codewords do not enhance the classical capacity of these channels.

Next we showed that the formula for the product-state capacity of the periodic channel, which is given by the supremum of the average of the Holevo quantities for the channel branches, cannot be written as the average of the Holevo capacities evaluated for each channel branch, when the channel branches consist of amplitude damping channels. This result has an important implication in the next chapter. 
\chapter{Strong converse to the channel coding theorem for a \\periodic quantum channel} We introduce the channel coding theorem and concentrate, in particular, on the strong converse to the coding theorem for quantum channels. We discuss the fact that the strong converse does not hold for the product-state capacity of the periodic channel introduced in Chapter 4 and we demonstrate a, so-called, ``weakened'' strong converse for this channel. See \cite{HolevoSurvey98} for a survey of quantum coding theorems.

\begin{remark}
Wehner and K\"onig \cite{KW09} recently proved the fully general strong converse theorem for a family of channels, that is, they proved that the strong converse theorem holds for a family of quantum channels even in the case when entangled state inputs are allowed.
\end{remark}

\section{Coding theorems and quantum channels}

The channel coding theorem is comprised of two parts, namely the direct part of the theorem, which refers to the construction of the code, and the converse to the theorem. Shannon \cite{Shannon48} proposed the theorem for classical channels and the first rigorous proof was provided by Feinstein \cite{Feinstein}.

The capacity of a quantum channel $\Phi$ provides a limit on the amount of information which can be transmitted reliably per channel use. The direct part of the quantum channel coding theorem states that using $n$ copies of the channel, we can code with exponentially small probability of error at a rate $R =\frac{1}{n} \log |\mathcal{M}|$ if and only if $R \leq C$, in the asymptotic limit, where $\cal M$ denotes the set of possible codewords to be transmitted.
If the rate at which classical information is transmitted over a quantum channel exceeds the capacity of the channel, i.e. if $R>C$, then the probability of decoding the information correctly goes to zero in the number of channel uses. This is known as the strong converse to the channel coding theorem.

The strong converse to the channel coding theorem of a, so-called, classical quantum channel was proved independently by Winter \cite{Winter99} and by Ogawa and Nagaoka \cite{ON99} using different methods. We will follow the method used by Winter, namely, the method of types which was used by Wolfowitz \cite{Wolfowitz}, to prove the strong converse for classical channels. See \cite{Csiszar98, Hayashi} for an introduction to the method of types.

We begin by introducing some notation. Recall that a memoryless channel is given by a completely positive trace-preserving map ${\Phi}: {\mathcal{B}}({\cal H}) \to {\cal B}({\cal K})$, where ${\cal B}({\cal H})$ and ${\cal B}({\cal K})$ denote the states on the input and output Hilbert spaces ${\cal H}$ and $\cal K$, respectively.

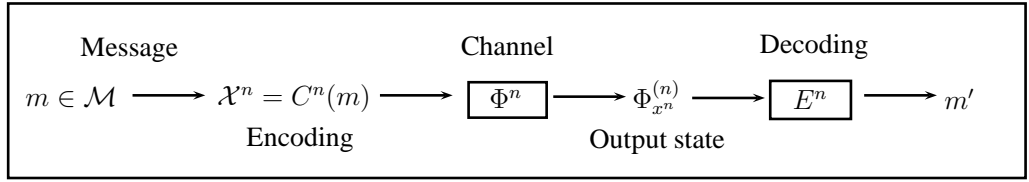
\begin{figure}[ht]
\begin{center}
\scalebox{.87} 
{
\begin{pspicture}(0,-1.35)(15.9609375,1.35)
\usefont{T1}{ptm}{m}{n}
\rput(4.882344,-0.12){$\mathcal{X}^n = C^n(m)$}
\psline[linewidth=0.04cm,arrowsize=0.05291667cm 2.0,arrowlength=1.4,arrowinset=0.4]{->}(2.4409375,-0.07)(3.5409374,-0.07)
\psframe[linewidth=0.04,dimen=outer](8.720938,0.15)(7.5009375,-0.43)
\usefont{T1}{ptm}{m}{n}
\rput(8.082344,-0.12){$\Phi^n$}
\usefont{T1}{ptm}{m}{n}
\rput(10.362344,-0.1){$\Phi_{x^n}^{(n)}$}
\psline[linewidth=0.04cm,arrowsize=0.05291667cm 2.0,arrowlength=1.4,arrowinset=0.4]{->}(6.1809373,-0.07)(7.2809377,-0.07)
\psline[linewidth=0.04cm,arrowsize=0.05291667cm 2.0,arrowlength=1.4,arrowinset=0.4]{->}(8.800938,-0.09)(9.900937,-0.09)
\psframe[linewidth=0.04,dimen=outer](13.320937,0.17)(12.060938,-0.45)
\usefont{T1}{ptm}{m}{n}
\rput(12.6823435,-0.14){$E^n$}
\psline[linewidth=0.04cm,arrowsize=0.05291667cm 2.0,arrowlength=1.4,arrowinset=0.4]{->}(13.500937,-0.07)(14.600938,-0.07)
\usefont{T1}{ptm}{m}{n}
\rput(2.3928125,0.62){Message}
\usefont{T1}{ptm}{m}{n}
\rput(4.973906,-0.74){Encoding}
\usefont{T1}{ptm}{m}{n}
\rput(8.110625,0.68){Channel}
\usefont{T1}{ptm}{m}{n}
\rput(10.369375,-0.78){Output state}
\usefont{T1}{ptm}{m}{n}
\rput(12.744532,0.68){Decoding}
\usefont{T1}{ptm}{m}{n}
\rput(14.942344,-0.12){$m'$}
\usefont{T1}{ptm}{m}{n}
\rput(1.5123438,-0.1){$m \in \mathcal{M}$}
\psframe[linewidth=0.04,dimen=outer](15.9609375,1.35)(0.5409375,-1.35)
\psline[linewidth=0.04cm,arrowsize=0.05291667cm 2.0,arrowlength=1.4,arrowinset=0.4]{->}(10.900937,-0.11)(12.000937,-0.11)
\end{pspicture}
}
\end{center}
\caption{Transmission over a classical-quantum channel.}
\label{CQchannel}
\end{figure}

Equivalently, we can describe a so-called \emph{classical-quantum} channel, usually denoted $W$, as a mapping from the classical message to the output state of the channel on $\mathcal{B}(\mathcal{K})$ as follows,
\be
W : \mathcal{X} \mapsto \mathcal{B}(\mathcal{K}),
\ee
where the message is first encoded into a sequence belonging the set $\mathcal{X}^n$, where $\cal X$ represents the input alphabet. The process is shown in Figure \ref{CQchannel} \cite{Hayashi}.

We can combine the two mapping descriptions as follows.
We wish to send classical information in the form of quantum states over a quantum channel $\Phi$.
A (discrete) memoryless quantum channel, $\Phi$, carrying classical information can be thought of as a map from a (finite) set, or alphabet, $\cal X$ into $\cal B(\cal K)$, taking each $x \in \cal X$ to $\Phi_{x}= \Phi(\rho_x)$, where the input state to the channel is given by $\{ \rho_x\}_{x \in \cal{X}}$ and each $\rho_x \in \cal B (\cal H)$. Let $d=\dim(\cal H)$ and $a=|\cal X|$.

For a probability distribution $P$ on the input alphabet $\mathcal{X}$, the average output state of a channel $\Phi$ is given by
\be
P \sigma = \sum_{x \in \mathcal{X}} P(x) \Phi(\rho_x).
\ee
The conditional von Neumann entropy of $\Phi$ given $P$ as defined by
\be
S(\Phi|P) = \sum_{x \in \mathcal{X}} P(x) S(\Phi(\rho_x)),
\ee
and the mutual information between the probability distribution $P$ and the channel $\Phi$ is defined as follows,
\be\label{mutInf}
I(P;\Phi)= S(P \sigma) - S(\Phi|P),
\ee
The product-state capacity of the channel $\Phi$ is given by the maximum of the mutual information (Equation \ref{mutInf}), taken over all possible probability distributions $P$, i.e.
\be
\chi^*(\Phi) = \max_P \, I(P;\Phi).
\ee
An \emph{$n$-block code} for a quantum channel $\Phi$ is a pair $(C^n,E^n)$, where $C^n$ is a mapping from a
finite set of messages $\cal M$, of length $n$, into $\mathcal{X}^n$, i.e. a sequence $x^n \in \cal X$ is assigned to each of the $|\cal M |$ messages, and $E^n$ is a POVM, i.e. a quantum measurement, on the output space $\mathcal{K}^{\otimes n}$ of the channel $\Phi_{x^n}^{(n)}$. This process is depicted in Figure \ref{CQchannel} below.

The maximum error probability of the code $(C^n,E^n)$ is defined as
\be
p_e(C^n,E^n) = \max \{ 1 - \tr (\Phi_{x^n}^{(n)} E^{n}_m) : m \in \cal M \}.
\ee
The code $(C^n,E^n)$ is called an $(n, \lambda)$-code, if
$p_e(C^n,E^n) \leq \lambda$.
The maximum size $| \cal M |$ of an $(n, \lambda)$-code is denoted $N(n,\lambda)$.

\section{Method of types}

Next we employ a technique known as the \emph{method of types} \cite{Csiszar98} to exploit the properties of \emph{variance-typical} sequences, leading to a sharp bound on the rate at which quantum information can be reliably transmitted over a memoryless quantum channel. This is achieved through the \emph{strong} converse theorem.

Define a finite alphabet $\cal X$ and sequences $x^n = x_1 , \dots, x_n \in \mathcal{X}^n$ and let
\be
N(x \big|x^n) = \big| \{ i \in \{1, \dots, n\} : x_i = x \} \big|
\ee
for $x \in \mathcal{X}$.

The \emph{type} of the sequence $x^n$ is given by the empirical distribution $P_{x^n}$ on $\cal X$ such that
\be\label{distP}
P_{x^n}(x) = \frac{N(x \big|x^n)}{n}.
\ee
Clearly, the number of types is upper bounded by $(n+1)^a$, where $a= \big| \cal X \big|$.

Now define the set of \emph{variance-typical sequences of length $n$ and of approximate type $P$}, for $\delta \geq 0$, as follows
\be\label{varTypSeq}
\mathcal{T}_{P,\delta}^n = \{ x^n \in \mathcal{X}: \forall x \in \mathcal{X} \; \big|N(x|x^n) - n P(x) \big| \leq \delta \sqrt{n} \sqrt{P(x)(1-P(x))} \}.
\ee
A set of type $P$ (rather than \emph{approximate} type P) is denoted $\mathcal{T}_{P,0}^n $, i.e. $\delta =0$.


\section{Theorems and lemmas}

Next we state the channel coding theorem for memoryless quantum channels.
We obtain the theorem by combining the direct part, given by Theorem \ref{Direct}, and the strong converse, given by Theorem \ref{strongConv}. The theorem is stated below.

\begin{theorem}(\rm{Coding theorem for memoryless quantum channels})\\
For every $\lambda \in (0,1)$ there exists a constant $K(\lambda, a,d)$ such that for all memoryless quantum channels $\Phi$,
\be
\big| \log N(n, \lambda) - n \chi^*(\Phi) \big| \leq K(\lambda,a,d) \sqrt{n}.
\ee
\end{theorem}

The direct part of the coding theorem for memoryless quantum channels is given by the following theorem. It was proved by Holevo \cite{Hol98} and Schumacher and Westmoreland \cite{SW97}, who built of the ideas of Hausladen et al. \cite{HJSWW96}.

\begin{theorem}\label{Direct}
\rm{(Code construction)}\\
 Given $\epsilon > 0$, there exists $n_0 \in \NN$, such that for all $n \geq n_0$ there exists $N(n,\epsilon)=N_n \geq 2^{n(\chi^*(\Phi) - \epsilon)}$, and there exist product states
$\rho_1^{(n)}, \dots, \rho_{N_n}^{(n)} \in \mathcal{B}(\mathcal{H}^{\otimes n})$ and positive operators $E_1^{(n)}, \dots, E_{N_n}^{(n)} \in \mathcal{B}(\mathcal{K}^{\otimes n})$, such that
$\sum_{m=1}^{N_n} E_m^{(n)} \leq I_n$ and
\be
\tr \left( \Phi^{(n)} \left( \rho_m^{(n)}\right) E_m^{(n)} \right) > 1-\epsilon,
\ee
for each $m$.
\end{theorem}

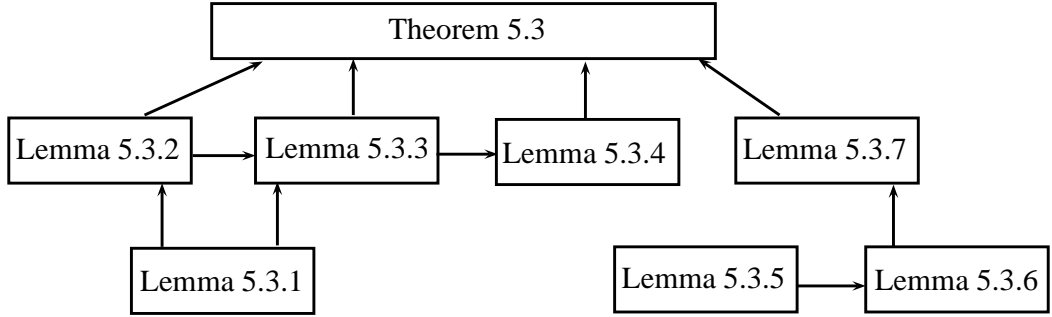
\begin{figure}
\begin{center}
\scalebox{0.96} 
{
\begin{pspicture}(0,-2.16)(14.28,2.16)
\psframe[linewidth=0.04,dimen=outer](9.22,0.56)(6.68,-0.38)
\psframe[linewidth=0.04,dimen=outer](12.5,0.58)(9.96,-0.36)
\usefont{T1}{ptm}{m}{n}
\rput(11.200937,0.13){Lemma \ref{L9}}
\psframe[linewidth=0.04,dimen=outer](10.84,-1.2)(8.3,-2.14)
\psframe[linewidth=0.04,dimen=outer](14.28,-1.2)(11.74,-2.14)
\usefont{T1}{ptm}{m}{n}
\rput(9.540937,-1.65){Lemma \ref{L7}}
\usefont{T1}{ptm}{m}{n}
\rput(13.000937,-1.67){Lemma \ref{L8}}
\psframe[linewidth=0.04,dimen=outer](5.92,0.58)(3.38,-0.36)
\usefont{T1}{ptm}{m}{n}
\rput(4.6609373,0.15){Lemma \ref{L5}}
\psframe[linewidth=0.04,dimen=outer](2.54,0.58)(0.0,-0.36)
\usefont{T1}{ptm}{m}{n}
\rput(1.2409375,0.13){Lemma \ref{L3}}
\psframe[linewidth=0.04,dimen=outer](4.22,-1.22)(1.68,-2.16)
\usefont{T1}{ptm}{m}{n}
\rput(2.9409375,-1.69){Lemma \ref{L4}}
\psframe[linewidth=0.04,dimen=outer](9.72,2.16)(2.78,1.36)
\usefont{T1}{ptm}{m}{n}
\rput(6.2990627,1.79){Theorem \ref{strongConv}}
\usefont{T1}{ptm}{m}{n}
\rput(7.9009376,0.09){Lemma \ref{L6}}
\psline[linewidth=0.04cm,arrowsize=0.05291667cm 2.0,arrowlength=1.4,arrowinset=0.4]{->}(5.88,0.06)(6.72,0.06)
\psline[linewidth=0.04cm,arrowsize=0.05291667cm 2.0,arrowlength=1.4,arrowinset=0.4]{->}(10.58,0.6)(9.46,1.38)
\psline[linewidth=0.04cm,arrowsize=0.05291667cm 2.0,arrowlength=1.4,arrowinset=0.4]{->}(3.7,-1.2)(3.7,-0.32)
\psline[linewidth=0.04cm,arrowsize=0.05291667cm 2.0,arrowlength=1.4,arrowinset=0.4]{->}(12.14,-1.22)(12.14,-0.36)
\psline[linewidth=0.04cm,arrowsize=0.05291667cm 2.0,arrowlength=1.4,arrowinset=0.4]{->}(2.12,-1.22)(2.12,-0.34)
\psline[linewidth=0.04cm,arrowsize=0.05291667cm 2.0,arrowlength=1.4,arrowinset=0.4]{->}(1.88,0.6)(3.5,1.34)
\psline[linewidth=0.04cm,arrowsize=0.05291667cm 2.0,arrowlength=1.4,arrowinset=0.4]{->}(2.54,0.04)(3.4,0.04)
\psline[linewidth=0.04cm,arrowsize=0.05291667cm 2.0,arrowlength=1.4,arrowinset=0.4]{->}(10.82,-1.76)(11.76,-1.76)
\psline[linewidth=0.04cm,arrowsize=0.05291667cm 2.0,arrowlength=1.4,arrowinset=0.4]{->}(7.92,0.56)(7.92,1.34)
\psline[linewidth=0.04cm,arrowsize=0.05291667cm 2.0,arrowlength=1.4,arrowinset=0.4]{->}(4.74,0.6)(4.74,1.38)
\end{pspicture}
}
\end{center}
\caption{Map of the proof of Theorem \ref{strongConv}.}
\label{map}
\end{figure}

The following lemmas are required in order to prove the strong converse theorem for a memoryless quantum channel (Theorem \ref{strongConv}).
To make the proof more clear, we provide a ``map" to the proof in Figure \ref{map}.

We start with some definitions.
We first define the trace norm for an operator $A$ as follows,
\be\label{TrNorm}
\big|\big| A \big|\big|_1 = \tr\sqrt{A^* A}.
\ee
If $A$ is Hermitian, i.e. if $A = A^*$, then
\be
\big|\big| A \big|\big|_1 = \tr |A|,
\ee
and the trace norm of A can therefore be written as the trace of the difference of projection operators $\Pi^{+}$ and $\Pi^{-}$, where $\Pi^{\pm}$ are the projections onto the eigenspace of $A$ corresponding to all non-negative and all negative eigenvalues of $A$, i.e.
\be
\big|\big| A \big|\big|_1 = \tr \left( \Pi^{+}A - \Pi^{-}A \right).
\ee
More precisely, if $A$ has spectral decomposition,
\be
A = \sum_{i=1}^d \lambda_i \ket{u_i} \bra{u_i},
\ee
and therefore
\be
|A| = \sum_{i=1}^d |\lambda_i| \ket{u_i} \bra{u_i},
\ee
then the projections $\Pi^{\pm}$ can be expressed as follows
\be
\Pi^{+} = \sum_{\lambda_i \geq 0} \ket{u_i} \bra{u_i} \qquad \Pi^{-} = \sum_{\lambda_i < 0} \ket{u_i} \bra{u_i}.
\ee
Next, for a state $\rho$ we choose a diagonalisation
\be\label{stateDiag}
\rho = \sum_j R(j) \, \pi_j.
\ee
Clearly the list of eigenvalues $R(j)$ form a probability distribution and therefore $S(\rho) = H(R)$, where $S(\cdot)$ represents the von Neumann entropy of a state and $H(\cdot)$ represents the Shannon entropy of a probability distribution.

We can now define the \emph{variance-typical} projector of the state $\rho$ with constant $\delta \geq 0$ as follows,
\be
\Pi^{n}_{\rho,\delta} = \sum_{j^n \in \mathcal{T}^{n}_{R,\delta}} \pi_{j_1} \otimes \cdots \otimes \pi_{j_n}.
\ee
An operator $0 \leq B \leq \mathds{1}$, is said to be an \emph{$\eta$- shadow} of the state $\rho$ if
\be\label{shadow}
\tr (\rho B) \geq \eta.
\ee

The following lemma provides bounds on an operator $\Lambda$ with the constraint $0 \leq \Lambda \leq \mathds{1}$. These bounds will be used to prove subsequent lemmas.

\begin{lemma}\label{L4}
Let $0 \leq \Lambda \leq \mathds{1}$ and $\rho$ a state commuting with $\Lambda$ such that for some $\lambda, \mu_1, \mu_2 >0$, and let the following relations hold
\be
\tr(\rho \Lambda) \geq 1- \lambda  \qquad and \qquad \mu_1 \Lambda \leq \sqrt{\Lambda} \, \rho \, \sqrt{\Lambda} \leq \mu_2 \Lambda.
\ee
Then we obtain the following bounds
\be (1 - \lambda) \mu_2^{-1} \leq \tr (\Lambda) \leq \mu_1^{-1},
\ee
and for an $\eta$-shadow $B$ of $\rho$,
\be\label{lastL4}
\tr (B) \geq (\eta - \lambda) \mu_2^{-1}.
\ee
\end{lemma}
\begin{proof}
We first show that $(1- \lambda) \mu_2^{-1} \leq \tr(\Lambda) \leq \mu_1^{-1}$.
Using the inequalities $\tr(\rho \Lambda) \geq 1- \lambda$ and $\sqrt{\Lambda} \rho \sqrt{\Lambda} \leq \mu_2 \Lambda$,
\be
\tr (\Lambda) \geq \mu_2^{-1} (1 - \lambda).
\ee
Next, using $\tr(\rho \Lambda) \leq 1$ and $\tr(\rho \Lambda) \geq \mu_1 \tr(\Lambda)$,
\be
\tr(\Lambda) \leq \mu_1^{-1}.
\ee
Finally, we show Equation (\ref{lastL4}), as follows
\bea
\mu_2 \tr(B) &\geq& \tr(\mu_2 \Lambda B)  \non \\
&\geq& \tr(\sqrt{\Lambda} \rho \sqrt{\Lambda} B)  \non \\
&=& \tr(\rho B) - \tr( (\rho - \sqrt{\Lambda} \rho \sqrt{\Lambda}) B) \non \\
&\geq& \eta - \big|\big| \rho - \sqrt{\Lambda} \rho \sqrt{\Lambda} \big|\big|_1  \non \\
&=& \eta - ( \tr(\rho) - \tr (\rho \Lambda)) \non \\
&\geq& \eta - \lambda.
\eea
The first inequality above is due to $\Lambda \leq \mathds{1}$, the second one is due to $\mu_2 \Lambda \geq \sqrt{\Lambda} \rho \sqrt{\Lambda}$. The next inequality holds since $\tr(\rho B) \geq \eta$ and $0 \leq B \leq \mathds{1}$ and the final one is by $\tr (\rho \Lambda) \geq 1 - \lambda$.
\end{proof}
Next define $K = 2( \frac{\log(e)}{e})$. When calculating $P_{x^n}(x)$ (Equation \ref{distP}) for a particular symbol $x$ we are interested in whether or not the symbol $x$ appears in the sequence $x^n$. We therefore define Bernoulli random variables $X_i$ taking the value $1$ if and only if $x_i=x$, with probability $P(x)$.

\begin{lemma}\label{L3}
For every state $\rho$ and positive integer $n$, the following three inequalities hold.

Firstly, the probability that the state $\rho^{\otimes n}$ is typical, with respect to the set of variance typical sequences $\mathcal{T}_{R,\delta}^n$, is given by
\be\label{first3}
\tr(\rho^{\otimes n} \Pi^n_{\rho,\delta}) \geq 1 - \frac{d}{\delta^2}.
\ee
Secondly, the following relation holds,
\be\label{second3}
\Pi^n_{\rho,\delta}  \, \rho^{\otimes n}  \, \Pi^n_{\rho,\delta} \geq \Pi^n_{\rho,\delta} \, 2^{(- n S(\rho) - K \delta d\sqrt{n})}.
\ee
The above bound will be used directly to prove both Lemma \ref{L5} and the strong converse theorem.

Finally, the size of the projection $\Pi^{n}_{\rho,\delta}$ is upper bounded as follows,
\be\label{third3}
\tr( \Pi^{n}_{\rho,\delta}) \leq 2^{(nS(\rho) + Kd\delta\sqrt{n})}.
\ee
\end{lemma}
\begin{proof}
Observe that,
\be
\tr (\rho^{\otimes n}\Pi^n_{\rho,\delta}) = R^{\otimes n} (\mathcal{T}^n_{R,\delta}).
\ee
Chebyshev's inequality states that,
\be
P\left( \big| X - \mu \big| \geq k \sigma \right) \leq \frac{1}{k^2},
\ee
where $X$ is a random variable, $\mu$ is the associated mean, $k >0$ and $\sigma^2$ is the variance of $X$ with respect to $\mu$.

The set of variance typical sequences $\mathcal{T}_{R,\delta}^n$ (Equation (\ref{varTypSeq})) is the intersection of $d= \dim(\mathcal{H})$ events. For each $j$ (appearing in the sum in Equation (\ref{stateDiag})), the random variable $X = \frac{N( j \big| j^n)}{n} = \frac{1}{n} \sum_{i=1}^n \delta_{j,j_i}$ must deviate from its expectation $R(j)$ by at most $\frac{\delta \sqrt{R(j) (1 - R(j))}}{\sqrt{n}}$, by definition of $\mathcal{T}_{R,\delta}^n$.

Using Chebyshev's inequality,
\be
P \biggl( \bigg| \frac{N( j \big| j^n)}{n} - R(j) \bigg|  \geq \frac{\delta \sqrt{R(j) (1 - R(j))}}{\sqrt{n}}\biggr) \leq \frac{1}{\delta^2},
\ee
for a given $j$. Therefore, the \emph{union} of $d$ such events is less than or equal to $\frac{d}{\delta^2}$.
But $\mathcal{T}_{R,\delta}^n$ is the \emph{intersection} of the complementary events, therefore
\be
P^{\otimes n} (\mathcal{T}_{R,\delta}^n) \geq 1 - \frac{d}{\delta^2}.
\ee
Let $\pi^n = \pi_{j_{1}} \otimes \cdots \pi_{j_{n}}$ be one of the eigenprojections of the tensor product state $\rho^{\otimes n}$ constituting $\Pi^n_{\rho,\delta}$. Then,
\bea\label{Rprod}
\tr (\rho^{\otimes n}\pi^n) = R(j_1) \dots R(j_n) = \prod_{j=1}^d R(j)^{N(j|j^n)},
\eea
and given $\big|N(j|j^n) - nR(j) \big| \leq \delta \sqrt{n} \sqrt{R(j)(1-R(j))}$,
\bea
\big| -\log \tr(\rho^{\otimes n}\pi^n) - n S(\rho) \big| &=& \bigg| \sum_{j} - N(j | j^n) \log R(j) - n S(\rho)\bigg| \non \\
&=& \bigg| \sum_{j} -N(j | j^n) \log R(j) + n R(j) \log R(j) \bigg| \non \\
&\leq& -\sum_{j=1}^d \log R(j) \bigg| N(j | j^n) \log R(j) - n R(j) \bigg| \non \\
&\leq& \sum_{j=1}^d - \delta \sqrt{n} \sqrt{R(j)} \log R(j) \non \\
&=& - 2 \delta \sqrt{n} \sum_{j=1}^d \sqrt{R(j)} \log \sqrt{R(j)} \non \\
&\leq& K d \delta \sqrt{n},
\eea
since $- 2 \sqrt{R(j)} \log \sqrt{R(j)} \leq K$.
Therefore
\be
2^{(-n S(\rho) - K d \delta \sqrt{n})} \leq \tr(\rho^{\otimes n}\pi^n) \leq 2^{(-n S(\rho) + K d \delta \sqrt{n})}.
\ee
We have $\Pi^n_{\rho,\delta} = \sum_{j^n \in \mathcal{T}^n_{R,\delta}} \pi_{j_{1}} \otimes \cdots \otimes \pi_{j_n}$ and $\rho=\sum_j R(j)\pi_j$, therefore using Equation (\ref{Rprod}) and the lower bound on $\tr(\rho^{\otimes n}\pi^n)$ i.e. $\tr(\rho^{\otimes n}\pi^n) \geq 2^{(-n S(\rho) - K d \delta \sqrt{n})}$, we obtain the following,
\bea
\Pi^n \rho^{\otimes n} \Pi^n &=& \sum_{j^n \in \mathcal{T}^n_{R,\delta}} (\pi_{j_{1}} \otimes \cdots \otimes \pi_{j_{n}}) \, \, \rho^{\otimes n} \sum_{j^n \in \mathcal{T}^n_{R,\delta}} (\pi_{j_{1}} \otimes \cdots \otimes \pi_{j_{n}}) \non\\
&=& \sum_{j^n \in \mathcal{T}^n_{R,\delta}} \prod_j R(j)^{N(j|j^n)} (\pi_{j_{1}} \otimes \cdots \otimes \pi_{j_{n}})\non\\
&=& \sum_{j^n \in \mathcal{T}^n_{R,\delta}} \tr(\rho^{\otimes n} \pi^n) (\pi_{j_{1}} \otimes \cdots \otimes \pi_{j_{n}})\non\\
&\geq& \sum_{j^n \in \mathcal{T}^n_{R,\delta}} 2^{(-n S(\rho) - K d \delta \sqrt{n})} (\pi_{j_{1}} \otimes \cdots \otimes \pi_{j_{n}}) \non\\
&=& \Pi^n 2^{(-n S(\rho) - K d \delta \sqrt{n})}.
\eea
Therefore by Lemma \ref{L4} (taking $\mu_1 = 2^{-n S(\rho) - K d \delta \sqrt{n}}$),
\be
\tr( \Pi^{n}_{\rho,\delta}) \leq 2^{(nS(\rho) + K\delta\sqrt{n})}.
\ee
\end{proof}

We now fix diagonalisations $\Phi_x = \sum_{j=1}^d Q (j \big|x) \, (\pi_x)_j$, where $Q(
\cdot \big| \cdot)$ is a \\ stochastic matrix, and define the \emph{conditional variance-typical projector of $\Phi$ \\given $x^n$ with constant $\delta$} to be
\be
\Pi^n_{\Phi,\delta}(x^n) = \bigotimes_{x \in \cal X} \Pi^{I_{x}}_{\Phi_{x}, \, \delta}
\ee
where $I_{x} = \{ i \in \{ 1, \dots, n \} : x_i = x \}$.
With $\Phi_{x^n}^{(n)} = \Phi_{x_1} \otimes \Phi_{x_2} \otimes \dots \otimes \Phi_{x_n}$, we then have the following lemma,

\begin{lemma}\label{L5}

For all $x^n \in \mathcal{X}^{n}$ of type $P$, the probability that the output state $\Phi_{x^n}^{(n)}$ is typical with respect to the conditional variance-typical sequences $\mathcal{T}^{I_x}_{Q,\delta}$, is bounded below as follows
\be\label{first5}
\tr \left(\Phi_{x^n}^{(n)} \, \Pi^n_{\Phi,\delta}(x^{n}) \right) \geq 1 - \frac{ad}{\delta^2},
\ee
and with $\Pi^n = \Pi^n_{\Phi,\delta}(x^{n})$ the following inequality holds, and will subsequently be used to prove Lemma \ref{L6} and the strong converse theorem,
\be\label{second5}
\Pi^n  \Phi^{(n)}_{x^n}  \Pi^n \leq \Pi^n 2^{(-nS(\Phi| P) + K \delta d\sqrt{an})}.
\ee
Every $\eta$-shadow $B$ of $\Phi^{(n)}_{x^n}$ satisfies,
\be\label{third5}
\tr(B) \geq \left( \eta - \frac{ad}{\delta^2}\right)2^{(nS(\Phi |P) - Kd\sqrt{a}\delta\sqrt{n})}.
\ee
\end{lemma}
\begin{proof}
The first inequality (\ref{first5}), is obtained by applying Lemma \ref{L3} $a$ times.
The inequality given by expression (\ref{third5}) follows from the inequalities (\ref{first5}) and (\ref{second5}), using Lemma \ref{L4}. Next we prove the inequality (\ref{second5}).
Using $\Pi^n_{\Phi,\delta}(x^n) = \bigotimes_{x \in \cal X} \Pi^{I_{x}}_{\Phi_{x}, \, \delta}$, we have,
\bea
\Pi^n  \Phi^{(n)}_{x^n}  \Pi^n  &=& \bigotimes_{x \in \cal X} \Pi^{I_{x}}_{\Phi_{x}, \, \delta} \, \, \Phi_x^{\otimes I_x} \, \, \bigotimes_{x \in \cal X} \Pi^{I_{x}}_{\Phi_{x}, \, \delta}\non \\
&=& \bigotimes_{x \in \cal X} \, \sum_{j^{I_x} \in \mathcal{T}^{I_x}_{Q,\delta}} (\pi_{j_1} \otimes \cdots \otimes \pi_{j_{I_{x}}}) \, \Phi_x^{\otimes I_x} \sum_{j'^{I_{x}} \in \mathcal{T}^{I_x}_{Q,\delta}} (\pi_{j'_1} \otimes \cdots \otimes \pi_{j'_{I_{x}}}) \non \\
&=& \bigotimes_{x \in \mathcal{X}} \sum_{j^{I_{x}} \in \mathcal{T}^{I_x}_{Q,\delta}} Q(j_1 |x) \cdots Q(j_{I_x} |x) (\pi_{j_1} \otimes \cdots \otimes \pi_{j_{I_{x}}}) \non \\
&=& \bigotimes_{x \in \mathcal{X}} \sum_{j^{I_{x}} \in \mathcal{T}^{I_x}_{Q,\delta}} \prod_{j=1}^d Q(j|x)^{N(j|j^{I_x})} (\pi_{j_1} \otimes \cdots \otimes \pi_{j_{I_{x}}}),
\eea
but, since $j^{I_{x}} \in \mathcal{T}^{I_{x}}_{Q,\delta}$, and $Q(j|x) <1$
\be
N(j|j^{I_{x}}) \geq |I_x| Q(j|x) - \delta \sqrt{|I_x|} \sqrt{Q(j|x)},
\ee
and therefore
\be
\Pi^n  \Phi^{(n)}_{x^n} \Pi^n \leq \bigotimes_{x \in \mathcal{X}} \sum_{j^{I_{x}} \in \mathcal{T}^{I_x}_{Q,\delta}} \prod_{j=1}^d Q(j|x)^{(|I_x| Q(j|x) - \delta \sqrt{I_x} \sqrt{Q(j|x)})} (\pi_{j_1} \otimes \cdots \otimes \pi_{j_{I_{x}}}) .
\ee
Recall that we are assuming $x^n$ to be of type $P$, therefore $|I_x| = n P(x)$ and
\bea
\sum_{x \in \mathcal{X}} \sum_{j=1}^d  2^{|I_x| Q(j|x) \log Q(j|x)}
&=& \sum_{x \in \mathcal{X}} \sum_{j=1}^d 2^{n P(x) Q(j|x) \log Q(j|x)} \non \\
&=& \sum_{x \in \mathcal{X}} 2^{- n P(x) S(\Phi_x)},
\eea
and
\be
2^{-2 \delta \sqrt{|I_x|} \sqrt{Q(j|x)} \log \sqrt{Q(j|x)}} \leq 2^{Kd \delta \sqrt{|I_x|}}.
\ee
Using the Cauchy-Schwartz inequality
\be
|\langle{a} | b\rangle| \leq || a ||_2 . || b ||_2
\ee
i.e. $\left(\sum_k a_k b_k \right)^2 \leq \left(\sum_k a_k^2 \right) \left(\sum_k b_k^2\right)$
it is clear that $\sum_{x \in \mathcal{X}}  \sqrt{|I_x|} \leq \sqrt{an}$ and therefore,
\bea
\Pi^n  \Phi^{(n)}_{x^n}  \Pi^n &\leq&  \sum_{x \in \mathcal{X}}  \sum_{j^{I_{x}} \in \mathcal{T}^{I_x}_{Q,\delta}} 2^{-n P(x) S(\Phi_x) + Kd \delta \sqrt{|I_x|}} \non \\
&\leq& \Pi^n 2^{-n S(\Phi|P) + Kd \delta \sqrt{an}},
\eea
as required.

Now we can use Lemma \ref{L4}, taking $\Lambda = \Pi^n_{\Phi,\delta}(x^n)$ and $\rho = \Phi^{(n)}_{x^n}$. We can therefore conclude, using $\tr (B) \geq (\eta - \lambda) \mu_2^{-1}$ from Lemma \ref{L4}, that
\be
\tr(B) \geq \left( \eta - \frac{ad}{\delta^2}\right)2^{(nS(\Phi |P) - Kd\sqrt{a}\delta\sqrt{n})}.
\ee
\end{proof}

\begin{lemma}\label{L6}
Let $x^n$ be of type $P$. Then, the probability that the output state $\Phi_{x^n}^{(n)}$ is typical with respect to the variance typical projection $\Pi^n_{P\sigma, \delta \sqrt{a}}$ is lower bounded as follows,
\be
\tr \left( \Phi_{x^n}^{(n)} \Pi^n_{P\sigma, \delta \sqrt{a}} \right) \geq 1 - \frac{ad}{\delta^2}.
\ee
\end{lemma}
\begin{proof}
First diagonalise
\be
P\sigma = \sum_{j=1}^d q_j \bar{\pi_j}
\ee
and define the CPT map $\Psi: \mathcal{B}(\mathcal{H}) \mapsto \mathcal{B}(\mathcal{K})$ by,
\be\label{Psi}
\Psi (\omega) = \sum_{j=1}^d \bar{\pi_j} \, \omega \, \bar{\pi_j}.
\ee
We now show the following,
\be\label{ineqL6}
\Pi^n_{P\sigma, \delta \sqrt{a}} \geq \Pi^n_{\Psi \Phi,\delta} (x^n).
\ee
Let $\bar{\pi}^n = \bar{\pi}_{j_{1}} \otimes \dots \otimes \bar{\pi}_{j_{n}}$ be one of the product states comprising
\be
\Psi \Phi_x = \sum_{j=1}^d q(j|x) \bar{\pi_j}
\ee
and consequently
\be
\Pi^n_{\Psi \Phi,\delta} (x^n) = \bigotimes_{x \in \mathcal{X}} \Pi^n_{\Psi \Phi_x, \delta}
\ee
and
\be
\bigg| N(j|j^{I_x}) - |I_x| q_{j|x} \bigg| \leq \delta \sqrt{|I_x|} \sqrt{q_{j|x} ( 1 - q_{j|x})}.
\ee
Since $x^n$ is of type $P$ we have $|I_x| = n P(x)$ and $q_j = \sum_{x \in \mathcal{X}} P(x) q_{j|x}$, then
\bea
\big| N(j|j^{n}) - n q_{j} \big| &\leq&  \sum_{x \in \mathcal{X}} \bigg| N(j|j^{I_x}) - |I_x| q_{j|x} \bigg| \non \\
&\leq& \sum_{x \in \mathcal{X}} \delta \sqrt{n} \sqrt{P(x)} \sqrt{q_{j|x} ( 1 - q_{j|x})} \non \\
&\leq&  \delta \sqrt{n} \sqrt{a} \sqrt{\sum_{x \in \mathcal{X}} P(x) q_{j|x} ( 1 - q_{j|x})} \non \\
&\leq&  \delta \sqrt{n} \sqrt{a} \sqrt{q_{j} ( 1 - q_{j})},
\eea
using the Cauchy-Schwartz inequality and concavity of $x \mapsto x(1-x)$ and $q_j = \sum_{x \in \mathcal{X}} P(x) q_{j|x}$.

We can conclude that $\bar{\pi}^n = \bar{\pi}_{j_{1}} \otimes \dots \otimes \bar{\pi}_{j_{n}}$ contributes to $\Pi^n_{P\sigma, \delta \sqrt{a}}$ and therefore $\Pi^n_{P\sigma, \delta \sqrt{a}} \geq \Pi^n_{\Psi \Phi,\delta} (x^n)$, as required.

Using the definition of the CPT map $\Psi$ given by Equation (\ref{Psi}) and the trace preserving property of quantum channels, we obtain the following
\bea
\tr (\Phi_{x^n}^{(n)} \Pi_{P \sigma,\delta \sqrt{a}}^n ) &=& \tr ( \Psi^{\otimes n} (\Phi_{x^n}^{(n)} \Pi^n_{P \sigma, \delta \sqrt{a}})) \non \\
&=& \tr\left( \sum ({\bar{\pi}}_{j_1} \otimes \dots \otimes \bar{\pi}_{j_n})  \, \Phi_{x^n}^{(n)} \Pi_{P \sigma , \delta \sqrt{a}}^n \, ({\bar{\pi}}_{j_1} \otimes \dots \otimes \bar{\pi}_{j_n})\right) \non \\
&=& \tr\left( \sum ({\bar{\pi}}_{j_1} \otimes \dots \otimes \bar{\pi}_{j_n})  \, \Phi_{x^n}^{(n)} \, ({\bar{\pi}}_{j_1} \otimes \dots \otimes \bar{\pi}_{j_n}) \Pi_{P \sigma , \delta \sqrt{a}}^n \right) \non \\
&=& \tr(\Psi^{\otimes n} (\Phi_{x^n}^{(n)}) \Pi^n_{P \sigma, \delta \sqrt{a}}) \non \\
&\geq& \tr(\Psi^{\otimes n} (\Phi_{x^n}^{(n)}) \Pi^n_{\Psi \Phi,\delta} (x^n)) \non \\
&\geq& 1 -\frac{ad}{\delta^2},
\eea
where the first inequality is by Equation \ref{ineqL6} and the final inequality is by Lemma \ref{L5}.
\end{proof}

Next we introduce two fidelity lemmas. In the following $\rho$ is taken to be a state pure and $\sigma$ may be a mixed state.
The trace norm distance $D(\rho, \sigma)$ is defined for $\rho$ and $\sigma$ as follows
\be
D(\rho, \sigma) = \frac{1}{2} \big| \big| \rho - \sigma \big| \big|_1,
\ee
and the pure state fidelity
\be
F(\rho, \sigma) = \tr (\rho \,\sigma).
\ee

\begin{lemma}\label{L7}
Let $\rho = \ket{\psi} \bra{\psi}$ and $\sigma = \ket{\phi} \bra{\phi}$ be pure states. Then
\be
1 - F(\rho, \sigma) = D(\rho, \sigma)^2.
\ee
\end{lemma}
\begin{proof}
Take $\ket{\psi} = \ket{0} = \left( \begin{array}{c}
1 \\
0 \\
\end{array} \right)$ and $\ket{\phi} = \alpha \ket{0} + \beta \ket{1} = \left( \begin{array}{c}
\alpha \\
\beta \\
\end{array} \right)$, where $|\alpha|^2 + |\beta|^2 = 1$. Therefore
\be
\rho = \left(
         \begin{array}{cc}
           1 & 0 \\
           0 & 0 \\
         \end{array}
       \right),
\qquad
\sigma = \left(
         \begin{array}{cc}
           |\alpha|^2 & \alpha \beta \\
           \alpha \beta & |\beta|^2 \\
         \end{array}
       \right).
\ee
The fidelity $F(\rho, \sigma)$ can now be calculated,
\be
F(\rho, \sigma) =  \tr(\rho \sigma) = |\alpha|^2 .
\ee
The trace distance, $D(\rho,\sigma)=\frac{1}{2} \big| \big| \rho - \sigma \big| \big|_1 = \frac{1}{2} \tr \sqrt{(\rho - \sigma)^* (\rho - \sigma)}$ and therefore, using $|\alpha|^2 + |\beta|^2 =1$,
\be
D(\rho,\sigma) = |\beta|,
\ee
and therefore
\be
1 - F(\rho,\sigma) = D(\rho,\sigma)^2.
\ee
\end{proof}

In the following lemma we relax the assumption that \emph{both} states $\sigma$ and $\rho$ must be pure states.

\begin{lemma}\label{L8}
Let $\sigma$ be any arbitrary mixed state, and $\rho$ a pure state. Then
\be
D(\rho,\sigma)^2 \leq 1 - F(\rho,\sigma).
\ee
\end{lemma}
\begin{proof}
Write $\sigma = \sum_j q_j \pi_j$, where $\pi_j$ are pure states. Then, using Lemma \ref{L7},
\bea
1 - F(\rho, \sigma) &=& \sum_j q_j (1 - F(\rho, \pi_j))\non \\
&=& \sum_j q_j D(\rho, \pi_j)^2, \non \\
&\geq& \left(\sum_j q_j D(\rho, \pi_j)\right)^2 \non \\
&\geq& D(\rho, \sigma)^2.
\eea
The first inequality above is due to the convexity of $f(x) = x^2$. The second inequality is due to the convexity of $D(\rho, \sigma)$ (triangle inequality).
\end{proof}

Informally, the following lemma states that, under the trace norm (defined by Equation (\ref{TrNorm})), the state $\rho$ is disturbed by at most $\sqrt{8 \lambda}$ by the operator $X$, provided that $ 1- \tr(\rho X) \leq \lambda \leq 1$. In the proof of the strong converse theorem, the positive operator $X$ above will be replaced by the projector onto the typical subspace for $P \sigma$, denoted $\Pi^n_{P \sigma, \delta \sqrt{a}}$, and the state $\rho$ will be replaced by the output state $\Phi^{(n)}_{x^n}$. In this case the above lemma takes on the following important interpretation. If the probability that the output state is not typical (with respect to the typical subspace $P\sigma$) is less than $\lambda$, i.e. if $1- \tr(\Phi^{(n)}_{x^n} \, \Pi^n_{P \sigma, \delta \sqrt{a}}) \leq \lambda$, then under the trace norm, the state $\Phi^{(n)}_{x^n}$ is disturbed by at most $\sqrt{8 \lambda}$, when projected onto the typical subspace for the average output state $P \sigma$.

\begin{lemma}\label{L9}
Let $\rho$ be a state and $X$ a positive operator with $X \leq \mathds{1}$ and \\$ 1- \tr(\rho X) \leq \lambda \leq 1$. Then
\be
\big| \big|\rho - \sqrt{X} \; \rho \; \sqrt{X} \big| \big|_1 \leq \sqrt{8\lambda}.
\ee
\end{lemma}
\begin{proof}
Let $Y = \sqrt{X}$ and $\rho = \sum_k p_k \pi_k$, where $\pi_k$ are pure states and $p_k \geq 0$. We then have the following
\bea\label{L9ineq}
\big| \big| \rho - Y \rho Y \big|\big|_1^2 &\leq& \left( \sum_k p_k \big| \big| \pi_k - Y\pi_k Y \big| \big|_1\right)^2 \non \\
&\leq& \sum_k p_k \big| \big| \pi_k - Y \pi_k Y \big| \big|_1^2 \non \\
&\leq& 4 \sum_k p_k (1 - \tr(\pi_k Y \pi_k Y)) \non \\
&\leq& 8 \sum_k p_k (1 - \tr(\pi_k Y))\non \\
&\leq& 8 (1 - \tr(\rho Y)) \leq 8 \lambda.
\eea
The first inequality is by the triangle inequality,
\be
\big| \big| x + y \big| \big|_1 \leq \big| \big| x \big| \big|_1 + \big| \big| y \big| \big|_1.
\ee
The second is due to the convexity of $x \mapsto x^2$ and the third is due to Lemma \ref{L8}.

The next inequality is shown as follows. Since $\pi_k$ is a projection $\pi_k = (\pi_k)^2$, and
\bea
1 - \tr(\pi_k Y \pi_k Y) &=& \tr(\pi_k - \pi_k Y \pi_k Y) \non \\
&=& \tr(\pi_k -\pi_k Y) + \tr(\pi_kY - \pi_k Y \pi_k Y) \non \\
&=& 1 - \tr(\pi_k Y) + \tr(\pi_k Y \pi_k (\pi_k - \pi_k Y)).
\eea
But $|| Y || \leq 1$ and therefore $\tr(\pi_k Y \pi_k (\pi_k - \pi_k Y)) \leq \tr(\pi_k - \pi_k Y)$ and we have
\bea
1 - \tr(\pi_k Y \pi_k Y) \leq 2 (1 - \tr(\pi_k Y)).
\eea
The final inequality in (\ref{L9ineq}) uses $Y \geq X$ and $1 - \tr(\rho X) \leq \lambda$.
\end{proof}

Next we state and prove the strong converse theorem for memoryless quantum channels (Winter) \cite{Winter99}.

\subsection{Strong converse for a memoryless quantum channel}

\begin{theorem}\label{strongConv}
\rm{(Strong converse for memoryless quantum channels)}\\
For $\lambda \in (0,1)$ there exits a constant $K(\lambda,a,d)$ such that for every quantum channel $\Phi$ and $(n,\lambda)$-code
\be\label{theorem}
|\mathcal{M}| \leq 2^{\,(n \chi^*(\Phi) + K(\lambda,a,d) \sqrt{n})}.
\ee
\end{theorem}
\begin{proof}
Note that if we assume all codewords to be of the same type $P$, then we may tighten the above bound. We first show that the number of codewords is upper bounded as follows
\be\label{addAssumption}
|\mathcal{M}_P| \leq \frac{4}{1 - \lambda} \, 2^{\, (n I(P;\Phi) + 2 K(\lambda,a,d) \sqrt{n})},
\ee
taking $\delta = \frac{\sqrt{32 ad}}{1 - \lambda}$. The theorem then follows since there are at most $(n+1)^a$ possible types, where $a$ is the length of the alphabet $\mathcal{X}$, i.e. $a=\big| \mathcal{X} \big|$. We will demonstrate this once we have proved expression (\ref{addAssumption}).

First take an $(n, \lambda)$-code with the decoding operators $E_m^{n}$, i.e. 
\be
\tr( \Phi^{(n)} (\rho_m^{(n)}) E_m^{n}) > 1- \lambda,
\ee
where $\Phi^{(n)} (\rho_m^{(n)})$ is the output from a memoryless quantum channel acting on the input state $\rho_m^{(n)}$ and $\tr( \Phi^{(n)} (\rho_m^{(n)}) E_m^{n})$ is the probability of successful decoding.

To prove inequality (\ref{theorem}) we now construct the following new decoding operators $E_m^{n'}$
\be
E_m^{n'} = \Pi_{P \sigma, \delta \sqrt{a}}^{n} \; E_m^{n} \; \Pi_{P \sigma, \delta \sqrt{a}}^{n},
\ee
where $\Pi_{P \sigma, \delta \sqrt{a}}^{n}$ is the projection onto the typical subspace for $P\sigma$, the average output state of the channel $\Phi$.

Let $C^n_m = C^{n}(m) = x^n$, for $m \in \mathcal{M}$.
Then, $(C^n_m,E_m^{n'})$ is an $(n, \frac{1+\lambda}{2})$-code, since for $m \in \{1, \dots N_n \}$ the probability $\tr \left( \Phi^{(n)} (\rho_m^{(n)}) E_m^{n'} \right)$ can be written as follows,
\bea
\tr \left( \Phi^{(n)} (\rho_m^{(n)}) E_m^{n'}  \right) &=& \tr\left( \Phi^{(n)} (\rho_m^{(n)}) E_m^{n}  \right) \non \\
&-& \tr \left( \left(\Phi^{(n)} (\rho_m^{(n)}) - \Pi_{P \sigma, \delta \sqrt{a}}^{n} \Phi^{(n)} (\rho_m^{(n)}) \Pi_{P \sigma, \delta \sqrt{a}}^{n} \right) E_m^{n} \right) \non \\
&\geq&  1 - \lambda - \big| \big| \Phi^{(n)} (\rho_m^{(n)}) - \Pi_{P \sigma, \delta \sqrt{a}}^{n} \Phi^{(n)} (\rho_m^{(n)}) \Pi_{P \sigma, \delta \sqrt{a}}^{n} \big| \big|_1 \non\\
&\geq&  1 - \lambda - \sqrt{\frac{8ad}{\delta^2}} \non \\
&=& \frac{1 - \lambda}{2}.
\eea
The first inequality holds since $(C^n_m, E_m^n)$ is assumed to be an $(n, \lambda)$-code and since $E^{n}_m \leq \mathds{1}$.
The second inequality uses Lemma \ref{L6} and Lemma \ref{L9} as follows. Lemma \ref{L9} states that, for $1 - \tr(\rho X) \leq \lambda' \leq 1$, the following inequality holds
\be
\big| \big|\rho - \sqrt{X} \; \rho \; \sqrt{X} \big| \big|_1 \leq \sqrt{8\lambda'}.
\ee
In our case, we put $X = \Pi^{n}_{P \sigma, \delta \sqrt{a}}$ and $\rho= \Phi^{(n)}(\rho_m^{(n)})$ and by Lemma \ref{L6}
\be
1 - \tr(\Phi^{(n)}(\rho_m^{(n)}) \Pi^{n}_{P\sigma, \delta \sqrt{a} }) \leq \frac{ad}{\delta^2}.
\ee
The second inequality then follows from Lemma \ref{L9}, taking $\lambda' = \frac{ad}{\delta^2}$.
The code $(C^n_m, E_m^n)$ is therefore an $(n, \frac{1+\lambda}{2})$, since
\bea
p_e(C^n_m, E_m^{n'}) &=& 1 - \tr \left( \Phi^{(n)} (\rho_m^{(n)}) E_m^{n'}  \right) \non \\
&\leq& \frac{1+\lambda}{2}.
\eea
Since $\tr \left( \Phi^{(n)} (\rho_m^{(n)}) E_m^{n'}  \right) \geq \frac{1 - \lambda}{2}$, the operator $E_m^{n'}$ is an $\eta = \frac{1 - \lambda}{2}$ shadow
of $\Phi^{(n)} (\rho_m^{(n)})$ (see Equation (\ref{shadow}) for definition of $\eta$ shadow) and by Lemma \ref{L5},
\be\label{trD}
\tr \left( E_m^{n'} \right) \geq \frac{1 - \lambda}{4} \, 2^{( S(\Phi |P) - Kd\sqrt{a}\delta\sqrt{n})}.
\ee
The sequence $x^n$ is of type $P$ so we can use Lemma \ref{L3} as follows,
\be
\sum_{m \in \mathcal{M}} \tr \, E_m^{n'} \leq \tr \, \Pi^{n}_{P \sigma, \delta \sqrt{a}}
\leq 2^{(nS(P \sigma)+ Kd\sqrt{a}\delta \sqrt{n})}.
\ee
Therefore,
\be
\big| \mathcal{M} \big| \, \tr \, E_m^{n'}
\leq 2^{(nS(P \sigma)+ Kd\sqrt{a}\delta \sqrt{n})}
\ee
and using Equation (\ref{trD}) we obtain an upper bound on the number of codewords,
\be
\big| \mathcal{M} \big| \leq \frac{4}{1 - \lambda} 2^{(-nS(\Phi|P) + Kd \sqrt{a} \delta \sqrt{n})}.2^{(nS(P \sigma) + Kd \sqrt{a} \delta \sqrt{n})}
\ee
and using the defintion of mutual information (Equation \ref{mutInf})
\be
\big| \mathcal{M} \big| \leq \frac{4}{1 - \lambda} 2^{(nI(P;\Phi) +2Kd\sqrt{a}\delta \sqrt{n})},
\ee
using $I(P;\Phi)= S(P \sigma) - S(\Phi|P)$. This proves Equation (\ref{addAssumption}).

Next we make use of the fact that the number of types is upper bounded by $(1 + n)^a$ where $a$ is the length of the input alphabet. This result has the interpretation that the number of types increase only polynomially with $n$. 

With the additional assumption that all the codewords are of the same type, we have 
\bea
\sum_{P} \big| \mathcal{M}_P \big| &\leq&  k  \sum_{P} 2^{(nI(P;\Phi) +2Kd\sqrt{a}\delta \sqrt{n})} \non \\
&=& k \, (1+n)^a  \, 2^{(nI(P;\Phi) +2Kd\sqrt{a}\delta \sqrt{n})} \non \\
&\leq& k \, (1+n)^a  \, 2^{(n\chi^*(\Phi) +2Kd\sqrt{a}\delta \sqrt{n})}
\eea
where, $k = \frac{4}{1 - \lambda}$. Clearly, $2^{(2Kd\sqrt{a}\delta \sqrt{n})}$ dominates the $k \, (1+n)^a$ term for $n$ large, i.e.
\be
k \, (1+n)^a \leq 2^{(c \sqrt{n})}.
\ee
Therefore
\be
\sum_{P} \big| \mathcal{M}_P \big| \leq 2^{(n\chi^*(\Phi) + 2Kd\sqrt{a}\delta \sqrt{n})}.
\ee
\end{proof}

\section{Coding theorem for a periodic quantum channel}

Recall that a periodic channel acting on an $n$-fold density operator can be written in the following form
\be\label{periodicSect} \Phi^{\left(n \right)} \left( \rho
^{\left(n \right)} \right) = \frac{1}{L} \sum_{i=0}^{L-1} \left(
\Phi_i \otimes \Phi_{i+1} \otimes \cdots\otimes \Phi_{i+n-1}
\right) \left( \rho ^{\left(n \right)} \right), \ee
where $\Phi_i$ are CPT maps and the index is cyclic modulo the period $L$. The Holevo quantity for the $i$-th branch of the channel is denoted $\chi_i(\{p_j,\rho_j\})$.
The product-state capacity of the channel (\ref{periodicSect}) is given by
\be\label{cap_PerSect}
C_p \left( \Phi \right) = \frac{1}{L} \sup_{\{p_j, \rho_j\}}
\sum_{i=0}^{L-1} \chi_i(\{p_j, \rho_j\}). \ee
The proof for the direct part of this theorem is provided in Appendix B.

We show that the strong converse theorem \emph{does not} hold for the above expression. As a consequence it does not provide a sharp upper-bound on the rate at which classical information can be transmitted over the periodic channel.

Recall that the strong converse for a memoryless quantum channel $\Phi$ states that
\be
\log_2 \big| \mathcal{M} \big| \leq n\chi^*(\Phi) +K\sqrt{n}.
\ee
The strong converse for the periodic quantum channel does not hold since the capacity of the channel, $C_p$, is upper bounded as follows,
\be\label{capIneq}
C_p < \bar{C_p},
\ee
where,
\be\label{CBar}
\bar{C_p} = \frac{1}{L} \sum_{i=0}^{L-1} \sup_{\{p_j, \rho_j\}} \chi_i(\{p_j, \rho_j\}).
\ee
However, each branch of the periodic channel $\Phi^{(n)}$ can be written as a memoryless channel of dimension $d'=dL$ as follows,
\be
(\Phi_i \otimes \Phi_{i+1} \otimes \cdots \otimes \Phi_{i+L-1})^{\otimes \frac{n}{L}}.
\ee

\begin{remark}
Note that equality for expression (\ref{capIneq}) can be shown to hold for the depolarising channel and it is shown in Section \ref{ampDampBranch} that a strict inequality holds for the amplitude-damping channel.
\end{remark}
Since we are limited to using product-state inputs, the product-state capacity of each channel branch is additive and therefore equal, i.e.
\bea\label{capBranch}
\chi^*(\Phi_0 \otimes \Phi_1 \otimes \dots \otimes \Phi_{L-1})^{\otimes \frac{n}{L}} &=& \frac{n}{L} \left(\chi^*(\Phi_0) + \chi^*(\Phi_1) + \dots + \chi^*(\Phi_{L-1}) \right) \non \\
&=& \frac{n}{L} \sum_{i=0}^{L-1} \chi^*(\Phi_i).
\eea
\begin{remark}
Notice that, like Equation (\ref{capBranch}), $\bar{C_p}$ (Equation (\ref{CBar})) is the average of the Holevo capacities of the $L$ channel branches. Therefore, if we knew in advance which channel branch will be chosen, then we could take the rate $R=\bar{C_p}$ with the probability of error $p_e \leq \epsilon$ and the strong converse would immediately follow, using Theorem \ref{strongConv}.
However, we do not have this additional information. We compromise by assuming to know the channel branch in advance and then compensate for this assumption by taking $p_e \leq \frac{1}{L} + \epsilon$.
But, the rate $R=\bar{C_p}$ could be too high to take for certain channel branches, (branches consisting of the amplitude damping channels, for example).
We therefore must choose a rate in between $C_p$ and $\bar{C_p}$ with $p_e \leq \frac{1}{L} + \epsilon$.
\end{remark}
We choose a rate in between the two expressions above and we demonstrate the strong converse using this ``compromised'' rate. We refer to this result as the ``weakened'' strong converse to the coding theorem for the periodic quantum channel.

More precisely, we choose a rate $R$ such that
\be\label{rate}
C_p < R < \bar{C_p}.
\ee
The direct part of the theorem, i.e. with rate below the upper bound, may be argued as follows.

We choose a particular channel branch, say, $i=1$
and simply take the code \\$(C^n,E^n)$ for this (memoryless) product channel,
\be
(\Phi_1 \otimes \Phi_2 \otimes \cdots \otimes \Phi_L)^{\otimes n},
\ee
with rate $R  < \bar{C_p}$.
However, we must pay a penalty (thereby diluting the theorem and reducing the theorem for the periodic channel to that of a single branch) for fixing a particular branch and therefore must construct the code such that,
\be
p_e  \leq \frac{1}{L} + \epsilon,
\ee
where $\frac{1}{L}$ is the probability of choosing a particular branch.

Using Theorem \ref{strongConv} we now argue that the strong converse holds for the rate given by Equation (\ref{rate}).
Let $(C^n, E^n)$ be an $(n, \lambda)$-code. Therefore
\be
p_e = \frac{1}{L} \sum_{i=0}^{L-1} p_{e}^i \leq \lambda,
\ee
where $p_e$ denotes the average probability of error for the periodic channel and $p_e^i$ denotes the probability of error for the $i$-th channel branch.
Therefore we can apply Theorem \ref{strongConv} as follows
\bea
\log_2 \big| \mathcal{M}_i \big| &\leq& n C_p + K(\lambda,a,d) \sqrt{n} \non \\
&=&  \frac{n}{L} \sup_{\{p_j, \rho_j \}} \sum_{i=0}^{L-1} \chi_i(\{p_j, \rho_j \}) + K(\lambda,a,d) \sqrt{n}.
\eea
Also, since $p_e \leq \lambda$, $\exists i$ such that $p_e^i \leq \lambda$,
\bea
\log_2 \big| \mathcal{M}_i \big| &\leq&  \bar{C_p}(\Phi_i^{\otimes \frac{n}{L}}) + K(\lambda,a,d) \sqrt{n} \non \\
&=&  \frac{n}{L} \sum_{i=0}^{L-1} \chi^*(\Phi_i) + K(\lambda,a,d) \sqrt{n},
\eea
where,
\be
\Phi_i^{\otimes \frac{n}{L}} = (\Phi_i \otimes \Phi_{i+1} \otimes \cdots \otimes \Phi_{i+L-1})^{\otimes \frac{n}{L}}.
\ee

We can now conclude that, although the strong converse theorem does not hold for the product-state capacity of the periodic channel defined by Equation (\ref{periodicSect}), the ``weakened " strong converse does hold, as argued above. 

\section{Summary}

We introduced the method of types \cite{Csiszar98} and provided Winter's proof \cite{Winter99} of the strong converse theorem for memoryless quantum channels, updating the notation and providing detailed proofs for the lemmas used to prove the theorem.

Next we considered the strong converse theorem for the periodic quantum channel introduced in Chapter 4 and showed that the strong converse does not hold for this channel. This conclusion is drawn based on a result shown in Chapter 4, namely that due to the fact that the average and the supremum cannot be interchanged in the formula for calculating the product state capacity of the periodic channel with the amplitude damping channel branches, this formula cannot be re-written in a way which would lead to a direct application of the strong converse theorem. 

We do show, however, that if we weaken the scenario by assuming to know that the channel chosen is known in advance, then the strong converse theorem does hold.

The direct part of the channel coding theorem for the periodic quantum channel is provided in Appendix \ref{ProofPSC}.


\appendix
\addappheadtotoc

\chapter{Carath\'eodory's Theorem \& an application to minimal \\ optimal-ensembles} \section{Carath\'eodory's Theorem}\label{Cara}

Carath\'eodory's theorem is stated as follows
\begin{theorem}\label{CaraTheorem}
Let $\mathcal{S} \subset \mathbb{R}^d$ be a set. Then every point $x$ in the convex hull of $\cal S$ can be represented as a convex combination of $d +1$ points from $\cal S$, i.e.
\be
x = \sum_{i=0}^d \alpha_i x_i,
\ee
\end{theorem}
where $x_i \in \mathcal{S}$, $\alpha_i \geq 0$ and $\sum_i \alpha_i =1$.


\section{Application}\label{Datta}

Next we prove that ensembles containing  at most $d^2$ pure states are sufficient to maximise the Holevo quantity of a CPT map. This proof was provided by N. Datta \cite{Datta}. We first prove that $d^2+1$ states are sufficient, using Carath\'eodory's theorem \cite{Caratheodory, Steinitz}.

\begin{proof}

Note that the set of density operators is described by $d^2 -1$ parameters. Let
\be
f(\rho) = (f_1(\rho), \cdots, f_{d^2 -1}(\rho), f_{d^2}(\rho)) = (f_1(\rho), \cdots, f_{d^2 -1}(\rho), S(\Phi(\rho)),
\ee
be the vector-valued function, with the first $d^2-1$ components corresponding to the linear degrees of freedom of $\rho$, for every density operator $\rho$.

Consider the set of images $f(\mathcal{P}) \subset \mathbb{R}^{d^2}$ of pure states $\cal P$. To every ensemble of pure states $\mathcal{E} := \{ q_i, \ket{\psi_i}\bra{\psi_i} \}$, we can associate the point
\be
\vec{f_{\cal E}} := \sum_i q_i \, f(\ket{\psi_i}\bra{\psi_i}) \subset \mathbb{R}^{d^2}
\ee
in the convex hull of $f(\cal P)$. Moreover, the Holevo quantity
\be
\chi(\mathcal{E}) = S\left(\sum_i q_i \Phi (\ket{\psi_i}\bra{\psi_i})\right) - \sum_i q_i S(\Phi (\ket{\psi_i}\bra{\psi_i}))
\ee
is a function of this vector only, i.e.
\be
\chi(\mathcal{E}) = G(\vec{f_{\mathcal{E}}})
\ee
for some function $G$.

To see this note that the average input state $\sum_i q_i \ket{\psi_i}\bra{\psi_i}$ and thus the corresponding entropy are completely specified by the first $d^2 -1$ components of $\ket{\psi_i}\bra{\psi_i}$ because of linearity of the function $f(\cdot)$, and $\sum_i q_i \, S (\ket{\psi_i}\bra{\psi_i})$ is the last entry of $\vec{f_{\cal E}}$.

Now, let $\mathcal{E}$ be an ensemble maximising $\chi(\cal E)$, then by Carath\'eodory's theorem, $\vec{f_{\cal E}}$ can be represented as
\be
\vec{f_{\cal E}} = \sum_{i=0}^{d^2 -1} p_i \, f(\ket{\phi_i}\bra{\phi_i}).
\ee

We then define the pure state ensemble $\mathcal{E}' := \{ p_i, \ket{\phi_i}\bra{\phi_i} \}_{i=0}^{d^2 -1}$ which consists of only $d^2 +1$ pure states. By definition ,
\be
\vec{f_{\cal E}} = \vec{f_{\mathcal{E}'}},
\ee
hence, $\chi(\mathcal{E}') = \chi(\cal E)$, as desired.

The above proof shows that $d^2+1$ states are sufficient. The stronger statement can be obtained using a strengthening of Carath\'eodory's theorem by Fenchel and Eggleston (\cite{Eggleston}, Theorem 18).

This states that if $\mathcal{S} \subset \mathbb{R}^m$ is the union of at most $m$ connected subsets, then every $x$ in the convex hull of $\mathcal{S}$ can be represented as a convex combination of at most $m$ points in $\mathcal{S}$.

Since $f$ is continuous and the set of pure states $\cal P$ is a compact, connected set, the image $f(\mathcal{P}) \subset \mathbb{R}^{d^2}$ is also compact and connected. Hence the union $f(\mathcal{P})$ is the union of only one connected set and we obtain the desired result.
\end{proof}

The above result, that it is sufficient to consider ensembles containing just $d^2$ pure states when maximising the Holevo quantity, was first shown by Davies \cite{Davies78}. Note that his proof also utilises Carath\'eodory's Theorem (Theorem \ref{CaraTheorem}).
\chapter{Product-state capacity of a periodic quantum channel} \section{Proof of the product-state capacity of a periodic quantum channel}\label{ProofPSC}

The following proof is a special case of the proof, given by Datta and Dorlas in
\cite{DD07Per}, for the product-state capacity of a channel with
\emph{arbitrary} Markovian noise correlations.

\section{Preliminaries}

A general quantum channel is given by completely
positive trace-preserving (CPT) maps $\Phi^{(n)}: {\cal B}({\cal
H}^{\otimes n}) \to {\cal B}({\cal K}^{\otimes n})$, where ${\cal
H}$ and ${\cal K}$ are the input and output Hilbert spaces of the
channel. Here we consider a periodic channel of the following
form
\begin{equation}  \Phi^{(n)}(\rho^{(n)}) = \frac{1}{L}
\sum_{i=0}^{L-1} (\Phi_i \otimes \Phi_{i+1} \otimes \dots \otimes
\Phi_{i+n-1})(\rho^{(n)}), \label{perch}\end{equation}  where we
assume that a set of CPT maps $\Phi_i:{\cal B}({\cal H}) \to {\cal
B}({\cal K})$ ($i=0,\dots,L-1$) is given, and the index is cyclic
modulo the period $L$.

If we denote the Holevo quantity for the $i$-th branch by
$\chi_i$, i.e. $$ \chi_i(\{p_j,\rho_j\}) = S \left( \sum_j p_j
\Phi_i(\rho_j) \right) - \sum_j p_j S(\Phi_i(\rho_j)), $$ then we
shall prove that the product capacity of the channel (\ref{perch})
is given by \begin{equation} C_p(\Phi) = \sup_{\{p_j,\rho_j\}}
\frac{1}{L} \sum_{i=0}^{L-1} \chi_i(\{p_j,\rho_j\}).
\end{equation}

\section{The Quantum Feinstein Lemma}
\label{qfein}

The direct part of the theorem follows from
\begin{theorem}
\label{SuccProbDirect} Given $\epsilon > 0$, there exists $n_0 \in \NN$
such that for all $n\geq n_0$ there exists $N_n \geq
2^{n(C(\Phi)-\epsilon)}$ and there exist product states ${\tilde
\rho}^{(n)}_1, \dots, {\tilde \rho}^{(n)}_{N_n} \in {\cal S}({\cal
H}^{\otimes n})$ and positive operators $E^{(n)}_1,\dots,
E^{(n)}_{N_n} \in {\cal B}({\cal K}^{\otimes n})$ such that
$\sum_{k=1}^{N_n} E^{(n)}_k \leq \one$ and
\begin{equation} \tr  \Phi^{(n)} \left( {\tilde
\rho}^{(n)}_k \right) E^{(n)}_k  > 1-\epsilon, \label{error}
\end{equation} for each $k$.
\end{theorem}

\begin{proof} We first construct a preamble to the
code which serves to identify the first branch $i$ chosen. To
distinguish the initial branch, notice first of all that the
corresponding CPT maps $\Phi_i$ need not all be distinct! However,
we may assume that there is no internal periodicity of these maps;
otherwise the channel be contracted to a single such period. This
means, that for any two states $i,i' \in \{0,\dots,L-1\}$ ($i <
i'$) there exists $k \leq L-1$ such that $\Phi_{i+k} \neq
\Phi_{i'+k}$. Then choose $\omega = \omega_{i,i'}$ such that
\begin{equation} f := F(\Phi_{i+k}(\omega), \Phi_{i'+k}(\omega)) <
1. \end{equation} In the following we write $\Phi_i^{(n)}$ for the
branch of the channel with initial state $i$, i.e.
\begin{equation} \Phi_i^{(n)}(\rho^{(n)}) = (\Phi_i \otimes
\Phi_{i+1} \otimes \dots \otimes \Phi_{i+n-1})(\rho^{(n)}).
\label{i-branch} \end{equation}

\begin{lemma} For any $0 \leq i < i' \leq L-1$, let $\omega$ be
a state as above. Then \begin{equation} F \left(
\Phi^{(mL)}_i(\omega^{\otimes mL}),
\Phi_{i'}^{(mL)}(\omega^{\otimes mL}) \right) \to 0 \end{equation}
as $m \to \infty$. \label{AppL1}
\end{lemma}

\begin{proof}
\begin{eqnarray} \lefteqn{ F \left(
\Phi^{(mL)}_i(\omega^{\otimes mL}),
\Phi_{i'}^{(mL)}(\omega^{\otimes mL}) \right) } \non \\ &=& \left[
F \left( \Phi_i^{(L)}(\omega^{\otimes L}), \Phi_{i'}^{(L)}
(\omega^{\otimes L}) \right) \right]^m \non \\ &\leq & \left[
F(\Phi_{i+k}(\omega), \Phi_{i'+k}(\omega)) \right]^m = f^m \to 0.
\end{eqnarray}
\end{proof}

We now introduce, for any pair of states $\sigma, \sigma'$ on
$\cal K$, and $\gamma, \gamma' > 0$, the difference operators
\begin{equation} A^{(M)}_{\sigma,\sigma'} = \gamma
\sigma^{\otimes M} - \gamma' (\sigma')^{\otimes M}. \end{equation}
Let $\Pi^\pm$ be the orthogonal projections onto the eigenspaces
of $A_{\sigma,\sigma'}^{(M)}$ corresponding to all non-negative,
and all negative eigenvalues, respectively. In \cite{DD07Per} we proved
the following lemma
\begin{lemma} \label{AppL2} Suppose that for a given $\delta >0$,
\begin{equation} |\tr
[|A_{\sigma,\sigma'}^{(M)}|] - (\gamma+\gamma')| \leq \delta.
\end{equation} Then \begin{equation}
|\tr [\Pi^+ (\sigma)^{\otimes M}] - 1| \leq \frac{\delta}{2
\gamma}
\end{equation} and
\begin{equation}
|\tr [\Pi^- (\sigma')^{\otimes M}]-1| \leq \frac{\delta}{2
\gamma'}. \end{equation}
\end{lemma}

To compare the outputs of all the different branches of the
channel, we define projections ${\tilde \Pi}_i$ on the tensor
product space $\bigotimes_{0 \leq i < i'<L} \mathcal{K}^{\otimes M} =
\mathcal{K}^{\otimes ML_2}$ with $L_2 = \binom{L}{2}$ as follows
\begin{equation} {\tilde \Pi}_i = \bigotimes_{0 \leq i_1 < i_2 <L}
\Gamma_{i_1,i_2}^{(i)}
\end{equation}
where,
\begin{equation}
 \Gamma_{i_1,i_2}^{(i)} =
\left\{ \begin{array}{lcl} {\one}_dM &\mbox{ if } &i_1 \neq i
\mbox{
and } i_2 \neq i \\ \Pi_{i_1,i}^- &\mbox{ if } &i_2 = i \\
\Pi_{i,i_2}^+ &\mbox{ if } &i_1 = i. \end{array} \right.
\end{equation}

Notice that it follows from the fact that $\Pi_{i,i'}^+
\Pi_{i,i'}^- = 0$, that the projections ${\tilde \Pi}_i$ are also
disjoint,
\begin{equation} {\tilde \Pi}_i {\tilde \Pi}_{i'} = 0 \quad
{\hbox{for }} \, i \ne i'.
\end{equation}

It now follows easily with the help of the previous lemma and the
inequalities \cite{NC}
\begin{equation} \tr (A_1) +
\tr (A_2) - 2 F(A_1,A_2) \leq || A_1 - A_2 ||_1 \leq \tr
(A_1) + \tr (A_2)
\end{equation} for any two positive operators $A_1$ and $A_2$,
that these projections distinguish the relevant initial branches.
Indeed, if we introduce the corresponding preamble state
\begin{equation} \omega^{(ML_2)} = \left(\bigotimes_{i_1 < i_2}
\omega_{i_1,i_2}^{\otimes M} \right), \label{preamble}
\end{equation} then we have

\begin{lemma}
For all $i \in \{0,\dots,L-1\}$, \begin{equation} \lim_{M \to
\infty} \tr \left[ {\tilde \Pi}_i\, \Phi_i^{(ML_2)} \left(
\omega^{(ML_2)} \right) \right] = 1. \end{equation}
\end{lemma}
In the following we fix $M$ so large that \begin{equation}
\tr \left[ {\tilde \Pi}_i \,\Phi_i^{\otimes ML_2} \left(
\omega^{(ML_2)} \right) \right] > 1-\delta \label{Piper}
\end{equation} for all $i \in \{0,\dots,L-1\}$. We also assume that
$M$ is a multiple of $L$ so that \ref{AppL1} applies. The product
state $\omega^{(ML_2)}$, defined through \ref{preamble} is used
as a preamble to the input state encoding each message, and serves
to distinguish between the different branches, $\Phi_i^{(n)}$, of
the channel. If $\rho_k^{(n)} \in {\cal{B}}({\cal{H}}^{\otimes
n})$ is a state encoding the $k^{th}$ classical message in the set
${\cal{M}}_n$, then the $k^{th}$ codeword is given by the product
state $$ \omega^{(ML_2)} \otimes \rho_k^{(n)}.$$ Note that, since
$M$ is a multiple of $L$, the index of the first channel branch
applying to $\rho_k$ is also $i$.

Continuing with the proof of Theorem \ref{SuccProbDirect}, let the
maximum of the mean Holevo quantity $\frac{1}{L} \sum_{i=0}^{L-1}
\chi_i$ be attained for an ensemble $\{ p_{j},
\rho_{j}\}_{j=1}^{J}$. Denote $\sigma_{i,j} = \Phi_i(\rho_{j})$,
${\bar \sigma}_i = \sum_{j=1}^{J} p_{j} \Phi_i(\rho_{j})$.

Choose $\delta > 0$. We will relate $\delta$ to $\epsilon$ at a
later stage. Consider the typical subspaces ${\overline{\cal
T}}_{i,\epsilon}^{(n)}$ of ${\cal{K}}^{\otimes n}$, with
projection ${\bar P}_{i,n}$ such that if ${\bar\sigma}_i$ has a
spectral decomposition
\begin{equation} {\bar \sigma}_i = \sum_{k} {\bar
\lambda}_{i,k} |\psi_{i,k} \rangle \langle \psi_{i,k}|
\end{equation} then if $\uk = (k_1,\dots,k_n)$,
$|\psi_{i,k_1} \rangle \otimes \dots \otimes |\psi_{i,k_n} \rangle
\in \overline{{\cal T}}_{i,\epsilon}^{(n)}$ if and only if
\begin{equation} \left| \frac{1}{n} \sum_{j=1}^n \log {\bar
\lambda}_{i,k_j} + S({\bar \sigma}_i) \right| <
\frac{\epsilon}{4}.
\end{equation}  Then, for $n$ large enough,
\begin{equation} \tr ({\bar P}_{i,n} {\bar \sigma}_i^{\otimes n})
> 1-\delta^2. \end{equation}
For any given initial index $i$, we let ${\overline{\cal
V}}_{i,\epsilon}^{(n)}$ be the subspace of ${\cal{K}}^{\otimes n}$
spanned by the vectors $|\psi_{i,k_1} \rangle \otimes
|\psi_{i+1,k_2} \rangle \otimes |\psi_{i+n-1,k_n}\rangle$, where
$|\psi_{i,k_1} \rangle \otimes |\psi_{i,k_{L+1}} \rangle \otimes
\dots \otimes |\psi_{i,k_{[(n-1)/L]L+1}}  \rangle \in
{\overline{\cal T}}_{i,\epsilon}^{([(n-1)/L]+1)}$, etc. Clearly,
if we denote ${\bar P}_i^{(n)}$ the projection onto
${\overline{\cal V}}_{i,\epsilon}^{(n)}$, then for $n$ large
enough,
\begin{equation} \tr ({\bar P}_{i}^{(n)} {\bar \sigma}_i
\otimes {\bar \sigma}_{i+1} \otimes \dots \otimes {\bar
\sigma}_{i+n-1}) > 1-\delta^2. \label{Pn} \end{equation} Moreover,
if $|\psi_{i,k_1} \rangle \otimes |\psi_{i+1,k_2} \rangle \otimes
|\psi_{i+n-1,k_n}\rangle \in {\overline{\cal
V}}_{i,\epsilon}^{(n)}$ then \begin{equation} \left| \frac{1}{n}
\sum_{j=1}^n \log {\bar \lambda}_{i+j-1,k_j} + \frac{1}{L}
\sum_{i=0}^{L-1} S({\bar \sigma}_i) \right| < \frac{\epsilon}{4}.
\label{Tbar}
\end{equation} Let $n_1$ be so large that (\ref{Pn}) and
(\ref{Tbar}) hold for $n \geq n_1$.

We need a similar result for the average entropy
\begin{equation} {\bar S} = \frac{1}{L} \sum_{i=0}^{L-1}
\sum_{j=1}^{J} p_{j}\, S(\sigma_{i,j}). \label{Sbar}
\end{equation}

\begin{lemma}\label{AppL4} Fix $i \in \{0,\dots,L-1\}$.
Given a sequence $\uj = (j_1,\dots,j_n)$ with $1 \leq j_r \leq
J(i+r-1)$, let $P_{i,\uj}^{(n)}$ be the projection onto the
subspace of ${\cal{K}}^{\otimes n}$ spanned by the eigenvectors of
$\sigma_{i,\uj}^{(n)} = \sigma_{i,j_1} \otimes \dots \otimes
\sigma_{i+n-1,j_n}$ with eigenvalues $ \lambda_{\uj, \uk}^{(n)} =
\prod_{r=1}^n \lambda_{i+r-1,j_r,k_r}$ such that
\begin{equation} \left| \frac{1}{n} \log \lambda_{\uj, \uk}^{(n)}
+ {\bar S} \right| < \frac{\epsilon}{4}. \end{equation} For any
$\delta > 0$ there exists $n_2 \in \NN$ such that for $n \geq
n_2$,
\begin{equation} \EE \left( \tr \left( \sigma_{i,\uj}^{(n)}
P_{i,\uj}^{(n)} \right) \right) > 1-\delta^2, \end{equation} where
$\EE$ denotes the expectation with respect to the probability
distribution $\{p_{\uj}^{(n)}\}$ on the states $\rho_{\uj}^{(n)}$.
\end{lemma}

\begin{proof}
Define i.i.d. random variables $X_1,\dots,X_n$
with distribution given by \begin{equation} {\rm Prob}\,(X_r =
\lambda_{i+r-1,j,k}) = p_{i+r-1,j}\,\lambda_{i+r-1,j,k}.
\end{equation}  By the Weak Law of Large Numbers,
\begin{eqnarray} \frac{1}{n} \sum_{r=1}^n \log X_r &\to& \frac{1}{L}
\sum_{i=0}^{L-1} \sum_{j=1}^{J} \sum_k p_{j}\,\lambda_{i,j,k} \log
\lambda_{i,j,k} \non\\ &=& - \frac{1}{L} \sum_{i=1}^L
\sum_{j=1}^{J(i)} p_{j}\,S(\sigma_{i,j}) = -{\bar S}.
\end{eqnarray} It follows that there exists $n_2$ such that for $n
\geq n_2$, the typical set $T_{i,\epsilon}^{(n)}$ of sequences of
pairs $((j_1,k_1), \dots,(j_n,k_n))$ such that  \begin{equation}
\left| \frac{1}{n} \sum_{r=1}^n \log \lambda_{i+r-1,j_r,k_r} +
{\bar S} \right| < \frac{\epsilon}{3} \end{equation} satisfies
\begin{equation} \PP \left( T_{i,\epsilon}^{(n)} \right) =
\sum_{((j_1,k_1),\dots, (j_n,k_n)) \in T_{\epsilon}^{(n)}}
\prod_{r=1}^n p_{j_r} \lambda_{i+r-1j_r,k_r} > 1-\delta^2.
\end{equation} Obviously,
\begin{equation} P_{i,\uj}^{(n)} \geq
\sum_{\substack{\uk= (k_1, \ldots, k_n)\\ ((j_1,k_1),\dots, (j_n,k_n))
\in T_{i,\epsilon}^{(n)}}} |\psi_{\uj,\uk}^{(n)} \rangle \langle
\psi_{\uj,\uk}^{(n)} | \end{equation} and \begin{equation} \EE
\left( \tr \left( \sigma_{\uj}^{(n)} P_{i,\uj}^{(n)} \right)
\right) \geq \PP \left( T_{i,\epsilon}^{(n)} \right) > 1-\delta^2.
\end{equation}
\end{proof}

The remainder of the proof is essentially the same as that in \cite{DD07Per}. Let $N={\tilde N}(n)$ be the maximal number of product
states ${\tilde \rho}_1^{(n)}, \dots, {\tilde \rho}_N^{(n)}$ on
${\cal H}^{\otimes n}$ (each of which is a tensor product of
states in the maximising ensemble $\{p_j, \rho_j\}_{j=1}^J$) for
which there exist positive operators $E_1^{(n)}, \dots, E_N^{(n)}$
on ${\cal K}^{\otimes ML_2} \otimes {\cal K}^{\otimes n}$ such
that
\begin{enumerate}
\item[(i)] $ E_k^{(n)} = \sum_{i=1}^L {\tilde \Pi}_i \otimes
E_{k,i}^{(n)} $ and $ \sum_{k=1}^N E_{k,i}^{(n)} \leq {\bar
P}_i^{(n)}$ and  \item[(ii)] $ \disp{ \frac{1}{L} \sum_{i=1}^L
\tr \left[\,\left({\tilde \Pi}_i \otimes E_{k,i}^{(n)}
\right)\, \Phi_i^{(ML_2 + n)}\left( \omega^{(ML_2)} \otimes
{\tilde\rho}_k^{(n)} \right) \right]
> 1-\epsilon} $ and
\item[(iii)] $ \disp{ \frac{1}{L} \sum_{i=1}^L \tr
\left[\,\left( {\tilde \Pi}_i \otimes E_{k,i}^{(n)} \right)
\Phi_i^{(ML_2 + n)}\left( \omega^{(ML_2)} \otimes {\bar
\rho}^{\otimes n} \right) \right] \leq 2^{-n[C(\Phi) - \half
\epsilon]} }$
\end{enumerate}
where ${\bar\rho} = \sum_{j=1}^{J} p_{j} \rho_{j}$.

For each $i=1,\dots,M$ and $\uj = (j_1,\dots,j_n)$ such that $1
\leq j_r \leq J$, we define, as before, \begin{equation}
V_{i,\uj}^{(n)} = \left( {\bar P}_i^{(n)} - \sum_{k=1}^N
E_{k,i}^{(n)} \right)^{1/2} {\bar P}_i^{(n)} P_{i,\uj}^{(n)} {\bar
P}_i^{(n)} \left( {\bar P}_i^{(n)} - \sum_{k=1}^N E_{k,i}^{(n)}
\right)^{1/2}.
\end{equation}
and we put \begin{equation} V_{\uj}^{(n)} := \sum_{i=1}^M {\tilde
\Pi}_i \otimes V_{i,\uj}^{(n)}.
\end{equation}
Clearly $ V_{i,\uj}^{(n)} \le {\bar P}_i^{(n)} - \sum_{k=1}^N
E_{k,i}^{(n)}$.

$V_{\uj}^{(n)}$ is a candidate for an additional measurement
operator, $E_{N+1}^{(n)}$, for Bob with corresponding input state
$ {\tilde \rho}_{N+1}^{(n)} = \rho_{\uj}^{(n)}= \rho_{j_1} \otimes
\rho_{j_2} \ldots \otimes \rho_{j_n}$. Clearly, the condition (i),
given above, is satisfied and we also have

\begin{lemma}
\begin{equation} \frac{1}{L} \sum_{i=1}^L
\tr \,\left[ \left( {\tilde \Pi}_i \otimes V_{i,\uj}^{(n)}
\right) \Phi_i^{(ML_2 + n)} \left( \omega^{(ML_2)} \otimes {\bar
\rho}^{\otimes n} \right)] \right] \leq 2^{-n[C(\Phi) - \half
\epsilon]}.
\end{equation}
\end{lemma}

\begin{proof}
Put $Q_{n,i} = \sum_{k=1}^{N} E_{k,i}^{(n)}$. Note
that $Q_{n,i}$ commutes with ${\bar P}_i^{(n)}$. Using the fact
that ${\bar P}_i^{(n)} \Phi^{(n)}_i ({\bar\rho}^{\otimes n}) {\bar
P}_i^{(n)} \leq 2^{-n [\frac{1}{L} \sum_{i=1}^L S({\bar
\sigma}_i)- \frac{1}{4}\epsilon]}$ by (\ref{Tbar}), we have,
denoting ${\bar\sigma}_i^{(n)} = \Phi^{(n)}_i({\bar\rho}^{\otimes
n})$,
\begin{eqnarray} \tr ( {\bar\sigma}_i^{(n)} V_{i,\uj}^{(n)})
&=& \tr \left[ {\bar \sigma}_i^{(n)} ({\bar P}_i^{(n)} -
Q_{n,i})^{1/2} {\bar P}_i^{(n)} P_{i,\uj}^{(n)} {\bar P}_i^{(n)}
({\bar P}_i^{(n)} - Q_{n,i})^{1/2} \right] \non
\\ &=& \tr \left[ {\bar P}_i^{(n)} {\bar \sigma}_i^{(n)}
{\bar P}_i^{(n)} ({\bar P}_i^{(n)} - Q_{n,i})^{1/2}
P_{i,\uj}^{(n)} ({\bar P}_i^{(n)} - Q_{n,i})^{1/2} \right] \non \\
&\leq & 2^{-n[\frac{1}{L} \sum_{i=1}^L S({\bar
\sigma}_i)-\frac{1}{4} \epsilon]} \tr \left[ ({\bar P}_i^{(n)}
- Q_{n,i})^{1/2} P_{i,\uj}^{(n)} ({\bar P}_i^{(n)} - Q_{n,i})^{1/2} \right] \non \\
&\leq & 2^{-n[\frac{1}{L} \sum_{i=1}^L S({\bar
\sigma}_i)-\frac{1}{4} \epsilon]} \tr\, (P_{i,\uj}^{(n)} ) \leq
2^{-n[\frac{1}{L} \sum_{i=1}^L S({\bar \sigma_i})- {\bar S} -
\frac{1}{2} \epsilon]},
\end{eqnarray} where, in the last inequality, we used the standard
upper bound on the dimension of the typical subspace, $
\tr(P_{i,\uj}^{(n)}) \leq 2^{n[{\bar S} + \frac{1}{4}
\epsilon]}$, which follows from Lemma \ref{AppL4}.
\end{proof}

By maximality of ${N}$ it now follows that the condition (ii)
above cannot hold, that is, \begin{equation} \frac{1}{L}
\sum_{i=1}^L \tr \left[\,\left({\tilde \Pi}_i \otimes
V_{i,\ul{j}}^{(n)} \right)\, \Phi_i^{(ML_2 + n)}\left(
\omega^{(ML_2)} \otimes \rho_{\ul{j}}^{(n)} \right) \right] \leq
1-\epsilon
\end{equation} for every $\uj$, and this yields the following
\begin{corollary}\label{corr1}
\begin{equation} \frac{1}{L} \sum_{i=1}^L \EE \left(
\tr \left[\,\left({\tilde \Pi}_i \otimes V_{i,\ul{j}}^{(n)}
\right)\, \Phi_i^{(ML_2 + n)}\left( \omega^{(ML_2)} \otimes
\rho_{\ul{j}}^{(n)} \right) \right] \right) \leq 1-\epsilon.
\end{equation}
\end{corollary}

\noindent We also need the following lemma
\begin{lemma} For all $\eta' > \delta^2 + 3 \delta$,
\begin{equation} \frac{1}{L} \sum_{i=1}^L \tr \left[\,
\left({\tilde \Pi}_i \otimes {\bar P}_i^{(n)} P_{i,\uj}^{(n)}
{\bar P}_i^{(n)} \right) \, \Phi_i^{(ML_2 + n)} \left(
\omega^{(ML_2)} \otimes \Phi_i^{(n)}(\rho_{\ul{j}}^{(n)}) \right)
\right] > 1-\eta'
\end{equation} if $n$ is large enough. \end{lemma}

\begin{proof}
This is proved as in \cite{DD07Per}.
\end{proof}

\begin{lemma} Assume $\eta' < \frac{1}{3} \epsilon$ and write
\begin{equation} Q_{n,i} = \sum_{k=1}^N E_{k,i}^{(n)}.
\end{equation}
Then for $n$ large enough,
\begin{equation} \frac{1}{L} \sum_{i=1}^L
\EE \left( \tr \left[\,\left({\tilde \Pi}_i \otimes Q_{n,i}
\right) \, \Phi_i^{(ML_2 + n)} \left( \omega^{(ML_2)} \otimes
\rho_{\ul{j}}^{(n)} \right) \right] \right) \geq \eta^{\prime 2}.
\end{equation}  \label{AppLemma7}
\end{lemma}

\begin{proof}
This follows as before from the previous lemma
using the Cauchy-Schwarz inequality.
\end{proof}

It now follows that for $n$ large enough, ${{\tilde N}(n)} \geq
(\eta')^2 \,2^{n[C(\Phi)- \half \epsilon]}.$ We take the
following states as codewords,
\begin{equation} \rho_k^{(ML_2+n)} = \omega^{(ML_2)} \otimes
{\tilde \rho}_k^{(n)}. \end{equation} For $n$ sufficiently large
we then have \begin{equation} {N_{n+ML_2}} = {\tilde N}(n) \geq
(\eta')^2\,2^{n[C(\Phi)- \half \epsilon]} \geq
2^{(ML_2+n)[C(\Phi)- \epsilon]}.
\end{equation}

To complete the proof we need to show that the set $E_k^{(n)}$
satisfies (\ref{error}). But this follows immediately from
condition (ii),
\begin{eqnarray} \lefteqn{\tr \left[ \Phi^{(ML_2+n)}
\left( \rho^{(ML_2+n)}_k \right) E_k^{(n)} \right] =} \non \\ &=&
\frac{1}{L} \sum_{i=0}^{L-1} \tr \left[\,\Phi_i^{\otimes
(ML_2+n)} \left( \omega^{(ML_2)} \otimes {\tilde \rho}^{(n)}_k
\right) \left( {\tilde \Pi}_i \otimes
E_{i,k}^{(n)}\right) \right] \non \\
&>& 1-\epsilon.
\end{eqnarray}

\end{proof}

We have now provided a proof for the direct part of the channel coding theorem for the periodic quantum channel introduced in Chapter 4. Note that in Chapter 5 we show that the strong converse theorem does not in fact hold for the periodic quantum channel.

\addcontentsline{toc}{chapter}{List of figures}
\listoffigures
\addcontentsline{toc}{chapter}{Bibliography}
\bibliographystyle{IEEEtran}
\bibliography{Bib}

\begin{thebibliography}{10}
\providecommand{\url}[1]{#1}
\csname url@samestyle\endcsname
\providecommand{\newblock}{\relax}
\providecommand{\bibinfo}[2]{#2}
\providecommand{\BIBentrySTDinterwordspacing}{\spaceskip=0pt\relax}
\providecommand{\BIBentryALTinterwordstretchfactor}{4}
\providecommand{\BIBentryALTinterwordspacing}{\spaceskip=\fontdimen2\font plus
\BIBentryALTinterwordstretchfactor\fontdimen3\font minus
  \fontdimen4\font\relax}
\providecommand{\BIBforeignlanguage}[2]{{%
\expandafter\ifx\csname l@#1\endcsname\relax
\typeout{** WARNING: IEEEtran.bst: No hyphenation pattern has been}%
\typeout{** loaded for the language `#1'. Using the pattern for}%
\typeout{** the default language instead.}%
\else
\language=\csname l@#1\endcsname
\fi
#2}}
\providecommand{\BIBdecl}{\relax}
\BIBdecl

\bibitem{Hol98}
A.~Holevo, ``The capacity of the quantum channel with general signal states,''
  \emph{IEEE Transactions on Information Theory}, vol.~44, pp. 269--273, 1998,
  arXiv:quant-ph/9611023.

\bibitem{SW97}
B.~Schumacher and M.~Westmoreland, ``Sending classical information via noisy
  quantum channels,'' \emph{Physical Review A}, vol.~56, pp. 131--138, 1997.

\bibitem{DD07Per}
N.~Datta and T.~Dorlas, ``The coding theorem for a class of quantum channels
  with long-term memory,'' \emph{Journal of Physics A}, vol.~40, pp.
  8147--8164, 2007, arXiv:quant-ph/0610049.

\bibitem{Winter99}
A.~Winter, ``Coding theorem and strong converse for quantum channels,''
  \emph{IEEE Transactions on Information Theory}, vol.~45, pp. 2481--2485,
  1999.

\bibitem{CoverThomas}
T.~M. Cover and J.~A. Thomas, \emph{Elements of Information Theory}.\hskip 1em
  plus 0.5em minus 0.4em\relax John Wiley and Sons, Inc., 1991.

\bibitem{Khinchin}
A.~I. Khinchin, \emph{Mathematical Foundations of Information Theory}.\hskip
  1em plus 0.5em minus 0.4em\relax Dover Publications, Inc., New York, 1957.

\bibitem{Shannon48}
C.~Shannon, ``A mathematical theory of communication,'' \emph{The Bell System
  Technical Journal}, vol.~27, pp. 379--423, 623--656, 1948.

\bibitem{Hastings09}
M.~Hastings, ``Superadditivity of communication capacity using entangled
  inputs,'' \emph{Nature Physics}, vol.~4, pp. 255--257, 2009, arXiv:0809.3972.

\bibitem{HW08}
P.~Hayden and A.~Winter, ``Counterexamples to the maximal p-norm
  multiplicativity conjecture for all $p > 1$,'' \emph{Communications in
  Mathematical Physics}, vol. 284, pp. 263--280, 2008, arXiv:0807.4753.

\bibitem{SY08}
G.~Smith and J.~Yard, ``Quantum communication with zero-capacity channels,''
  \emph{Science}, vol. 321, pp. 1812--1815, 2008, arXiv:0807.4935.

\bibitem{CCH09}
T.~Cubitt, J.~Chen, and A.~Harrow, ``Superactivation of the asymptotic
  zero-error classical capacity of a quantum channel,'' arXiv:0906.2547.

\bibitem{SS09}
G.~Smith and J.~Smolin, ``Can nonprivate channels transmit quantum
  information?'' \emph{Physical Review Letters}, vol. 102, p. 010501, 2009,
  arXiv:0810.0276.

\bibitem{LWZG09}
K.~Li, A.~Winter, X.~Zou, and G.~Guo, ``Private capacity quantum channels is
  not additive,'' \emph{Physical Review Letters}, vol. 103, p. 120501, 2009,
  arXiv:0903.4308.

\bibitem{NC}
M.~A. Nielsen and I.~L. Chuang, \emph{Quantum Computation and Quantum
  Information}.\hskip 1em plus 0.5em minus 0.4em\relax Cambridge: Cambridge
  University Press, 2000.

\bibitem{Feinstein}
A.~Feinstein, ``A new basic theorem of information theory,'' \emph{IRE
  Transactions on Information Theory}, vol. PGIT-4, pp. 2--22, 1954.

\bibitem{BB84}
C.~H. Bennett and G.~Brassard, ``Quantum cryptography: Public-key distribution
  and coin tossing,'' \emph{Proceedings of the IEEE International Conference on
  Computers, Systems and Signal Processing}, pp. 175--179, 1984.

\bibitem{LGM09}
C.~Lupo, V.~Giovannetti, and S.~Mancini, ``Capacities of lossy bosonic memory
  channels,'' arXiv:0903.2764.

\bibitem{BDF09}
G.~Benenti, A.~D'Arrigo, and G.~Falci, ``Enhancement of transmission rates in
  quantum memory channels with damping,'' \emph{Physical Review Letters}, vol.
  103, p. 020502, 2009, arXiv:0903.1424.

\bibitem{DM09}
T.~Dorlas and C.~Morgan, ``Classical capacity of quantum channels with
  memory,'' \emph{Physical Review A}, vol.~79, p. 032320, 2009,
  arXiv:0902.2834.

\bibitem{BNS98}
H.~Barnum, M.~Nielsen, and B.~Schumacher, ``Information transmission through a
  noisy quantum channel,'' \emph{Physical Review A}, vol.~57, pp. 4153--4175,
  1998, arXiv:quant-ph/9702049.

\bibitem{DM08}
T.~Dorlas and C.~Morgan, ``Calculating a maximizer for quantum mutual
  information,'' \emph{International Journal of Quantum Information}, vol.~6,
  pp. 745--750, 2008.

\bibitem{Kraus83}
K.~Kraus, \emph{States, Effects and Operations, Fundamental Notions of Quantum
  Theory}, ser. Lecture Notes in Physics.\hskip 1em plus 0.5em minus
  0.4em\relax Berlin Springer Verlag, 1983.

\bibitem{Hayashi}
M.~Hayashi, \emph{Quantum Information: An Introduction}.\hskip 1em plus 0.5em
  minus 0.4em\relax Springer-Verlag Berlin Heidelberg: Springer, 2006.

\bibitem{HHHH09}
R.~Horodecki, P.~Horodecki, M.~Horodecki, and K.~Horodecki, ``Quantum
  entanglement,'' \emph{Reviews of Modern Physics}, vol.~81, p. 865, 2009,
  arXiv:quant-ph/0702225.

\bibitem{SW98}
B.~Schumacher and M.~D. Westmoreland, ``Quantum privacy and quantum
  coherence,'' \emph{Physical Review Letters}, vol.~80, pp. 5695--5697, 1998.

\bibitem{Hol73}
A.~Holevo, ``Statistical problems in quantum physics,'' \emph{Proceedings of
  the Second Japan-USSR Symposium on Probability Theory, Springer Berlin/
  Heidelberg}, vol. 330, pp. 104--119, 1973.

\bibitem{SN96}
B.~Schumacher and M.~Nielsen, ``Quantum data processing and error correction,''
  \emph{Physical Review A}, vol.~54, pp. 2629--2635, 1996,
  arXiv:quant-ph/9604022.

\bibitem{Schumacher96}
B.~Schumacher, ``Sending entanglement through noisy quantum channel,''
  \emph{Physical Review A}, vol.~54, pp. 2614--2628, 1996.

\bibitem{FBS96}
C.~Bennett, C.~Fuchs, and J.~Smolin, ``Entanglement-enhanced classical
  communication on a noisy quantum channel,'' in \emph{Quantum communication
  and quantum measurement}, O.~Hirota, A.~Holevo, and C.~Caves, Eds.\hskip 1em
  plus 0.5em minus 0.4em\relax Plenum, New York, 1997, pp. 79--88,
  arXiv:quant-ph/9611006.

\bibitem{SW01}
B.~Schumacher and M.~Westmoreland, ``Optimal signal ensembles,'' \emph{Physical
  Review A}, vol.~63, p. 022308, 2001, arXiv:quant-ph/9912122.

\bibitem{Holevo73PD}
A.~Kholevo, ``Bounds for the quantity of information transmitted by a quantum
  communication channel,'' \emph{Problems in Information Transmission}, vol.~9,
  p. 177, 1973.

\bibitem{PW91}
A.~Peres and W.~Wooters, ``Optimal detection of quantum information,''
  \emph{Physical Review Letters}, vol.~66, p. 1119, 1991.

\bibitem{HJSWW96}
P.~Hausladen, R.~Jozsa, B.~Schumacher, M.~Westmoreland, and W.~K. Wootters,
  ``Classical information capacity of a quantum channel,'' \emph{Physical
  Review A}, vol.~54, pp. 1869--1876, 1996.

\bibitem{Davies78}
E.~Davies, ``Information and quantum measurement,'' \emph{IEEE Transactions on
  Information Theory}, vol.~24, pp. 596--599, 1978.

\bibitem{Eggleston}
H.~Eggleston, \emph{Convexity}.\hskip 1em plus 0.5em minus 0.4em\relax
  Cambridge University Press, 1958.

\bibitem{Grunbaum}
B.~Grunbaum, \emph{Convex polytopes}.\hskip 1em plus 0.5em minus 0.4em\relax
  Interscience Publishers, 1967.

\bibitem{BM04}
G.~Bowen and S.~Mancini, ``Quantum channels with a finite memory,''
  \emph{Physical Review A}, vol.~69, p. 012306, 2004, arXiv:quant-ph/0305010.

\bibitem{GM04}
V.~Giovannetti and S.~Mancini, ``Bosonic memory channels,'' \emph{Physical
  Review A}, vol.~71, p. 062304, 2004, arXiv:quant-ph/0410176.

\bibitem{Mancini06}
S.~Mancini, ``Models for quantum memory channels,'' \emph{Journal of Physics:
  Conference Series}, vol.~36, pp. 121--125, 2006.

\bibitem{Norris}
J.~Norris, \emph{Markov Chains}.\hskip 1em plus 0.5em minus 0.4em\relax
  Cambridge: Cambridge University Press, 1997.

\bibitem{Ahlswede68}
R.~Ahlswede, ``The weak capacity of averaged channels,'' \emph{Zeitschrift
  f\"ur Wahrscheinlichkeitstheorie und Verwandte Gebiete}, vol.~11, pp. 61--73,
  1968.

\bibitem{Jacobs62}
K.~Jacobs, ``Almost periodic channels,'' \emph{Colloquium on Combinatorial
  Methods in Probabilistic Theory (Ahrhus)}, 1962.

\bibitem{Fuchs97}
C.~Fuchs, ``Nonorthogonal quantum states maximize classical information
  capacity,'' \emph{Physical Review Letters}, vol.~79, pp. 1162--1165, 1997,
  arXiv:quant-ph/9703043.

\bibitem{KR01}
C.~King and M.~Ruskai, ``Minimal entropy of states emerging from noisy quantum
  channel,'' \emph{IEEE Transactions on Information Theory}, vol.~47, pp.
  192--209, 2001, arXiv:quant-ph/9911079.

\bibitem{RSW02}
M.~B. Ruskai, S.~Szarek, and E.~Werner, ``An analysis of completely-positive
  trace-preserving maps on $2 \times 2$ matrices,'' \emph{Linear Algebra and
  its Applications}, vol. 347, pp. 159--187, 2002, arXiv:quant-ph/0101003.

\bibitem{King03}
C.~King, ``The capacity of the quantum depolarizing channel,'' \emph{IEEE
  Transactions on Information Theory}, vol.~49, pp. 221--229, 2003,
  arXiv:quant-ph/0204172.

\bibitem{Shor04}
P.~W. Shor, ``Equivalence of additivity questions in quantum information
  theory,'' \emph{Communications in Mathematical Physics}, vol. 246, pp.
  453--472, 2004, arXiv:quant-ph/0305035.

\bibitem{Pomeransky03}
A.~A. Pomeransky, ``Strong superadditivity of the entanglement of formation
  follows from its additivity,'' \emph{Physical Review A}, vol.~68, p. 032317,
  2003, arXiv:quant-ph/0305056.

\bibitem{Fukuda07}
M.~Fukuda, ``Simplification of additivity conjecture in quantum information
  theory,'' \emph{Quantum Information Processing}, vol.~6, pp. 179--186, 2007,
  arXiv:quant-ph/0608010.

\bibitem{BDiVSWK96}
C.~Bennett, D.~DiVincenzo, J.~Smolin, and W.~Wootters, ``Mixed state
  entanglement and quantum error correction,'' \emph{Physical Review A},
  vol.~54, pp. 3824--3851, 1996, arXiv:quant-ph/9604024.

\bibitem{KingUnital}
C.~King, ``Additivity for unital qubit channels,'' \emph{Journal of
  Mathematical Physics}, vol.~43, pp. 4641--4653, 2002, arXiv:quant-ph/0103156.

\bibitem{Shor02}
P.~W. Shor, ``Additivity of the classical capacity of entanglement-breaking
  quantum channels,'' \emph{Journal of Mathematical Physics}, vol.~43, pp.
  4334--4340, 2002, arXiv:quant-ph/0201149.

\bibitem{Keyl02}
M.~Keyl, ``Fundamentals of quantum information theory,'' \emph{Physics
  Reports}, vol. 369, pp. 431--548, 2002, arXiv:quant-ph/0202122.

\bibitem{MP02}
C.~Macchiavello and G.~M. Palma, ``Entanglement enhanced information
  transmission over a quantum channel with correlated noise,'' \emph{Physical
  Review A}, vol.~65, p. 050301, 2002, arXiv:quant-ph/0107052.

\bibitem{Ham02}
M.~Hamada, ``A lower bound on the quantum capacity of channels with correlated
  errors,'' \emph{Journal of Mathematical Physics}, vol.~43, pp. 4382--4390,
  2002, arXiv:quant-ph/0201056.

\bibitem{HN03}
M.~Hayashi and H.~Nagaoka, ``General formulas for capacity of classical-quantum
  channels,'' \emph{IEEE Transactions ion Information Theory}, vol.~49, pp.
  1753--1768, 2003, arXiv:quant-ph/0206186.

\bibitem{MPV03}
C.~Macchiavello, G.~M. Palma, and S.~Virmani, ``Entangled states maximize the
  two qubit channel capacity for some pauli channels with memory,''
  \emph{Physical Review A}, vol.~69, p. 010303, 2004, arXiv:quant-ph/0307016.

\bibitem{DWMcI03}
S.~Daffer, K.~W\'odkiewicz, and J.~K. McIver, ``Quantum markov channels for
  qubits,'' \emph{Physical Review A}, vol.~67, p. 062312, 2003,
  arXiv:quant-ph/0211001.

\bibitem{CJCM04}
N.~J. Cerf, J.~Clavareau, C.~Macchiavello, and J.~Roland, ``Quantum
  entanglement enhances the capacity of bosonic channels with memory,''
  \emph{Physical Review A}, vol.~72, p. 042330, 2005, arXiv:quant-ph/0412089.

\bibitem{RSGM05}
G.~Ruggeri, G.~Soliani, V.~Giovannetti, and S.~Mancini, ``Information
  transmission through lossy bosonic memory channels,'' \emph{Europhysics
  Letters}, vol.~70, p. 719, 2005, arXiv:quant-ph/0502093.

\bibitem{Bose03}
S.~Bose, ``Quantum communication through an unmodulated spin chain,''
  \emph{Physical Review Letters}, vol.~91, pp. 207\,901--207\,905, 2003,
  arXiv:quant-ph/0212041.

\bibitem{SDK09}
J.~Sch\"{a}fer, D.~Daems, and E.~Karpov, ``Capacity of a bosonic memory channel
  with gauss-markov noise,'' \emph{Physical Review A}, vol.~80, p. 062313,
  2009, arXiv:0907.0982.

\bibitem{DD07CC}
N.~Datta and T.~Dorlas, ``Classical capacity of quantum channels with general
  markovian correlated noise,'' \emph{Journal of Statistical Physics}, vol.
  134, pp. 1173--1195, 2009, arXiv:0712.0722.

\bibitem{Hol72}
A.~S. Holevo, ``On the mathematical theory of quantum communication channels,''
  \emph{Problems in Information Transmission}, vol.~8, pp. 62--71, 1972.

\bibitem{Vedral02}
V.~Vedral, ``The role of relative entropy in quantum information theory,''
  \emph{Reviews of Modern Physics}, vol.~74, pp. 197--234, 2002,
  arXiv:quant-ph/0102094.

\bibitem{HolevoSurvey98}
A.~Holevo, ``Coding theorems for quantum channels,'' \emph{Tamagawa University
  Research Review}, vol.~4, pp. 1--33, 1998, arXiv:quant-ph/9809023.

\bibitem{KW09}
R.~K\"{o}nig and S.~Wehner, ``A strong converse for classical channel coding
  using entangled inputs,'' \emph{Physical Review Letters}, vol. 103, p.
  070504, 2009, arXiv:0903.2838.

\bibitem{ON99}
T.~Ogawa and H.~Nagaoka, ``Strong converse to the quantum channel coding
  theorem,'' \emph{IEEE Transactions on Information Theory}, vol.~45, pp.
  2486--2489, 1999, arXiv:quant-ph/9808063.

\bibitem{Wolfowitz}
J.~Wolfowitz, \emph{Coding Theorems of Inforation Theory}.\hskip 1em plus 0.5em
  minus 0.4em\relax Berlin, Germany: Springer, 1964.

\bibitem{Csiszar98}
I.~Csisz\'ar, ``The method of types,'' \emph{IEEE Transactions on Information
  Theory}, vol.~44, pp. 2505--2523, 1998.

\bibitem{Datta}
N.~Datta, ``Proof that $d^2$ states are sufficient,'' private communication.

\bibitem{Caratheodory}
C.~Caratheodory, ``\"{U}ber den variabilit\"{a}tsbereich der fourierschen
  konstanten von positiven harmonischen funktionen,'' \emph{Rendiconti del
  Circolo Matamatico di Palermo}, vol.~32, pp. 193--217, 1911.

\bibitem{Steinitz}
E.~Steinitz, ``Bedingt konvergente reihen und konvexe systeme,'' \emph{Journal
  f\"{u}r die reine und angewandte Mathematik}, vol. 143, pp. 128--175, 1913.

\end{thebibliography}

\end{document}